\documentclass[letter,12pt]{report}

\usepackage{amsmath,amssymb}             
\usepackage[latin1]{inputenc}
\usepackage[OT1]{fontenc}
\usepackage[left=2.5cm,right=2.5cm,top=2cm,bottom=2cm,includefoot,includehead,headheight=13.6pt]{geometry}
\usepackage{setspace}
\usepackage{lineno}
\usepackage{footmisc}
\usepackage{indentfirst}
\usepackage{siunitx}
\usepackage{lmodern}
\usepackage{bm}
\usepackage{float}
\usepackage{tabu}
\usepackage{multirow}

\usepackage{graphicx,type1cm,eso-pic,color}

\usepackage{color}
\definecolor{linkcol}{rgb}{0,0,0.4} 
\definecolor{citecol}{rgb}{0.5,0,0} 

\usepackage{array}
\newcolumntype{L}[1]{>{\raggedright\let\newline\\\arraybackslash\hspace{0pt}}m{#1}}
\newcolumntype{C}[1]{>{\centering\let\newline\\\arraybackslash\hspace{0pt}}m{#1}}
\newcolumntype{R}[1]{>{\raggedleft\let\newline\\\arraybackslash\hspace{0pt}}m{#1}}

\newcommand{\xb}{$x_{Bj}~$}

\usepackage{rotating}                    
\usepackage{fancyhdr}                    

\let\headruleORIG\headrule
\renewcommand{\headrule}{\color{black} \headruleORIG}

\usepackage{colortbl}
\arrayrulecolor{black}

\fancypagestyle{plain}{
  \fancyhead{}
  \fancyfoot{}
  
}

\makeatletter

\def\cleardoublepage{\clearpage\if@twoside \ifodd\c@page\else%
  \hbox{}%
  \thispagestyle{empty}
  \newpage%
  \if@twocolumn\hbox{}\newpage\fi\fi\fi}

\makeatother
 
{%

\hrulefill
\vspace*{0.5cm}%
\end{minipage}
}

{ \begin{list}%
	{$\bullet$}%
	{\setlength{\labelwidth}{25pt}%
	 \setlength{\leftmargin}{30pt}%
	 \setlength{\itemsep}{\parsep}}}%
{ \end{list} }

\renewcommand{\epsilon}{\varepsilon}

\usepackage{hyperref}
\hypersetup
{
bookmarksopen=true,
pdftitle="Tagged EMC - ALERT Run Group Proposal",
pdfauthor="Rapha\"el Dupr\'e", 
pdfsubject="Tagged EMC - ALERT Run Group Proposal", 
pdfmenubar=true, 
pdfhighlight=/O, 
colorlinks=true, 
pdfpagemode=UseNone, 
pdfpagelayout=SinglePage, 
pdffitwindow=true, 
linkcolor=linkcol, 
citecolor=citecol, 
urlcolor=linkcol 
}

\begin{document}

\begin{titlepage}
     \begin{center}
       \vspace*{-1.8cm}
       \noindent \Huge \textbf{Tagged EMC Measurements on Light Nuclei} \\
     \end{center}
   
\renewcommand{\thefootnote}{\fnsymbol{footnote}}
     \begin{center}
       \vspace*{1.0cm}
       \noindent {W.~R.~Armstrong, J~Arrington, I.~Clo\"{e}t, K.~Hafidi\footnote[2]{Spokesperson}, 
       M.~Hattawy, D.~Potteveld, P. Reimer, Seamus Riordan, Z.~Yi} \\
       \vspace*{0.2cm}
       \noindent \emph{Argonne National Laboratory, Lemont, IL 60439, USA} \\
       \vspace*{0.7cm}
       \noindent {J. Ball, M. Defurne, M. Gar\c{c}on, H. Moutarde, S. Procureur, F. Sabati\'e} \\
       \vspace*{0.2cm}
       \noindent \emph{CEA, Centre de Saclay, Irfu/Service de Physique Nucl\'eaire, 91191 Gif-sur-Yvette, France} \\
       \vspace*{0.7cm}
       \noindent {W. Cosyn} \\
       \vspace*{0.2cm}
       \noindent \emph{Department of Physics and Astronomy, Proeftuinstraat 86, Ghent University, 9000 Ghent, Belgium} \\
       \vspace*{0.7cm}
       \noindent {M. Mazouz} \\
       \vspace*{0.2cm}
       \noindent \emph{Facult\'e des Sciences de Monastir, 5000 Tunisia} \\
       \vspace*{0.7cm}
       \noindent {A. Accardi} \\
       \vspace*{0.2cm}
       \noindent \emph{Hampton University, Hampton, VA 23668, USA} \\
       \vspace*{0.7cm}
       \noindent {J.~Bettane, 
                  R.~Dupr\'{e}$^\dagger$\footnote[3]{Contact person: dupre@ipno.in2p3.fr}, 
                  M.~Guidal, D.~Marchand, C.~Mu\~noz, S.~Niccolai, E.~Voutier} \\
       \vspace*{0.2cm}
       \noindent \emph{Institut de Physique Nucl\'eaire, CNRS-IN2P3, Univ. Paris-Sud, 
                       Universit\'e Paris-Saclay, 91406 Orsay Cedex, France} \\
       \vspace*{0.7cm}
       \noindent {K. P. Adhikari, J. A. Dunne, D. Dutta,  L. El Fassi,  L. Ye} \\
       \vspace*{0.2cm}
       \noindent \emph{Mississippi State University, Mississippi State, MS 39762, USA} \\
       \vspace*{0.7cm}
       \noindent {M. Amaryan, G.~Charles$^\dagger$, G. Dodge} \\
       \vspace*{0.2cm}
       \noindent \emph{Old Dominion University, Norfolk, VA 23529, USA} \\
       \vspace*{0.7cm}
       \noindent {V. Guzey} \\
       \vspace*{0.2cm}
       \noindent \emph{Petersburg Nuclear Physics Institute, National Research Center "Kurchatov Institute", 
                                  Gatchina, 188300, Russia} \\
       \vspace*{0.7cm}
       \noindent {N. Baltzell$^\dagger$, F.~X.~Girod, C.~Keppel, S.~Stepanyan} \\
       \vspace*{0.2cm}
       \noindent \emph{Thomas Jefferson National Accelerator Facility, Newport News, VA 23606, USA} \\
       \vspace*{0.7cm}
       \noindent {B.~Duran, S.~Joosten, Z.-E.~Meziani, M.~Paolone, M.~Rehfuss, N.~Sparveris} \\
       \vspace*{0.2cm}
       \noindent \emph{Temple University, Philadelphia, PA 19122, USA} \\
       \vspace*{0.7cm}
       \noindent {F. Cao, K. Joo, A. Kim, N. Markov} \\
       \vspace*{0.2cm}
       \noindent \emph{University of Connecticut, Storrs, CT 06269, USA} \\
       \vspace*{0.7cm}
       \noindent {C. Ciofi degli Atti, S. Scopetta} \\
       \vspace*{0.2cm}
       \noindent \emph{Universit\`a di Perugia, INFN, Italy} \\
       \vspace*{0.7cm}
       \noindent {W. Brooks, A. El-Alaoui} \\
       \vspace*{0.2cm}
       \noindent \emph{Universidad T\'ecnica Federico Santa Mar\'ia, Valpara\'iso, Chile} \\
       \vspace*{0.7cm}
       \noindent {S. Liuti} \\
       \vspace*{0.2cm}
       \noindent \emph{University of Virginia, Charlottesville, VA 22903, USA} \\
       \vspace*{1.1cm}
       \noindent {\Large \textbf{a CLAS Collaboration Proposal} } \\
      \end{center}
\renewcommand*{\thefootnote}{\arabic{footnote}}

\date{\today}

\end{titlepage}
\sloppy

\titlepage

\setcounter{page}{3}
      \renewcommand{\thefootnote}{\fnsymbol{footnote}}  
     \begin{center}
       \vspace*{-1.0cm}
      \noindent {\Large \textbf{Jefferson Lab PAC 45}} \\
      \vspace*{0.8cm}
       \noindent \Huge \textbf{Nuclear Exclusive and Semi-inclusive Measurements with a New CLAS12 Low Energy Recoil Tracker} \\
       \vspace*{0.8cm}
       \noindent \Large \textbf{ALERT Run Group\footnote[2]{Contact Person: Kawtar Hafidi (kawtar@anl.gov)} } \\      
       \vspace*{2.0cm}
       {\large\textbf{EXECUTIVE SUMMARY}}
     \end{center}
 
 \vspace*{0.4cm}

In this run group, we propose a comprehensive physics program to investigate 
the fundamental structure of the $^4$He nucleus. 
An important focus of this program is on 
the coherent exclusive Deep Virtual Compton Scattering (DVCS) and Deep 
Virtual Meson Production (DVMP) with emphasis on $\phi$ meson production. These are 
particularly powerful tools enabling model-independent nuclear 3D tomography 
through the access of partons' position in the transverse plane. These 
exclusive measurements will give the chance to compare directly 
the quark and gluon radii of the helium nucleus. 
Another important measurement proposed in this program is the study of the 
partonic structure of bound nucleons. To this end, we propose next generation 
nuclear measurements in which low energy recoil nuclei are detected. The 
tagging of recoil nuclei in deep inelastic reactions is a powerful technique, 
which will provide unique information about the nature of medium modifications 
through the measurement of the EMC ratio and its dependence on the nucleon 
off-shellness. 
Finally, we propose to measure incoherent spectator-tagged DVCS 
on light nuclei (d, $^4$He) where the observables are sensitive to the
Generalized Parton Distributions (GPDs) of a quasi-free neutron for the case of the deuteron, and bound proton and neutron for the case of $^4$He. The objective is to 
study and separate nuclear effects and their manifestation in GPDs.
The fully exclusive kinematics provide a novel approach for studying
final state interactions in the measurements of the 
beam spin asymmetries and the off-forward EMC ratio.\\

At the heart of this program is the Low Energy Recoil Tracker (ALERT) 
combined with the CLAS12 detector. The ALERT detector is composed of a stereo 
drift chamber for track reconstruction and an array of scintillators for 
particle identification. Coupling these two types of fast detectors will allow 
ALERT to be included in the trigger for efficient background rejection, while 
keeping the material budget as low as possible for low energy particle 
detection. ALERT will be installed inside the solenoid magnet instead of the 
CLAS12 Silicon Vertex Tracker and Micromegas tracker. We will use an 11 GeV 
longitudinally polarized electron beam (80\% polarization) of up to 1000~nA on a gas target 
straw filled with deuterium or $^4$He at 3 atm to obtain a luminosity up to
$6\times10^{34}$~nucleon~cm$^{-2}$s$^{-1}$. In addition we will need to run 
hydrogen and $^4$He targets at different beam energies for detector 
calibration. The following table summarizes our beam time request: \\

\newcommand{\minitab}[2][l]{\begin{tabular}{#1}#2\end{tabular}}
\begin{table}[ht!]
\label{tab:beamTimeRequest}
\center
\bgroup
\def\arraystretch{1.2}%
\tabulinesep=1.5mm
\begin{tabu}{C{3.1cm}C{2.8cm}C{1.6cm}C{2.3cm}C{1.6cm}C{2.5cm}}
\tabucline[2pt]{-}
\bf Configurations  & \bf Proposals & \bf Targets   & \bf Beam time request  & \bf Beam current & \bf Luminosity$^*$ \\
                    &                    &               & days    & nA       & n/cm$^{2}$\!/s     \\
\tabucline[1pt]{-}                                                   
{Commissioning}     & All$^\dagger$      & $^1$H, $^4$He & 5       & Various  & Various            \\
A                   & Nuclear GPDs       & $^4$He        & 10      & 1000      & $6\times10^{34}$   \\
B                   & Tagged EMC \& DVCS & $^2$H         & 20      & 500      & $3\times10^{34}$   \\
C                   & All$^\dagger$      & $^4$He        & 20      & 500      & $3\times10^{34}$   \\
\tabucline[1pt]{-}                                                   
{\bf TOTAL}         &                    & \,            & \bf 55  & \,       & \,                 \\
\tabucline[2pt]{-}
\end{tabu}
\egroup
\end{table}

\footnotetext[1]{This luminosity value is 
   based on the effective part of the target. When accounting for the target's 
   windows, which are outside of the ALERT detector, it is increased by 60\%.}
   
\footnotetext[2]{``All'' includes the four proposals of the run group: Nuclear GPDs, Tagged EMC, Tagged DVCS and Extra Topics. Note that the beam time request is only driven by the three first proposals.}
   
\renewcommand*{\thefootnote}{\arabic{footnote}}

\date{\today}

\sloppy

\titlepage

\renewcommand{\baselinestretch}{1.10}

\setcounter{page}{5}
\addcontentsline{toc}{chapter}{Abstract}

     \begin{center}
{\large\textbf{Abstract}}
    \end{center}
\vspace*{0.4cm}
We propose to measure tagged deep inelastic scattering from light 
nuclei (deuterium and $^4$He) by detecting the low energy nuclear spectator 
recoil (p, $^3$H and $^3$He) in addition to the scattered electron. The 
proposed experiment will provide stringent tests leading to clear 
differentiation between the many models describing the EMC effect, by
accessing the bound nucleon virtuality through its initial momentum at the point of 
interaction. Indeed, conventional nuclear physics explanations of the EMC effect
mainly based on Fermi motion and binding effects 
yield very different predictions than more exotic 
scenarios, where bound nucleons basically loose their identity when embedded 
in the nuclear medium. By distinguishing events where the interacting 
nucleon was slow, as described by a mean field scenario, or fast, very likely 
belonging to a correlated pair, will clearly indicate which phenomenon is 
relevant to explain the EMC effect. An important challenge for such measurements
using nuclear spectators is the control of the theoretical framework and,
in particular, final state interactions. This 
experiment will directly provide the necessary data needed to test our 
understanding of spectator tagging and final state interactions in $^2$H and 
$^4$He and their impact on the 
semi-inclusive measurements of the EMC effect described above.

\newpage

\tableofcontents

\chapter*{Introduction\markboth{\bf Introduction}{}}
\label{chap:intro}
\addcontentsline{toc}{chapter}{Introduction}
Inclusive electron scattering is a simple and yet a powerful tool to probe the 
structure of the nucleus; in the Deep Inelastic Scattering (DIS) regime it 
allows to access the partonic structure of hadrons. Using the nucleus as a target 
permits to study how nucleons and their parton distributions are modified when 
embedded in the nuclear environment. The modification of quark distributions 
in bound nucleons was first observed through the modification of the 
per-nucleon cross section in nuclei, known as the "EMC effect"~\cite{Aubert:1983xm}.
For moderate Bjorken~$x$, 0.35 $\leqslant$ $x_B$ $\leqslant$ 0.7, the 
per-nucleon DIS structure function for nuclei with A $\geqslant$ 3 was found 
to be suppressed compared to that of deuterium, the historic measurement being
confirmed and refined in the past 30 years~\cite{Ashman:1988bf, Arneodo:1988aa, 
Arneodo:1989sy, Gomez:1993ri, Allasia:1990nt, Seely2009}. \\

Since its discovery, the EMC effect has been a subject of extensive theoretical 
investigations aimed at understanding its underlying physics. While progress 
has been made in interpreting the main features of the effect, no single model 
has been able to explain convincingly the effect for both its $x_B$ and $A$ dependencies~%
\cite{Geesaman1995, Norton2003, Malace:2014uea}. A unifying understanding of 
the physical picture is still under intense debate. Most models of the EMC 
effect can be classified into two main categories:
\begin{itemize}
 \item ``Conventional'' nuclear models \cite{Ericson1983,Dunne1985,
Akulinichev1985,Jung1988} in which the effect could be understood by a reduced 
effective nucleon mass due to the nuclear binding, causing a shift of $x_B$ to 
higher values ($x_B$-rescaling or binding models). In these models the mass 
shift is sometimes accompanied by an increased density of virtual pions 
associated with the nuclear force (pion cloud models). 
 \item Models involving the change of the quark confinement size in the 
nuclear medium \cite{Close1983,Nachtmann1984,Jaffe1984,Close1988} can be 
viewed, in the language of QCD, as Q$^2$ rescaling models. In some cases, a 
simple increase of the nucleon radius is assumed (nucleon swelling), while in 
others, quark deconfinement is invoked and the nucleon degrees of freedom are 
replaced by multi-quark clusters. 
 \item Some more elaborate models fall in between or give very different 
predictions. We note here in particular the Point Like Configurations (PLC) 
suppression model as it gives direct predictions as a function of the nucleon
off-shellness. It was argued  
in~\cite{Frankfurt:1985cv} that PLCs are suppressed in 
bound nucleons and that large $x_B$ configurations in nucleons have smaller than 
average size leading to the EMC effect at large $x_B$. The EMC effect in this 
model is predicted to be proportional to the off-shellness of the struck nucleon 
and hence dominated by the contribution of the short-range correlations.
\end{itemize}

\begin{figure}[tbp]
  \begin{center}
    \includegraphics[angle=0, width=0.6\textwidth]{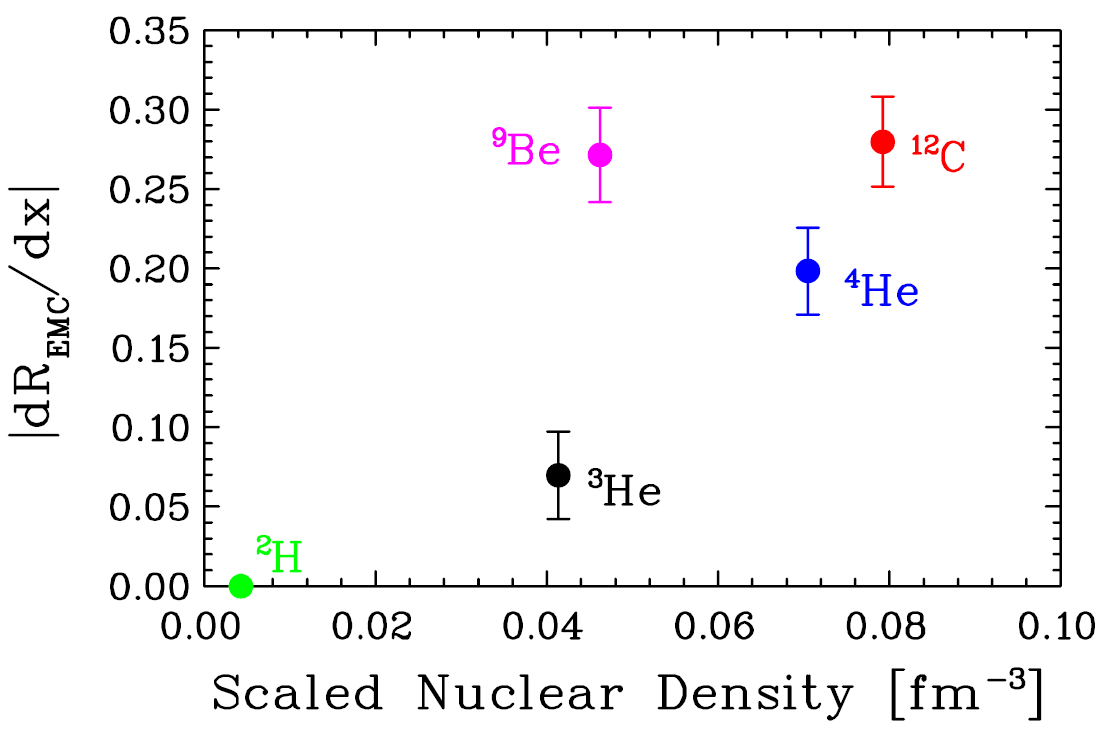}
    \caption{The slope of the isoscalar EMC ratio ({\it i.e.} the EMC ratio
    corrected for isospin asymmetry) for $0.35<x_B<0.7$ as a 
    function of nuclear density~\cite{Seely2009}.}
    \label{fig:ratio_hallc}
  \end{center}
\end{figure}

Recent experiment at Jefferson Lab measured the EMC effect for a series of 
light nuclei~\cite{Seely2009} changing significantly our understanding of
the EMC effect. The $^3$He EMC ratio was found to be 
roughly one third of the effect observed in $^4$He, violating the 
$A$-dependent fit to the SLAC data. Similarly, the density-dependent description
of the EMC effect has been contradicted by the large EMC effect 
found in $^9$Be (Figure~%
\ref{fig:ratio_hallc}). This suggests that the EMC effect may be sensitive to 
the local density experienced by a given nucleon or details of the nuclear 
structure, which has been first
suggested in~\cite{Frankfurt:1981mk}. Other models have also 
predicted a local EMC effect and describe the modification of the nucleons 
depending on their shells~\cite{Kumano1990,ciofiliuti1991,CiofidegliAtti1999}. 
The possibility of the EMC effect depending on the local environment of the 
nucleon also motivates the investigation of possible connections between the 
EMC effect and other density-dependent effects such as nucleon short-range 
correlations~\cite{Higinbotham:2010ye, Weinstein:2010rt}. \\

\begin{figure}[tb]
  \begin{center}
    \includegraphics[angle=0, width=0.5\textwidth]{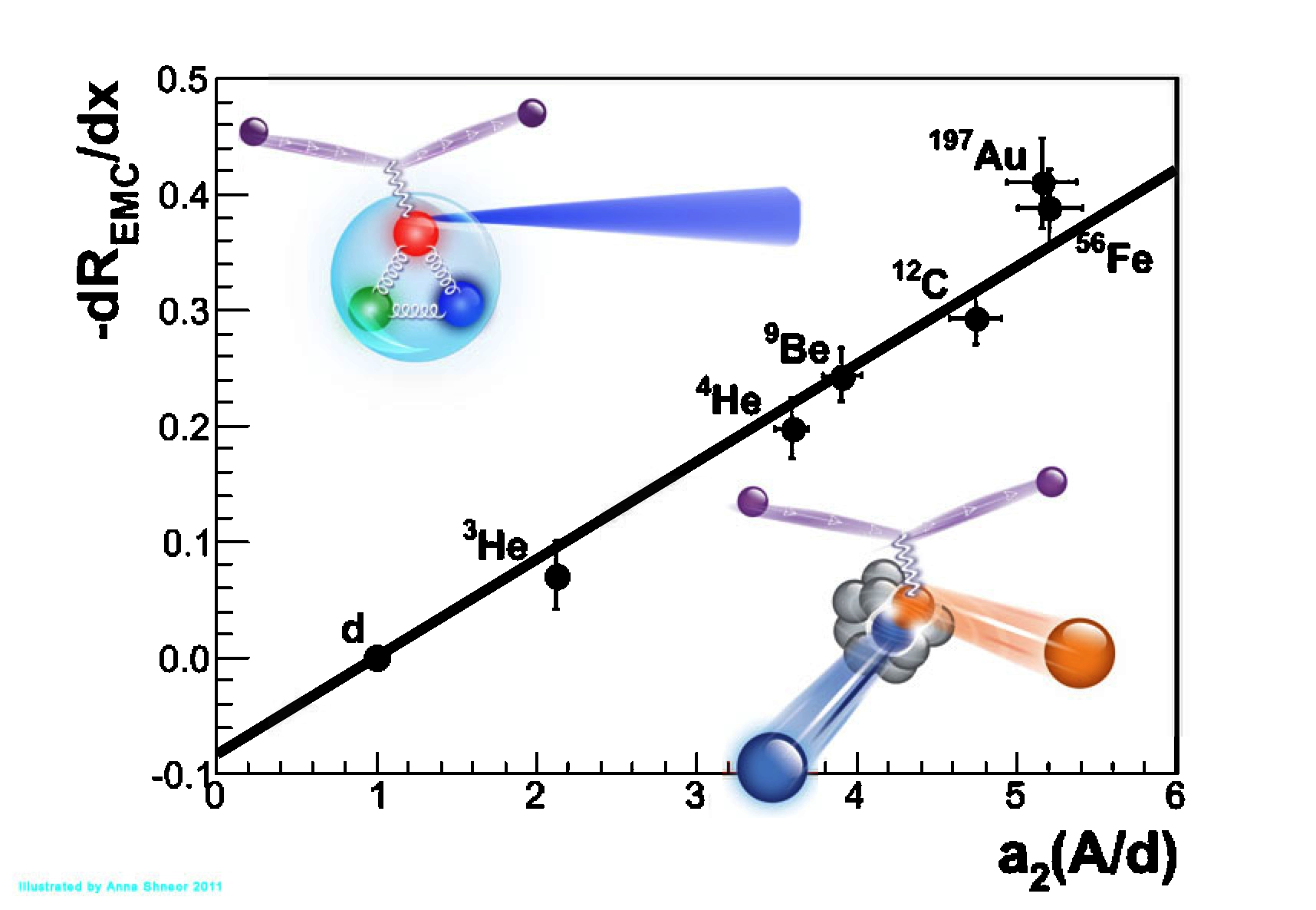}
    \caption{The EMC slopes versus the SRC scale factors 
    from~\cite{Geesaman:2015fha}. The uncertainties 
    include both statistical and systematic errors added in quadrature.}
    \label{fig:src-emc}
  \end{center}
\end{figure}

Short-range correlations (SRC) occur between nucleons located at less than the 
average inter-nucleon distance with high relative momentum~%
\cite{Frankfurt:1993sp, Egiyan:2003vg, Egiyan:2005hs}. Those pairs of nucleons 
carry 80\% of all nucleons kinetic energy inside the nucleus although they only 
represent about 20\% of the total number of nucleons~\cite{Hen:2014nza}. The 
plateau obtained in the inclusive 
cross section ratios of two nuclei (for example iron and deuterium) in the 
region $x_B > 1.5$ for $Q^2 > 1.5$ GeV$^2$ indicates that for nucleon momentum 
larger than Fermi momentum ($p\geqslant p_{N} \simeq$ 275 MeV/c), the nucleon 
momentum distributions in different nuclei have similar shapes but differ in 
magnitude. The ratio of the cross sections in the plateau region also called 
the "SRC scale factor $a_{2N}(A/d)$" was found to be linearly correlated with 
the slope of the EMC effect~\cite{Weinstein:2010rt}. This striking correlation 
shown in Figure~\ref{fig:src-emc} could indicate that high momentum bound 
nucleons are important players in the EMC effect. \\

To investigate the EMC effect further, we need to find ways of looking into the nuclei
with a sensitivity to the local density experienced by the nucleon we probe.  
This is possible by measuring the recoil fragment of the nucleus in addition 
to the scattered electron. Similarly to the SRC, the 
momentum of the recoil gives indications on the inter nucleon distance at 
the time of the interaction. Moreover, the momentum of this recoil can be linked to
the off-shellness of the interacting nucleon~\cite{CiofidegliAtti:2007vx} opening
new avenues to understand how the EMC effect arise in the nucleus. The two main challenges 
to perform such a measurement are, first, to be
able to detect the low energy nuclear fragments and, second, to understand the
Final State Interaction (FSI) effects on the observables. \\

The recent development of two small radial time projection chambers (RTPC) by
the CLAS collaboration for the measurement of the structure function of the 
neutron, by tagging the spectator proton from a deuterium target~\cite{Baillie:2011za},
and the measurement of coherent deep virtual Compton scattering off $^4$He~\cite{eg6_note}
has raised a lot of interest to use the same detector for this proposed tagged EMC 
measurement. However, it was found that the particle identification 
capabilities of the RTPC are not good enough to properly distinguish the 
different nuclear isotopes measured (in particular $^3$H from $^3$He). 
Therefore, we propose to use a different detector (a Low Energy Recoil 
Tracker - ALERT) based on a low gain drift chamber and a scintillator array for time
of flight measurements. Such a detector appears to be perfectly suited for our 
measurement as it offers low energy and large angle capabilities to measure slow recoils
similar to the RTPC. Moreover, such a detector will be much faster to collect the deposited charges
making it possible to include it in the trigger. This will allow to ignore 
the overwhelming majority of the events ($\sim90$\%) where the nuclear recoil
does not make it into the detection area and reduce the pressure on the DAQ. \\

Another important challenge for tagged measurements is to control the impact of
FSI on the observables. The large acceptance of both CLAS12 and ALERT is very
important in this regard as it allows to measure at the same time the regions
of the phase space expected to have negligible FSI and the regions where we expect a larger FSI effect.
The models used to correct for FSI~\cite{CiofidegliAtti2003,ciofi2004,
Alvioli:2006jd,Palli2009} can therefore be tested in
a wide kinematic range in the exact same conditions as the main measurement   
in order to ensure that the effect is understood properly. Then with the 
application of cuts to select the region where the effect is small, we can 
reduce the impact of FSI to a minimum. This
procedure allows to make sure FSI effects are small and under control to
minimize systematic uncertainty on our observables.

\chapter{The EMC effect in SIDIS}
\label{chap:physics}
At this point, it became clear that in order to advance our understanding of 
the EMC effect, it is necessary to study it as a function of new kinematic 
variables such as the nucleon 
off-shellness, which is accessible in semi inclusive measurements. Interests 
for slow nucleons and fragments tagging ($e+A \rightarrow e' + N + X$) studies 
are older than the EMC effect itself and has been identified earlier to be a 
promising tool to study nuclear effects~\cite{Frankfurt:1981mk,frankfurt1988,
CiofidegliAtti:1993ep}. In more recent work, Melnitchouk {\it et al.}~\cite{
Melnitchouk1997} showed that the tagged structure functions of deuteron in 
$(e,e'N_s)$ semi-inclusive reactions, where $N_s$ denotes the spectator 
nucleon, is a sensitive probe of the modification of the intrinsic structure 
of the bound nucleon allowing to discriminate between different EMC models. 
The extended case to heavier nuclei $A$, where the recoil nucleus $(A-1)$ is 
tagged was developed by Ciofi degli Atti {\it et al.}~\cite{CiofidegliAtti1999,
Palli2009,CiofidelgiAtti:2007qu,Atti:2010yf}, highlighting the importance of 
such measurements in the understanding the EMC effects. In this proposal, 
we want to investigate the origin of the medium induced modification of the nucleon 
structure function through several observables based on tagged DIS off deuterium
and helium targets.

\section{The spectator mechanism}
\begin{figure}[tbp]
  \begin{center}
    \includegraphics[angle=0, width=0.45\textwidth]{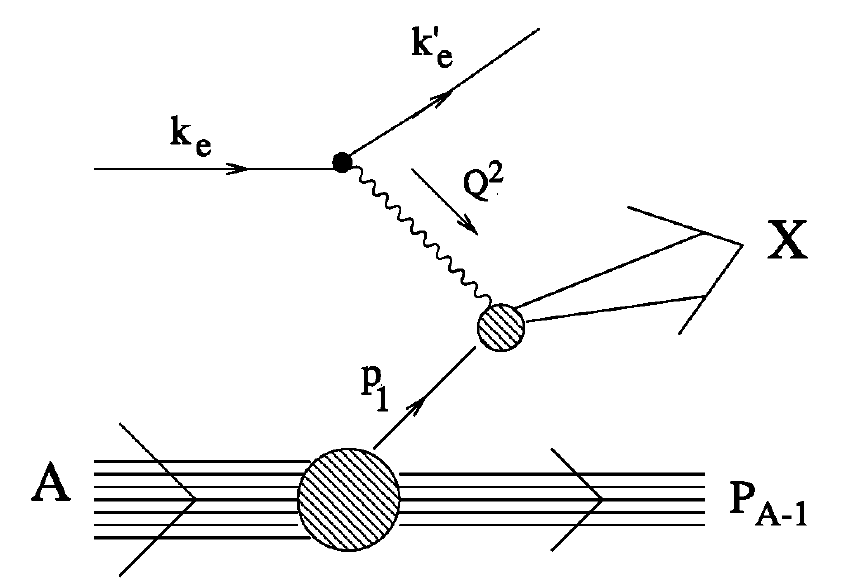}
    \caption{The process $A(e,e'(A-1))X$ within the impulse 
             approximation~\cite{CiofidegliAtti1999}.}
    \label{fig:spectator}
  \end{center}
\end{figure}

In the spectator mechanism or plane wave impulse approximation (PWIA), the DIS 
process corresponds to the absorption of the virtual photon by a quark inside 
a nucleon, followed by the recoil of the spectator nucleus $A-1$ without  
any final state interaction
(Figure~\ref{fig:spectator}). The differential semi-inclusive cross section can 
be written as~\cite{CiofidegliAtti1999}
\begin{equation}
\sigma_1^A (x_B,Q^2,\vec P_{A-1},y_A,z_1^A)= \frac{d^4\sigma}{dxdQ^2d\vec P_{A-1}}
 = K^A(x,y_A,Q^2,z_1^A)n_A(|\vec P_{A-1}|)z_1^AF_2^{N/A}(x_A,Q^2,p_1^2),
\label{eq:xs}
\end{equation}
where $Q^2 = -q^2 = -(k_e - k_e')^2 = \vec{q}^{\:2} - \nu$ is the four-momentum 
transfer, with $\vec{q} = \vec{k_e} - \vec{k_e'}$ and $\nu = E_e - E_e'$, 
$x_B = Q^2/2 M \nu$ is the Bjorken scaling variable, $p_1 \equiv (p_{10}, 
\vec{p_1})$, with $\vec{p_1} \equiv - \vec P_{A-1}$, is the four-momentum of the 
off-shell nucleon before its interaction with the virtual photon. $F_2^{N/A}$ is the DIS 
structure function of the nucleon $N$ in the nucleus $A$, $n_A(|\vec P_{A-1}|)$ 
is the three-momentum distribution of the bound nucleon, 
$z^{A}_{1} = {(p_1\cdot q)}/{M\nu}$ is the light cone momentum of the 
bound nucleon and $K^A$ is a kinematical factor given by 
\begin{equation}
K^A(x_B,y_A,Q^2,z_1^A) = { 4 \pi \alpha^2 \over Q^4 x_B } \cdot 
\left( {y \over y_A} \right) ^2 \times \left( {y_A^2 \over 2} + (1-y_A) - 
{p_1^2x_B^2y_A^2 \over (z_1^A)^2 Q^2 } \right),
\label{eq:ka}
\end{equation}
with $y= \nu / E_e$, $y_A = (p_1 \cdot q)/ (p_1 \cdot k_e)$ and 
$x_A = x_B / z_1^A$. \\

Nuclear effects in Eq.~\ref{eq:xs} are generated by the nucleon momentum 
distribution $n_A(|\vec P_{A-1}|)$ and by the quantities $y_A$ and 
$z^{A}_{1}$, which differ from the corresponding quantities for a free 
nucleon ($y= \nu / E_e$ and $z^{N}_{1} = 1$). In this framework the off-mass shellness of 
the nucleon ($p_1^2 \neq M^2$) generated by nuclear binding is taken into 
account within some small relativistic corrections when $A>2$~\cite{Melnitchouk:1993nk}.
In all the studies we propose here, it is important to ensure that the 
spectator mechanism is dominant and that scattering between spectator
nucleons and other reaction products is properly modeled. Our main goal here is to make sure
we understand the simple deuterium  case as well as the more complex helium
target. \\

\begin{figure}[tbp]
  \begin{center}
    \includegraphics[angle=0, width=0.5\textwidth]{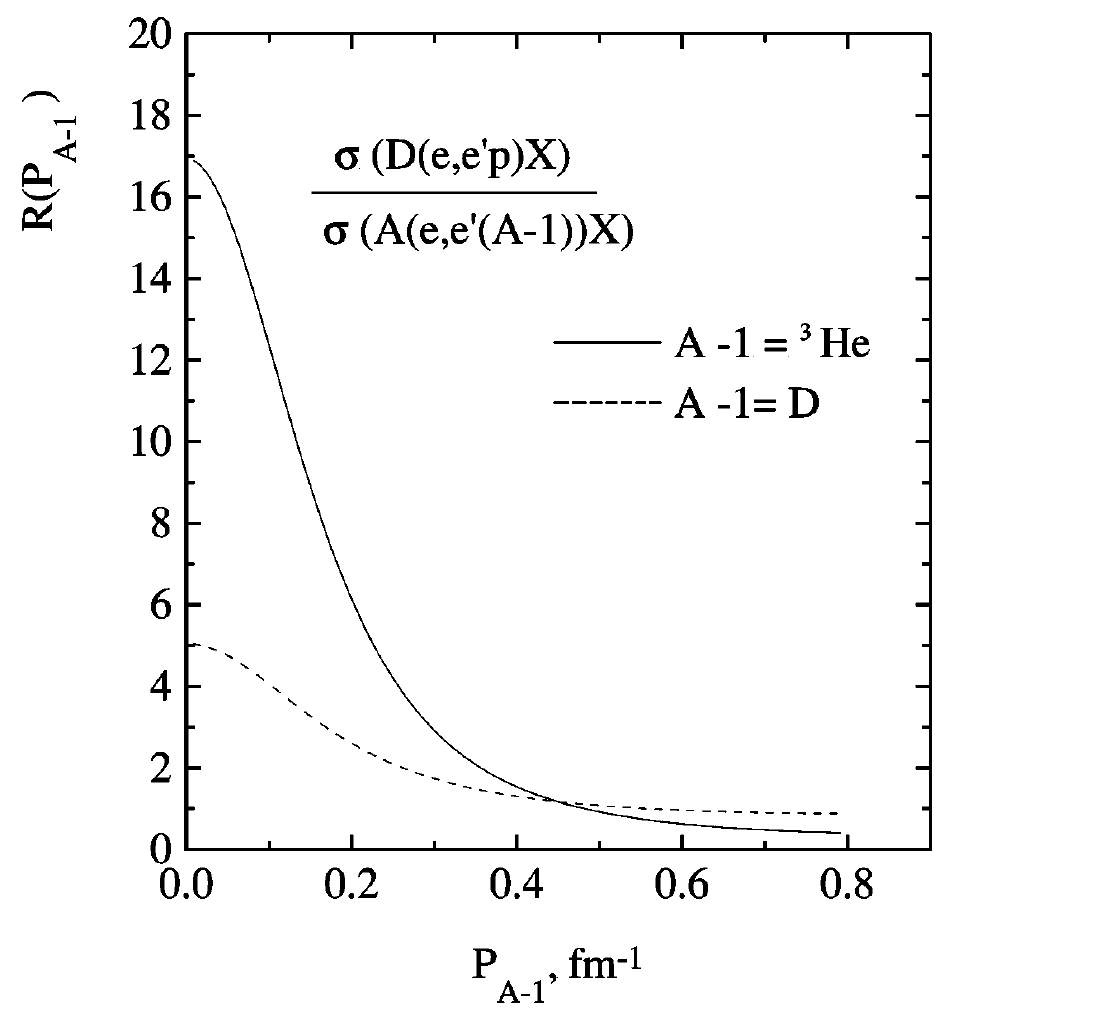}
    \caption{The ratio $R(A,A',|\vec P_{A-1}|)$ for the targets $^3$H (dashed) 
and $^4$He (full) as a function of the momentum of the backward emitted 
nucleus~\cite{CiofidegliAtti1999} relative to the virtual photon direction.}
    \label{fig:ratio_spec}
  \end{center}
\end{figure}

To test the spectator mechanism, we use the $\vec P_{A-1}$ dependence of 
semi-inclusive cross section ratio of different nuclei at the same values of 
$x_B$, $Q^2$ and with $|\vec P_{A'-1}| = |\vec P_{A-1}|$
\begin{equation}
R(x_B,Q^2,|\vec P_{A-1}|,z_1^A,z_1^{A'},y_A,y_{A'})\equiv 
\frac{\sigma_1^A(x_B,Q^2,|\vec P_{A-1}|,z_1^A,y_A)}
{\sigma_1^{A'}(x_B,Q^2,|\vec P_{A'-1}|,z_1^{A'},y_{A'})}.
\label{eq:ratio}
\end{equation}
In the Bjorken limit, the $A$ dependence of $R$ is expected to be entirely 
dominated by the nucleon momentum distribution 
$n_A(|\vec P_{A-1}|)$, which exhibits a strong $A$ dependence. Therefore, 
measurements of the $R$ ratio as a function of 
the recoil momentum $|\vec P_{A-1}|$ provide a strong test of the 
spectator mechanism independently of the model for $F_2^{N/A}$. Figure~%
\ref{fig:ratio_spec} illustrates the expected behavior of the ratio in Eq.~%
\ref{eq:ratio} from the processes D$(e, e' p)X$, $^3$He$(e, e' \mathrm{D})X$ and 
$^4$He$(e, e'^3 \mathrm{He})X$, as deep inelastic scattering off a bound 
neutron in different nuclei. \\

Detailed studies~\cite{%
CiofidegliAtti2003,ciofi2004,Alvioli:2006jd,Palli2009,Melnitchouk1997,
klimenko2006,Kaptari:2013dma} have shown 
that the FSI effects are minimized in the backward recoiling angle relative to 
the virtual photon direction and maximized in perpendicular kinematics.
From the previous discussion, it is clear that the observation of recoiling 
nuclei in the ground state, with a $|\vec P_{A-1}|$-dependence similar to the 
one predicted by the momentum distributions, would confirm these studies and the absence of 
significant FSI between the electroproduced hadronic states and the nuclear 
fragments.  

\section{EMC effect in deuteron}
\begin{figure}[tbp]
  \begin{center}
    \includegraphics[angle=0, width=0.5\textwidth]{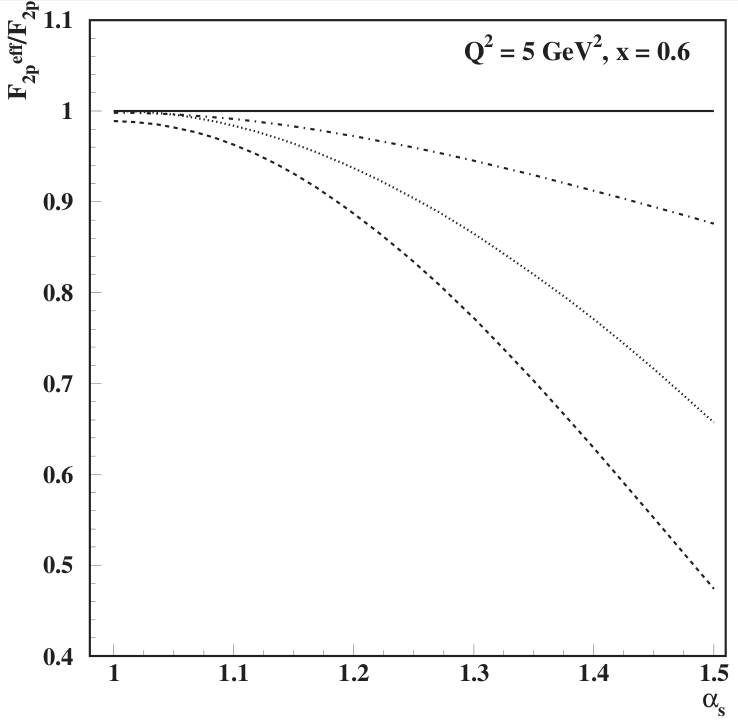}
    \caption{$F_{2p}^{eff}$ as a function of $\alpha_s$ for $x = 0.6$ and 
$p_T = 0$. Dashed line is a prediction for the PLC suppression model, dotted 
is for the $Q^2$-rescaling model, and dot-dashed for the binding/off-shell model~%
\cite{Melnitchouk1997}.}
    \label{fig:mel}
  \end{center}
\end{figure}
Recent work from Griffioen {\it et al.}~\cite{Griffioen:2015hxa} has shown that
the EMC effect is present, but very small, in the deuteron. However,
the recent finding of the possible dependence of the EMC effect on local nuclear
density raised the interest in using spectator nucleons to study the EMC 
effect. Indeed, the momentum of the spectator nucleon can be directly linked to 
the distance between the two nucleons in the deuteron~\cite{Melnitchouk:1993nk}
and therefore results in the enhancement of the EMC effect in certain configurations.\\

Melnitchouk {\it et al.} \cite{Melnitchouk1997} use the ratio of the effective 
$F^{eff}_{2p}$ measured in the deuterium, tagging the neutron, and compare it 
to the usual free $F_{2p}$. They predict significant effects for various 
models as a function of $\alpha \equiv \displaystyle{{E_s - p_z^s \over M}}$ 
(with $E_s$ and  $p_z^s$ the energy and longitudinal momentum of the 
spectator, respectively, and $M$ its mass), which characterizes the nucleon 
off-shellness. A semi-inclusive measurement will allow to discriminate between 
the very different model predictions (Figure \ref{fig:mel}).
\section{EMC effect in helium}
The process described for deuterium can be easily extended to heavier nuclei 
with several advantages. First, the nuclear effects in light nuclei, such as 
$^4$He, are much stronger, thus it enhances significantly the cross section 
for events with spectator momentum $\gtrsim 250$~MeV. Second, by detecting an 
intact light nucleus ($^3$H or $^3$He), we ensure that the final state 
interaction with the spectator is small and the contributions from the current 
or target fragmentation of the hard process are suppressed. On the down side, 
the theoretical calculations are more difficult, however recent theoretical 
progress indicates that these calculations although tedious could be 
performed~\cite{CiofidegliAtti2003,Alvioli:2006jd,Palli2009,CiofidegliAtti:1993ep,
Melnitchouk:1993nk}. \\

The quantity $R^A$ which is defined by:
\begin{equation}
R^A(x_B,x'_B,Q^2,|\vec P_{A-1}|) \equiv \frac{\sigma_1^A(x_B,Q^2,|\vec 
P_{A-1}|,z_1^{(A)},y_A)}{\sigma_1^A(x'_B,Q^2,|\vec P_{A-1}|,z_1^{(A)},y_A)},
\label{eq:ratio_a}
\end{equation}
represents the ratio between the cross sections on the nucleus $A$ at two 
different values of the Bjorken scaling variable. Due to the cancellation of 
all the other terms but the nucleon structure functions in Eq.~\ref{eq:ratio_a}, 
$R^A$ is highly sensitive to the nuclear effect. In the binding model 
($x$-rescaling), where the inclusive nuclear structure function is expressed 
through a convolution of the nuclear spectral function and the structure 
function of the bound nucleon, one has
\begin{equation}
R^A(x_B,x'_B,Q^2,|\vec P_{A-1}|) = \frac{x'_B}{x_B}\frac{F_2^{N/A}
(\frac{x_B}{z_1^A},Q^2)}{F_2^{N/A}(\frac{x'_B}{z_1^A},Q^2)}.
\label{eq:x_spec}
\end{equation}
In the $Q^2$-rescaling model~\cite{Close1988}, which is based on the medium 
modification of the $Q^2$-evolution equations of QCD and the assumption that 
the quark confinement radius for a bound nucleon is larger than the one for a free 
nucleon, the ratio becomes
\begin{equation}
R^A(x_B,x'_B,Q^2,|\vec P_{A-1}|) = \frac{x'_B}{x_B}\frac{F_2^{N/A}
(x_B,\xi_A(Q^2)Q^2)}{F_2^{N/A}(x'_B,\xi_A(Q^2)Q^2)}
\label{eq:Q_spec}
\end{equation}

\begin{figure}[tbp]
  \begin{center}
    \includegraphics[angle=0, width=0.4\textwidth]{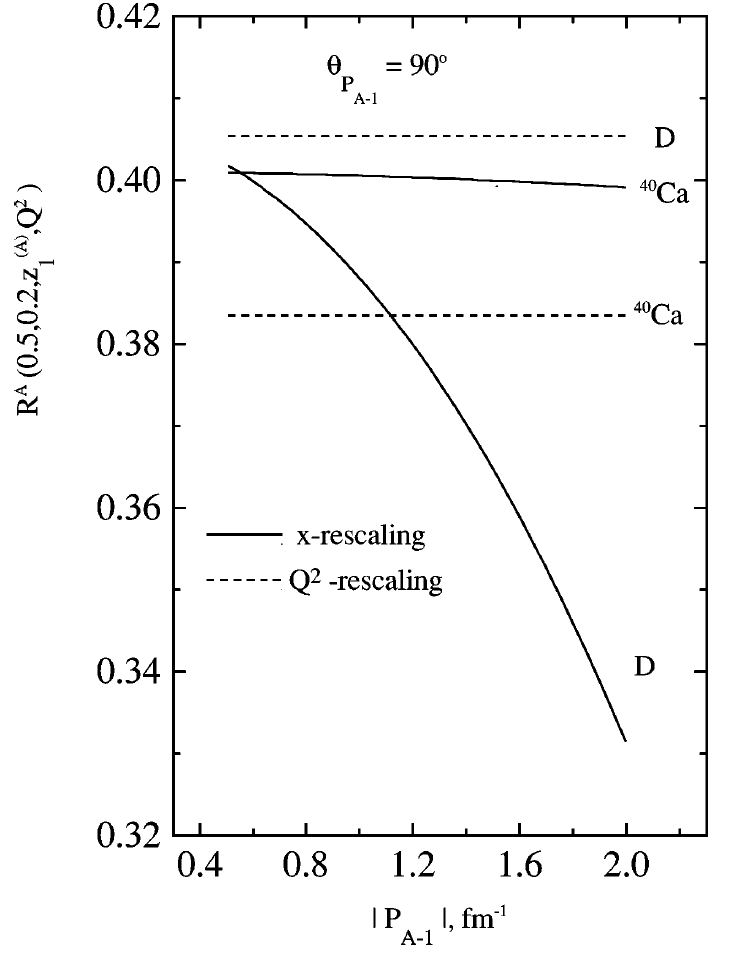}
    \includegraphics[angle=0, width=0.42\textwidth]{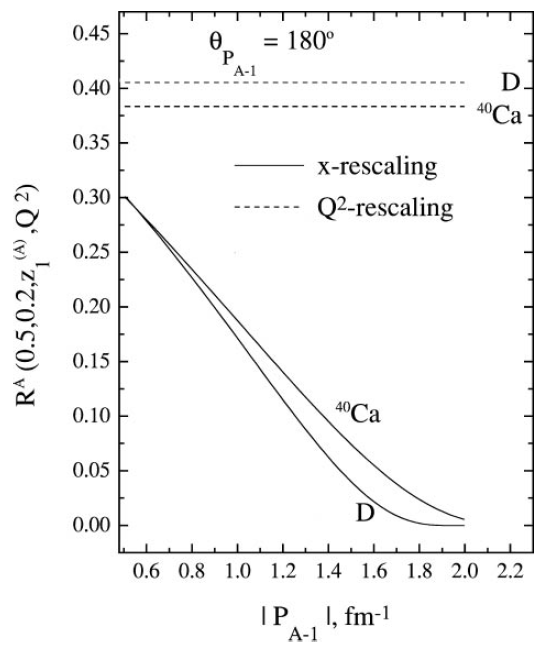}
    \caption{The ratio $R^A(x_B,x'_B)$ for $A = 2$ and $A = 40$, $x_B = 0.2$ 
and $x'_B = 0.5$, $Q^2 = 20$ GeV/$c$, plotted versus the momentum of the 
recoil nucleus ($A-1$) at perpendicular (left) and backward (right) angle 
$(\theta_{P_{A-1}} = 180^{\circ})$. The full and dashed curves are 
predictions of the $x_B$-rescaling (binding) and $Q^2$-rescaling models, 
respectively.}
    \label{fig:ratio_a}
  \end{center}
\end{figure}

While Eq.~\ref{eq:x_spec} is expected to depend both on $A$ and 
$|\vec P_{A-1}|$, Eq.~\ref{eq:Q_spec} would be a constant. By detecting nuclei 
with different recoil angles, this ratio would exhibit different behaviors, 
allowing a more detailed examination of the dynamics. Figure~\ref{fig:ratio_a}
shows theoretical predictions of the $R^A$ ratio in the $x$- and 
$Q^2$-rescaling models at both perpendicular and backward recoil kinematics.
We see there the power of discrimination of such measurement, between
a model that link the EMC effect directly to the spectator momentum and
one where it arises independently of it.
%
\section{Tagged EMC ratio}
\begin{figure}[tbp]
  \begin{center}
    \includegraphics[angle=0, width=0.5\textwidth]{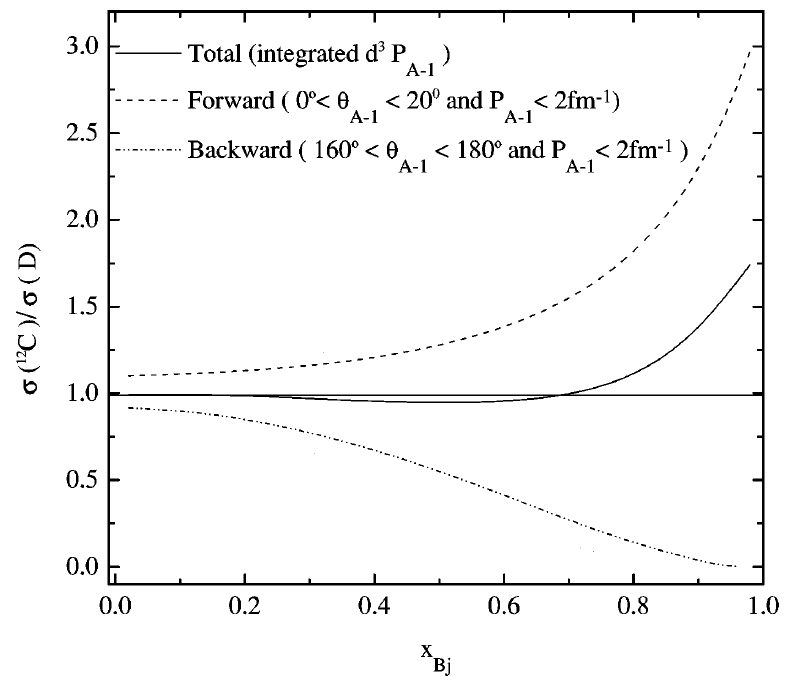}
    \caption{The semi-inclusive EMC ratio $R_0(x,Q^2)$ versus $x$ with 
nuclei emitted forward and backward, the full curve is the usual inclusive EMC 
ratio, the dashed and dotted curves are predictions for the local EMC effect 
for different spectator recoil angles~\cite{CiofidegliAtti1999} (see the 
legend).}
    \label{fig:r0}
  \end{center}
\end{figure}
Another observable used in theoretical calculations for the tagged EMC ratio is
\begin{equation}
R_0 (x,Q^2) =\frac{\int_a^b\sigma_1^Ad\vec P_{A-1}}{\int_a^b\sigma_1^Dd\vec P_{A-1}},
\label{eq:semi_emc}
\end{equation}
in which the cross section is integrated over a small momentum range of the 
recoil nucleus $\vec P_{A-1}$. In binding models it leads to opposite behavior 
for recoil nuclei emitted forward versus backward (Figure~\ref{fig:r0}) that 
cancels in the usual inclusive EMC ratio. These resulting deviations are much larger than the 
usual inclusive EMC effect and provides opportunity for a significant 
experimental test of the binding models.
\section{Flavor dependent parton distribution functions}
Measurements of tagged structure functions have been carried out in CLAS by the e6 
run group to study the EMC effect in deuteron \cite{klimenko2006} and later by 
the BONuS collaboration \cite{bonus6} to extract the $F^n_2$ structure 
function by tagging the low momentum recoil proton. The main goal of the BONuS 
measurements was to extract the ratio $F^n_2/F^p_2$ at high $x_B$ and therefore 
access the ratio of down to up quark distribution ($d/u$)~%
\cite{Baillie:2011za} at high $x_B$. By using $^4$He targets and tagging the recoiling 
$^3$He and $^3$H nuclei, one can select scattering off a weakly or deeply 
bound neutron and proton respectively depending on their off-shellness in the 
$^4$He nucleus. Since bound neutrons are always off-shell, even when 
$P_{^3He}=0$, an extrapolation procedure is needed to extract the free 
(i.e. on-shell) neutron structure function from the tagged recoil data~%
\cite{Sargsian:2005rm,Cosyn:2016oiq}. One could measure the $F_2$ structure functions of a 
weakly bound neutron in $^4$He and compare it to the $^2$H data to detect any 
nuclear dependence. This procedure is necessary for neutrons due to the 
absence of a free neutron target. It can also be quantitatively benchmarked 
using the $^4$He tagged data for scattering off a weakly bound proton, and 
comparing the results to the well measured free proton structure functions.\\

In addition, the ratio $\left(F_2^n/F_2^p \right)^{bound}/ \left(F_2^n/F_2^p 
\right)^{free}$ can be measured to extract the distributions of $d/u$ in a 
free nucleon and compare it to the same ratio for the bound nucleon. This is one 
way to explore the flavor dependent nuclear parton distributions which 
are little known experimentally. Such an effect, either in the anti-shadowing 
or EMC region, has been notably used to explain the NuTeV anomaly~\cite{
Brodsky:2004qa,Cloet2009}.
\section{Summary}
In this chapter, we have shown very strong theoretical motivations to measure 
the tagged structure function of nucleons in light nuclei such as deuterium and 
helium. The main difficulty being to properly handle the FSI. To solve this challenge, a 
large acceptance detector is necessary in order to demonstrate that data match 
models on a wide kinematic range in angle and momentum. Moreover, such large 
acceptance detector needs also to work at the lowest possible energy to ensure that 
quasi-free nucleons of low off-shellness can be effectively compared with the 
more virtual ones. Our proposition for such a detector is presented in the 
next chapter.

We presented theoretical work suggesting 
that a measurement on deuterium will already show an effect, however we have also showed
that higher nuclear masses provide much stronger signals and would ensure
a compelling measurement. Observing several nuclei in different kinematics 
lead to very different results for the classic pure \xb and $Q^2$-rescaling 
models. This will allow us to determine precisely which picture
or which combination of the two pictures is at the origin of the EMC effect.
This measurement will therefore provide a completely new insight into the
origin of the EMC effect and provide clear guidelines to build new models 
and better understand the partonic structure of nuclei.

In addition, our proposed experiment will allow to test the flavor symmetry
of the nuclear effects, which have been discussed by theoretical predictions in the
anti-shadowing and EMC regions. While this experiment is not dedicated to this 
question, for which additional isospin asymmetric targets would be necessary, 
we show that it can already provide a first test and pave the way for future works.

\setlength\parskip{\baselineskip}%
\chapter{Experimental Setup}
\label{chap:setting}
All the different measurements of the ALERT run group require, in addition to 
a good scattered electron measurement, the detection of low energy nuclear 
recoil fragments with a large kinematic coverage. Such measurements have been performed 
in CLAS (BONuS and eg6 runs), where the adequacy of a small additional detector
placed in the center of CLAS right around the target has shown to be the best 
solution. We propose here a similar setup using the CLAS12 spectrometer 
augmented by a low energy recoil detector. 

We summarize in Table~\ref{tab:req} the requirements for the different 
experiments proposed in the run group. By comparison with previous similar 
experiments, the proposed tagged measurements necessitate a 
good particle identification. Also, CLAS12 will be able to handle higher 
luminosity than CLAS so it will be key to exploit this feature in the future 
setting in order to keep our beam time request reasonable.

\begin{table}[ht!]
\centering
\footnotesize
\begin{tabu}{lccc}
\tabucline[2pt]{-}
Measurement  & Particles detected & $p$ range       & $\theta$ range                \rule[-7pt]{0pt}{20pt} \\
\tabucline[1pt]{-}                                                   
Nuclear GPDs & $^4$He             & $230 < p < 400 MeV/c$ & $\pi/4 < \theta < \pi/2$ rad  \rule[-7pt]{0pt}{20pt} \\
Tagged EMC   & p, $^3$H, $^3$He   & As low as possible    & As close to $\pi$ as possible \rule[-7pt]{0pt}{20pt} \\
Tagged DVCS  & p, $^3$H, $^3$He   & As low as possible    & As close to $\pi$ as possible \rule[-7pt]{0pt}{20pt} \\
\tabucline[2pt]{-}
\end{tabu}
\caption{Requirements for the detection of low momentum spectator fragments of the proposed measurements.}
\label{tab:req}
\end{table}

This chapter will begin with a brief description of CLAS12.  
After presenting the existing options for recoil detection and recognize that 
they will not fulfill the needs laid out above, we will describe the design of
the proposed new recoil detector ALERT. We will then present the reconstruction 
scheme of ALERT and show the first prototypes built by our technical 
teams. Finally, we specify the technical contributions of the different 
partners.

%
\section{The CLAS12 Spectrometer}
\begin{figure}
  \begin{center}
    \includegraphics[angle=0, width=0.75\textwidth]{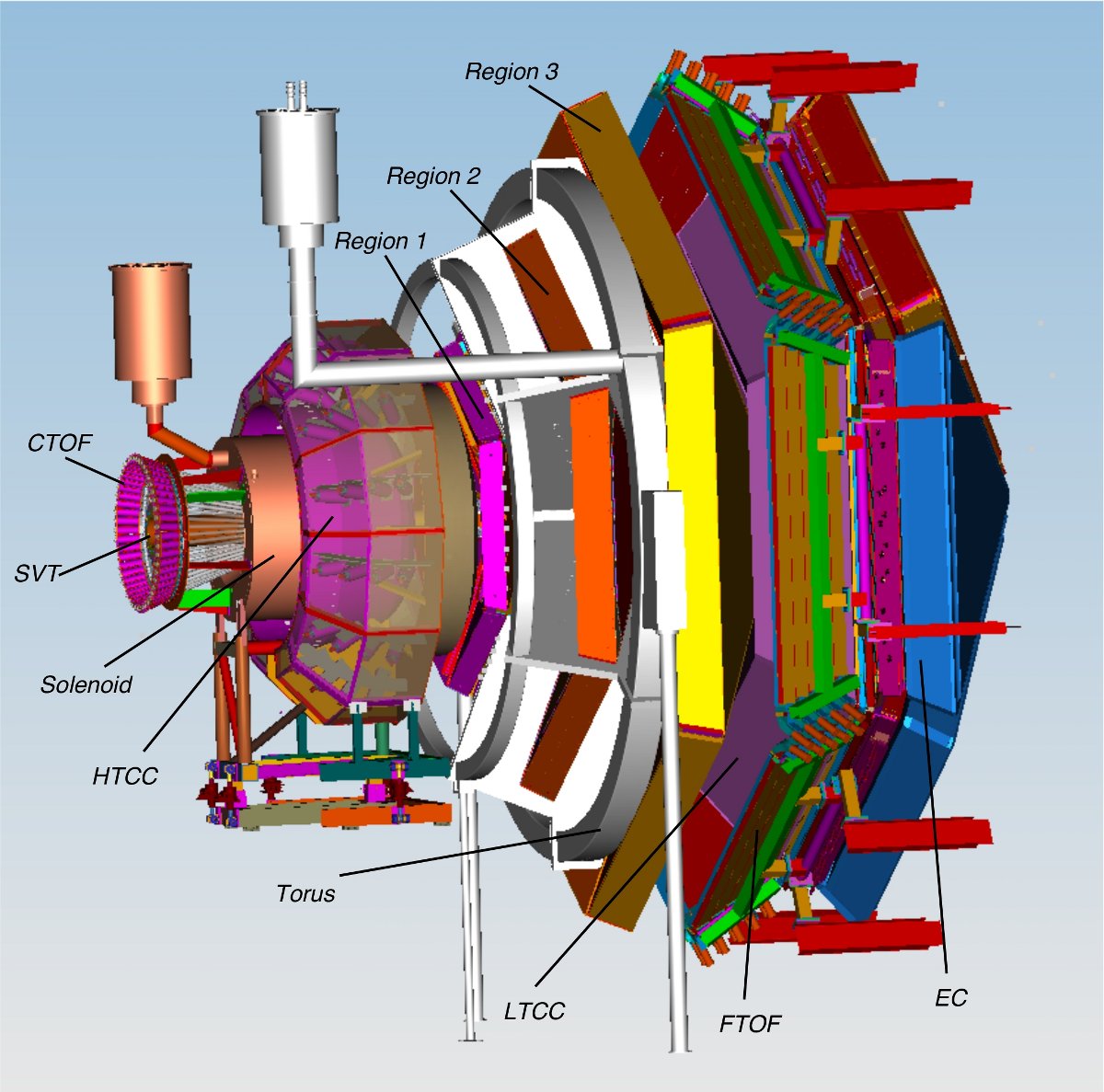}
    \caption{The schematic layout of the CLAS12 baseline design.}
    \label{fig:fd}
  \end{center}
\end{figure}

The CLAS12 detector is designed to operate with 11~GeV beam at an 
electron-nucleon luminosity of $\mathcal{L} = 
1\times10^{35}~$cm$^{-2}$s$^{-1}$. The baseline configuration of the CLAS12 
detector consists of the forward detector and the central detector 
packages~\cite{CD} (see Figure~\ref{fig:fd}). We use the forward detector
for electron detection in all ALERT run group proposals, while DVCS centered
proposals also use it for photon detection. The central
detector's silicon tracker and micromegas will be removed to leave room for
the recoil detector. 

The scattered electrons and photons will be detected in the forward detector which consists 
of the High Threshold Cherenkov Counters (HTCC), Drift Chambers (DC), the Low 
Threshold Cherenkov Counters (LTCC), the Time-of-Flight scintillators (TOF), 
the Forward Calorimeter and the Preshower Calorimeter. The charged particle 
identification in the forward detector is achieved by utilizing the combination 
of the HTCC, LTCC and TOF arrays with the tracking information from the Drift 
Chambers. The HTCC together with the Forward Calorimeter and the Preshower 
Calorimeter will provide a pion rejection factor of more than 2000 up to a 
momentum of 4.9~GeV/c, and a rejection factor of 100 above 4.9 GeV/c. The photons
are detected using the calorimeters.

\section{Available options for a Low Energy Recoil Detector}
We explored available solutions for the low-energy recoil tracker with 
adequate momentum and spatial resolution, and good particle identification for 
recoiling light nuclei (p, $^3$H and $^3$He). After investigating the 
feasibility of the proposed measurements using the CLAS12 Central Detector and 
the BONuS Detector~\cite{bonus6,bonus12}, we concluded that we needed to build 
a dedicated detector. We summarize in the
following the facts that led us to this conclusion.
\subsection{CLAS12 Central Detector}
The CLAS12 Central Detector~\cite{CD} is designed to detect various charged 
particles over a wide momentum and angular range. The main detector package 
includes:
\begin{itemize}
 \item Solenoid Magnet: provides a central longitudinal magnetic field up to 
5~Tesla, which serves to curl emitted low energy M{\o}ller electrons and determine 
particle momenta through tracking in the central detector.
 \item Central Tracker: consists of 3 double layers of silicon strips and 6 
    layers of Micromegas. The thickness of a single silicon layer is  
    \SI{320}{\um}.
 \item Central Time-of-Flight: an array of scintillator paddles with a 
cylindrical geometry of radius 26 cm and length 50 cm; the thickness of the 
detector is 2 cm with designed timing resolution of $\sigma_t = 50$ ps, used 
to separate pions and protons up to 1.2 GeV/$c$.
\end{itemize}

The current design, however, is not optimal for low energy particles 
($p<300$~MeV/$c$) due to the energy loss in the first 2 silicon strip layers. 
The momentum detection threshold is $\sim 200$ MeV/$c$ for protons, $\sim 
350$~MeV/$c$ for deuterons and even higher for $^3$H and $^3$He. These values 
are significantly too large for any of the ALERT run group proposals.

\subsection{BONuS12 Radial Time Projection Chamber}
The original BONuS detector was built for Hall B experiment E03-012 to study 
neutron structure at high $x_B$ by scattering electrons off an almost on-shell 
neutron inside deuteron. The purpose of the detector was to tag the low energy 
recoil protons ($p>60$ MeV/$c$). The key component for detecting the slow 
protons was the Radial Time Projection Chamber (RTPC) based on Gas Electron 
Multipliers (GEM). A later run period (eg6) used a 
newly built RTPC with a new design to detect recoiling $\alpha$ particles in 
coherent DVCS scattering. The major improvements of the eg6 RTPC were full 
cylindrical coverage and a higher data taking rate.

The approved 12~GeV BONuS (BONuS12) experiment is planning to use a similar 
device with some upgrades. The target gas cell length will be doubled, and the 
new RTPC will be longer as well, therefore doubling the luminosity and 
increasing the acceptance. Taking advantage of the larger bore ($\sim 700$ mm) of 
the 5~Tesla solenoid magnet, the maximum radial drift length will be increased 
from the present 3 cm to 4 cm, improving the momentum resolution by 
50\%~\cite{bonus12} and extending the momentum coverage. The main features of 
the proposed BONuS12 detector are summarized in Table~\ref{tab:comp}.

\begin{table}[tbp]
\bgroup
\def\arraystretch{1.1}%
\tabulinesep=1.2mm
\begin{tabu}{lcc}
\tabucline[2pt]{-}                                                   
\textbf{Detector Property}  & \textbf{RTPC}        & \textbf{ALERT}\\
\tabucline[1pt]{-}                                                   
Detection region radius & 4 cm                & 5 cm\\
Longitudinal length & $\sim$ 40 cm         & $\sim$ 30 cm \\
Gas mixture         & 80\% helium/20\% DME & 90\% helium/10\% isobutane \\
Azimuthal coverage  & 360$^{\circ}$               & 340$^{\circ}$\\
Momentum range      & 70-250 MeV/$c$ protons & 70-250 MeV/$c$ protons\\
Transverse mom. resolution & 10\% for 100~MeV/c protons & 10\% for 100~MeV/c protons\\
$z$ resolution & 3~mm & 3~mm \\
Solenoidal field    & $\sim 5$ T           & $\sim 5$ T \\
ID of all light nuclei & No                    & Yes \\
Luminosity      &$3\times10^{33}$ nucleon/cm$^{2}$\!/s & $6\times10^{34}$ nucleon/cm$^{2}$\!/s\\
Trigger             & can not be included  & can be included \\
\tabucline[1pt]{-}                                                   
\end{tabu}
\egroup
\caption{\label{tab:comp}Comparison between the RTPC (left column) and the new tracker (right column).}
\end{table}

\begin{figure}
  \begin{center}
    \includegraphics[angle=0, width=0.5\textwidth]{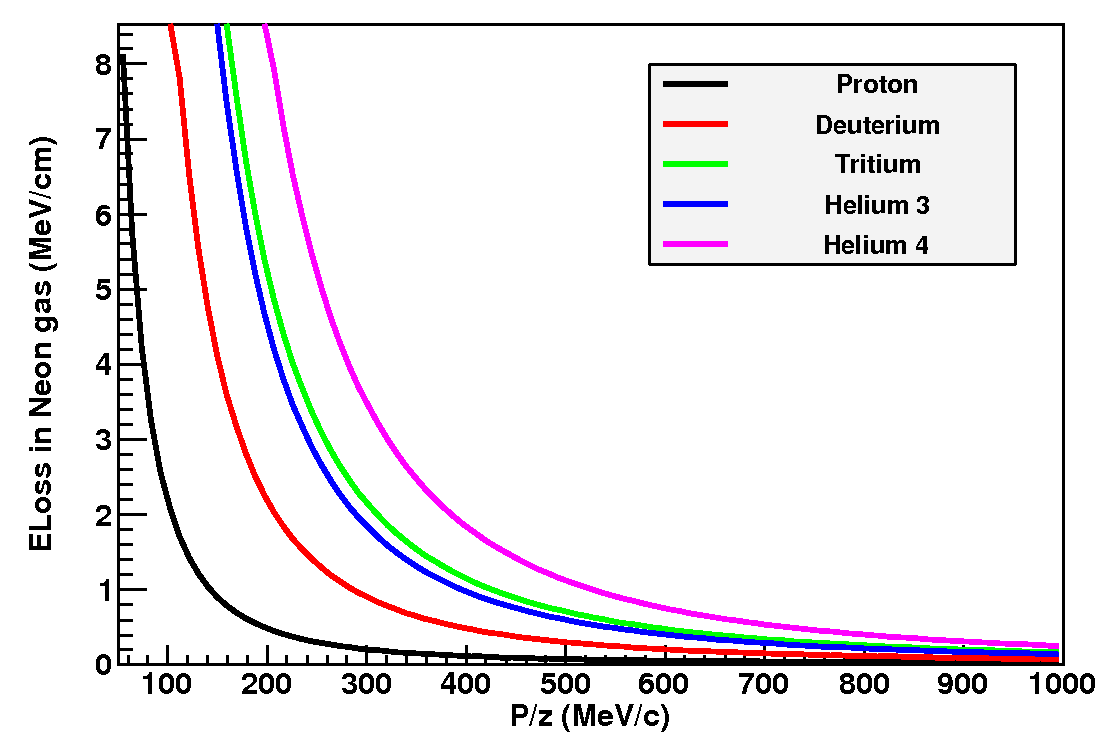}
    \caption{Calculation of energy loss in Neon gas as a function of the particle 
momentum divided by its charge for different nuclei. }
    \label{fig:eloss}
  \end{center}
\end{figure}

In principle, particle identification can be obtained from the RTPC through the 
energy loss $dE/dx$ in the detector as a function of the particle momentum (see 
Figure~\ref{fig:eloss}). However, with such a small difference between $^3$H and 
$^3$He, it is nearly impossible to discriminate between them
on an event by event basis because of the intrinsic width of the $dE/dx$ 
distributions. This feature is not problematic when using deuterium target, 
but makes the RTPC no longer a viable option for our tagged EMC and tagged DVCS 
measurements which require a $^4$He target and the differentiation of $^4$He, 
$^3$He, $^3$H, deuterons and protons.

Another issue with the RTPC is its slow response time due to a long drift 
time ($\sim5~\mu$s). If a fast recoil detector could be included in the trigger 
it would have a significant impact on the background rejection. Indeed, in
about 90\% of DIS events on deuteron or helium, the spectator fragments have too low energy 
or too small angle to get out of the target and be detected. By including
the recoil detector in the trigger, we would not be recording these events anymore.
Since the data acquisition speed was the main limiting factor for 
both BONuS and eg6 runs in CLAS, this 
would be a much needed reduction of the pressure on the DAQ.

\subsection{Summary}
In summary, we found that the threshold of the CLAS12 inner 
tracker is significantly too high to be used for our measurements. On the other hand, the 
recoil detector planned for BONuS12, a RTPC, is not suitable due to its 
inability to distinguish all kind of particles we need to measure.  
Moreover, as the RTPC cannot be efficiently included in the trigger, a lot of  
background events are sent to the readout electronics, which will cause its saturation and 
limit the maximum luminosity the detector can handle. Therefore, we propose 
a new detector design.

\section{Design of the ALERT Detector}
We propose to build a low energy recoil detector consisting of two sub-systems: 
a drift chamber and a scintillator hodoscope.
The drift chamber will be composed of 8 layers of sense wires to provide tracking 
information while the scintillators will provide particle 
identification through time-of-flight and energy measurements. To reduce the 
material budget, thus reducing the threshold to detect recoil particles 
at as low energy as possible, the scintillator 
hodoscope will be placed inside the gas chamber, just outside of the last 
layer of drift wires.

The drift chamber volume will be filled with a light gas mixture (90\% He and 
10\% C$_4$H$_{10}$) at atmospheric pressure. The amplification potential will
be kept low enough in order to not be sensitive to relativistic particles 
such as electrons and pions. Furthermore, a light 
gas mixture will increase the drift speed of the electrons from 
ionization. This will allow the chamber to withstand higher rates and 
experience lower hit occupancy. The fast signals from the chamber and 
the scintillators will be used in coincidence with electron trigger 
from CLAS12 to reduce the overall DAQ trigger rate and 
allow for operation at high luminosity.

\begin{figure}[tbp]
  \begin{center}
    \includegraphics[angle=0, width=0.75\textwidth]%
                    {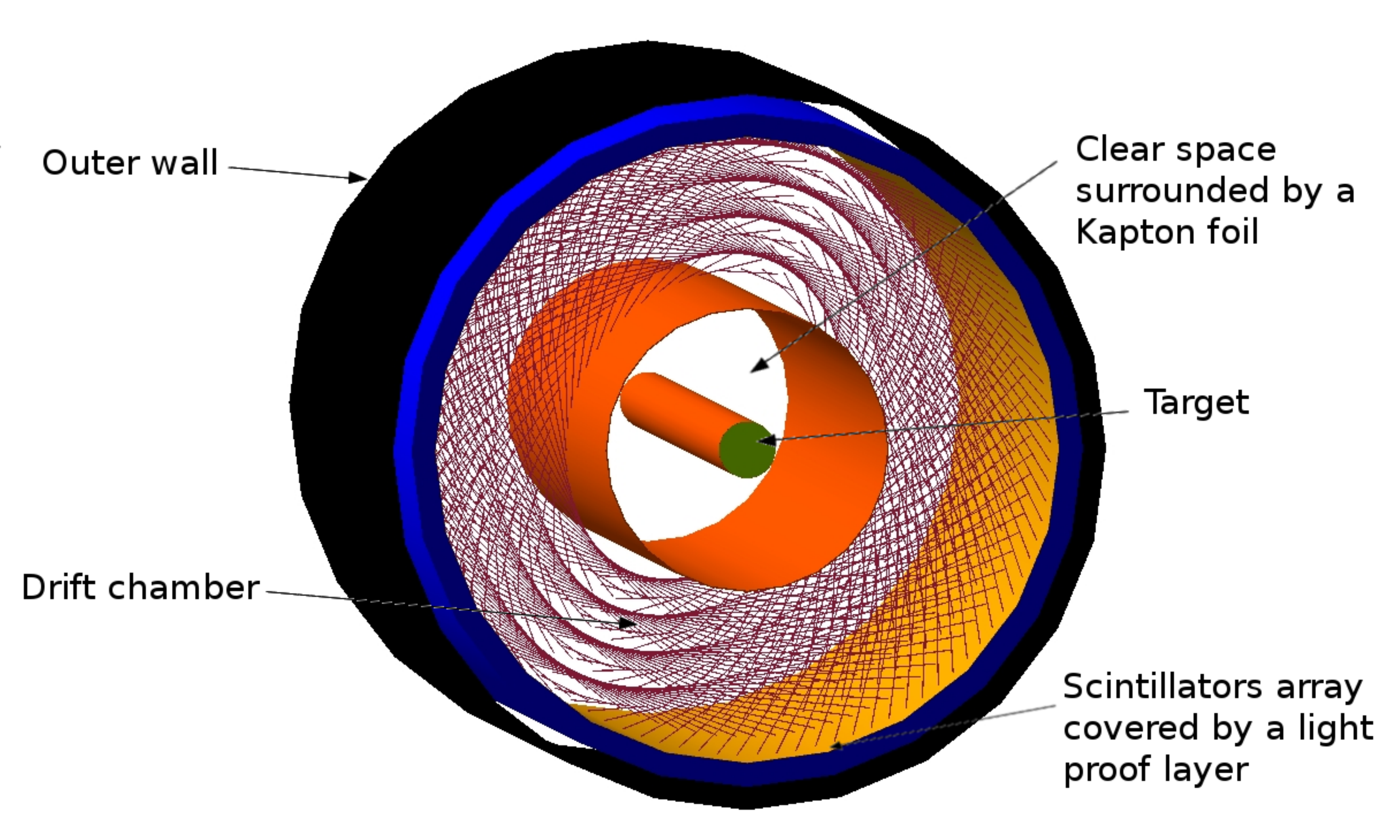}
    \caption{The schematic layout of the ALERT detector design, viewed 
from the beam direction.}
    \label{fig:new_lay}
  \end{center}
\end{figure}

The detector is designed to fit inside the central TOF of CLAS12; the 
silicon vertex tracker and the micromegas vertex tracker (MVT) will be 
removed. The available space has thus an outer radius of slightly more 
than 20~cm. A schematic 
layout of the preliminary design is shown in Figure~\ref{fig:new_lay} and its
characteristics compared to the RTPC design in Table~\ref{tab:comp}. The 
different detection elements are covering about $340^{\circ}$ of the polar 
angle to leave room for mechanics, and are 30~cm long with an effort made to 
reduce the particle energy loss through the materials. From the inside out,
it is composed of:
\begin{itemize}
\item a 30~cm long cylindrical target with an outer radius of 6~mm and target 
   walls \SI{25}{\um} Kapton filled with 3~atm of helium;
\item a clear space filled with helium to reduce secondary scattering from
   the high rate M\o{}ller electrons with an outer radius of 30~mm;
\item the drift chamber, its inner radius is 32~mm and its outer radius is 
85~mm;
\item two rings of plastic scintillators placed inside the gaseous chamber, 
   with total thickness of roughly 20~mm.
\end{itemize}

\subsection{The Drift Chamber}
While drift chambers are very useful to cover large areas at a moderate price, 
huge progress has been made in terms of their ability to withstand higher rates 
using better electronics, shorter distance between wires and optimization of 
the electric field over pressure ratio. Our design is based on other chambers 
developed recently. For example for the dimuon arm of ALICE at CERN, drift 
chambers with cathode planes were built in Orsay~\cite{AliceMuonArmChamber}. 
The gap between sense wires is 2.1~mm and the distance between two cathode 
planes is also 2.1~mm, the wires are stretched over about 1~m. Belle II is 
building a cylindrical drift chamber very similar to what is needed for this 
experiment and for which the space between wires is around 
2.5~mm~\cite{BelleIItdr}. Finally, a drift chamber with wire gaps of 1~mm is 
being built for the small wheel of ATLAS at CERN~\cite{ATLASChamber}. The 
cylindrical drift chamber proposed for our experiment is 300~mm long, and we 
therefore considered that a 2~mm gap between wires is technically a rather 
conservative goal. Optimization is envisioned based on experience with 
prototypes. 

The radial form of the detector does not allow for 90 degrees x-y wires in the 
chamber. Thus, the wires of each layer are at alternating angle of $\pm$ 
10$^{\circ}$, called the stereo-angle, from the axis of the drift chamber.  We 
use stereo-angles between wires to determine the coordinate along the beam 
axis ($z$). This setting makes it possible to use a thin forward end-plate to 
reduce multiple scattering of the outgoing high-energy electrons. A rough 
estimate of the tension due to the $\sim$2600 wires is under 600~kg, 
which appears to be reasonable for a composite end-plate. 

\begin{figure}
  \begin{center}
    \includegraphics[angle=0, width=0.7\textwidth,clip, trim=0mm 0mm 4cm 10cm]{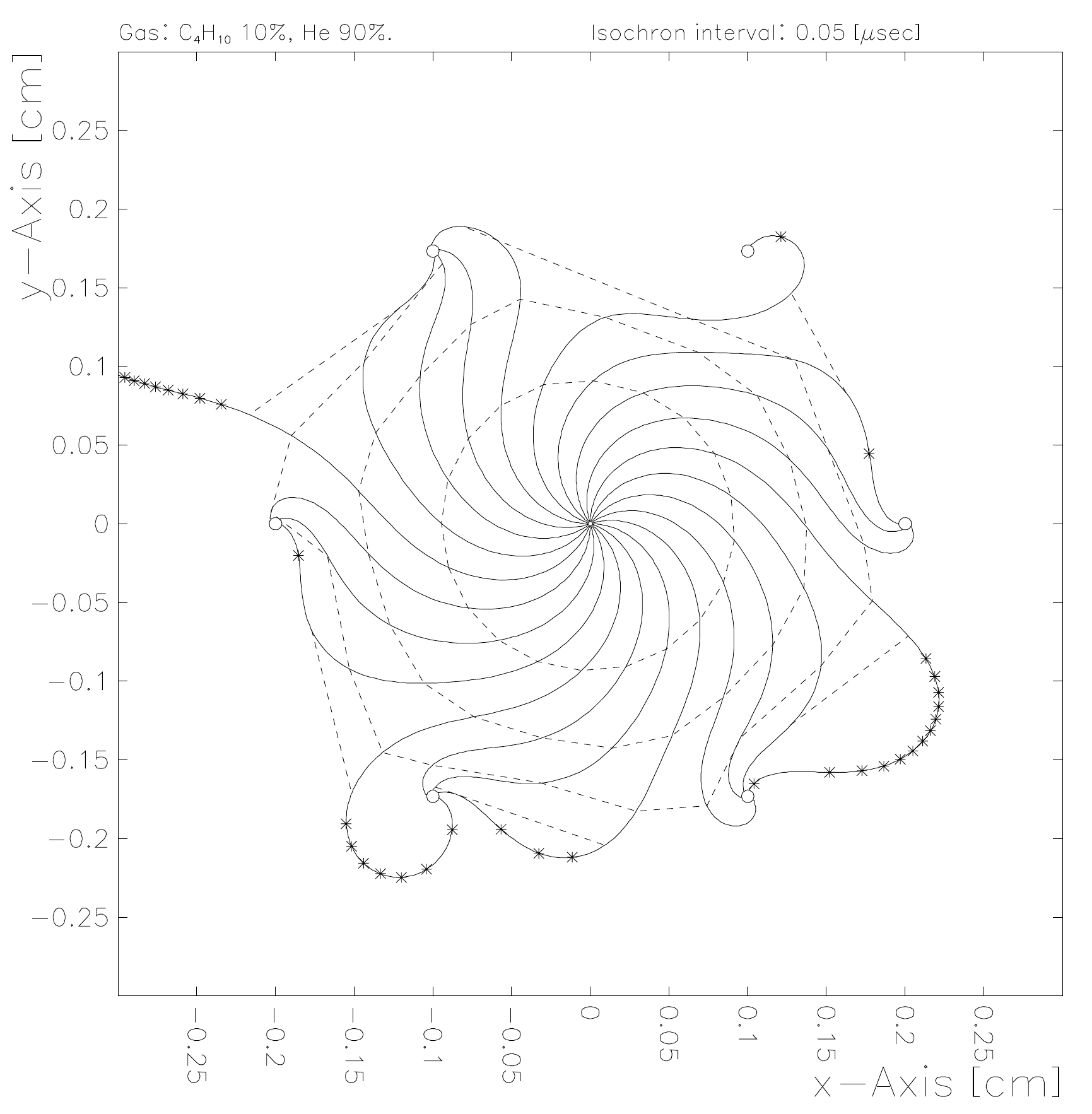}
    \caption{Drift lines simulated using MAGBOLTZ \cite{Magboltz} for one 
sense wire (at the center) surrounded by 6 field wires. The two electric field 
lines leaving the cell disappear when adjusting the voltages on the wires. 
Dashed lines are isochrones spaced by 50~ns. This shows that the maximum drift 
time is about 250~ns.}
    \label{fig:drift_cell}
  \end{center}
\end{figure}

The drift chamber cells are composed of one sense wire made of gold plated 
tungsten surrounded by field wires, however the presence of the 5~T magnetic 
field complicates the field lines. Several cell configurations have been studied with 
MAGBOLTZ~\cite{Magboltz}, we decided to choose a conservative 
configuration as shown in Figure~\ref{fig:drift_cell}. The sense wire is 
surrounded by 6 field wires placed equidistantly from it in a hexagonal 
pattern. The distance between the sense and field wires is constant and equal 
to 2~mm. Two adjacent cells share the field wires placed between them. The 
current design will have 8 layers of cells of similar radius. 
The simulation code MAGBOLTZ is calculating the drift speed and drift paths of 
the electrons (Figure~\ref{fig:drift_cell}). With a moderate electric field, the 
drift speed is around 10~microns/ns, the average drift time expected is thus 
250~ns (over 2~mm). Assuming a conservative 10~ns time resolution, the spatial 
resolution is expected to be around 200~microns due to field distortions and 
spread of the signal.

\begin{figure}
  \begin{center}
    \includegraphics[angle=0, width=0.5\textwidth, trim=5mm 5mm 5mm 15mm, 
    clip]{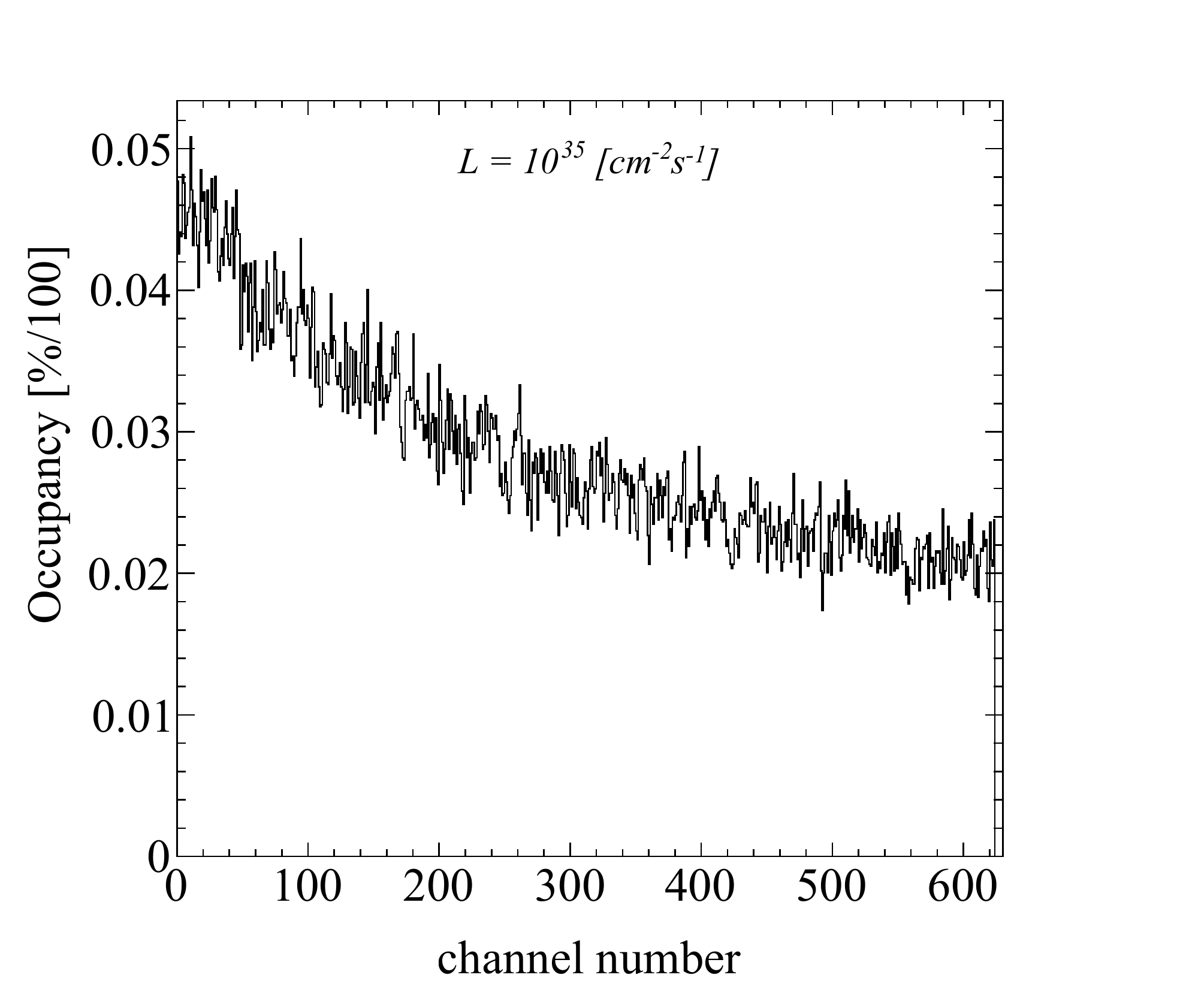}
    \caption{\label{fig:RCoccupancy}A full Geant4 simulation of the ALERT drift 
       chamber hit occupancy
       at a luminosity of $10^{35}$ cm$^{-2}$s$^{-1}$. The channel numbering 
    starts with the inner most wires and works outwards.}
  \end{center}
\end{figure}

\begin{figure}
  \begin{center}
    \includegraphics[angle=0, width=0.7\textwidth, trim=5mm 5mm 5mm 10mm, 
    clip]{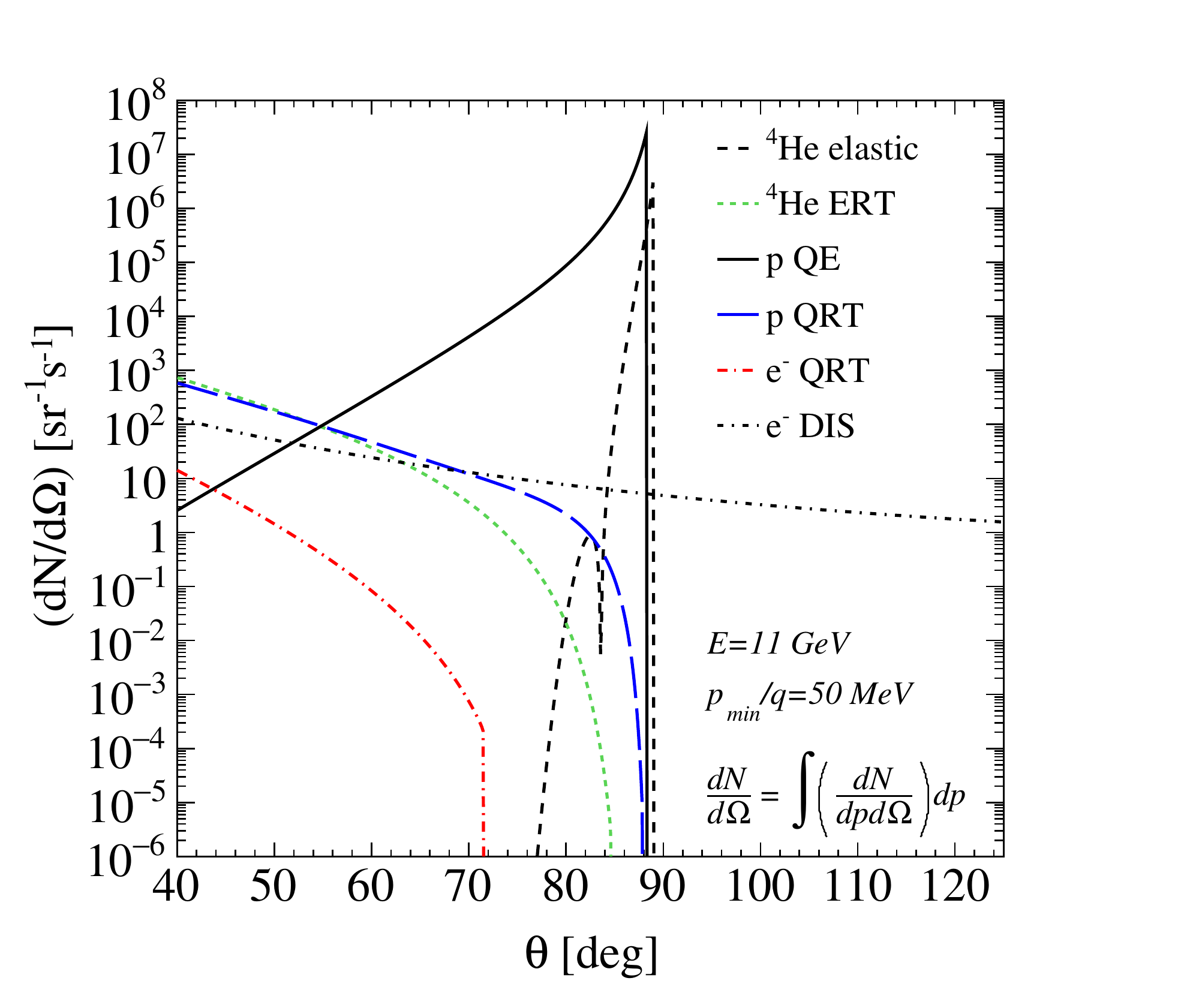}
    \caption{\label{fig:ALERTrates}The rates for different processes as 
    function of angle. The quasi-elastic radiative tails (QRT), $^4$He elastic 
 radiative tail (ERT), and DIS contributions have been integrated over momenta 
 starting at $p/q$ = 50~MeV/c, where $q$ is the electric charge of the particle 
 detected.}
  \end{center}
\end{figure}

The maximum occupancy, shown in Figure~\ref{fig:RCoccupancy},
is expected to be around 5\% for the inner most wires at $10^{35}$~cm$^{-2}$s$^{-1}$
(including the target windows). This is the maximum available luminosity for the 
baseline CLAS12 and is obtained based on the physics channels depicted 
in Figure~\ref{fig:ALERTrates}, assuming an integration time of 200~ns and 
considering a readout wire separation of 4~mm. This amount of accidental hits 
does not appear to be reasonable for a good tracking quality, we therefore 
decided to run only at half this luminosity for our main production runs. This 
will keep occupancy below 3\%, which is a reasonable amount for a drift chamber 
to maintain high tracking efficiency. When running the coherent processes with 
the $^4$He target, it is not necessary to detect the protons\footnote{This 
   running condition is specific to the proposal ``Partonic Structure of Light 
Nuclei'' in the ALERT run group.}, so the rate of accidental hits can then be 
highly reduced by increasing the detection threshold, thus making the chamber 
blind to the protons\footnote{The CLAS {\it eg6} run period was using the RTPC in 
the same fashion.}. In this configuration, considering that our main 
contribution to occupancy are quasi-elastic protons, we are confident that the 
ALERT can work properly at $10^{35}$~cm$^{-2}$s$^{-1}$.

We are currently planning to use the electronics used by the MVT of CLAS12, 
known as the DREAM chip \cite{7097517}. Its dynamic range and time resolution 
correspond to the needs of our drift chamber. To ensure that it is the 
case, tests with a prototype will be performed at the IPN Orsay (see 
section~\ref{sec:proto}).

\subsection{The Scintillator Array} \label{sec:scint}
The scintillator array will serve two main purposes. First, it will provide a 
useful complementary trigger signal because of its very fast response time, 
which will reduce the random background triggers. Second, it will provide 
particle identification, primarily through a time-of-flight measurement, but 
also by a measurement of the particle total energy deposited and path length in 
the scintillator which is important for doubly charged ions.

The length of the scintillators cannot exceed roughly 40~cm to keep the time 
resolution below 150~ps. It must also be segmented to match with tracks 
reconstructed in the drift chamber. Since $^3$He and $^4$He will travel at 
most a few mm in the scintillator for the highest anticipated momenta 
($\sim$~400~MeV/c), a multi-layer scintillator design provides an extra handle on 
particle identification by checking if the range exceeded the thickness of 
the first scintillator layer.

The initial scintillator design consists of a thin (2~mm) inner layer of 60 
bars, 30~cm in length, and 600 segmented outer scintillators (10 segments 
3~cm long for each inner bar) wrapped around the drift chamber. Each of these 
thin inner bars has SiPM\footnote{SiPM: silicon photomultiplier.} detectors 
attached to both ends. A thicker outer layer (18~mm) will be further segmented 
along the beam axis to provide position information and maintain good time 
resolution.

\begin{figure}
  \begin{center}
    \includegraphics[width=0.48\textwidth]{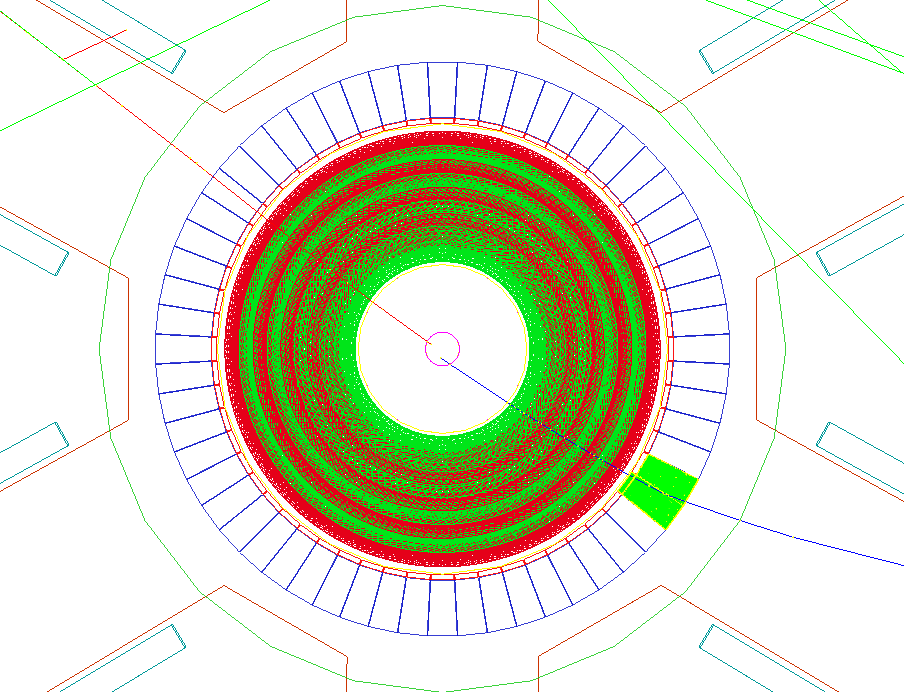}
    \includegraphics[width=0.48\textwidth]{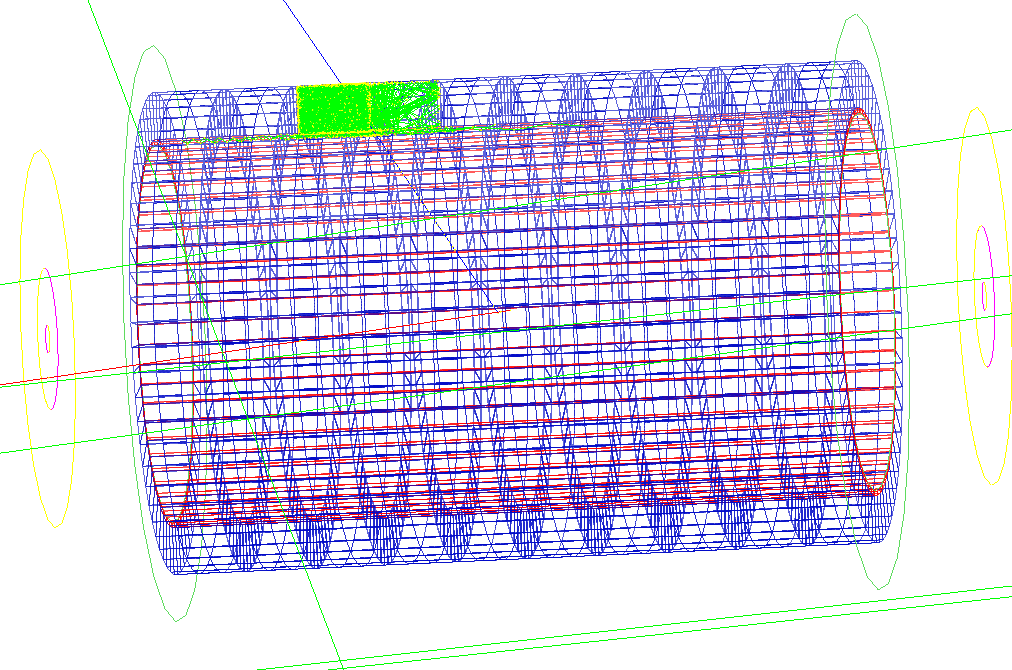}
    \caption{\label{fig:scintHodoscopeDesign}Geant4 simulation of a proton 
    passing through the recoil drift chamber and scintillator hodoscope. The 
 view looking downstream (left) shows the drift chamber's eight alternating 
 layers  of wires (green and red) surrounded by the two layers of scintillator 
 (red and blue). Simulating a proton through the detector, photons (green) are 
 produced in a few scintillators. On the right figure, the dark blue rings are graphical feature showing the contact between the adjacent outer scintillators.}
  \end{center}
\end{figure}

For the outer layer, a dual ended bar design and a tile design with embedded 
wavelength shifting fiber readouts similar to the forward tagger's hodoscope for 
CLAS12~\cite{FThodo} were considered. After simulating these designs, it was 
found that the time resolution was insufficient except only for the smallest 
of tile designs (15$\times$15$\times$7~mm$^3$). Instead of using fibers, a 
SiPM will be mounted directly on the outer layer of a keystone shaped 
scintillator that is 30~mm in length and 18~mm thick. This design can be seen 
in Figure~\ref{fig:scintHodoscopeDesign} which shows a full Geant4 simulation of 
the drift chamber and scintillators. By directly mounting the SiPMs to the 
scintillator we collect the maximum signal in the shortest amount of time.  
With the large number of photons we expect, the time resolution of SiPMs will 
be a few tens of ps, which is well within our target.

\begin{figure}
  \begin{center}
    \includegraphics[angle=0, 
    width=0.48\textwidth]{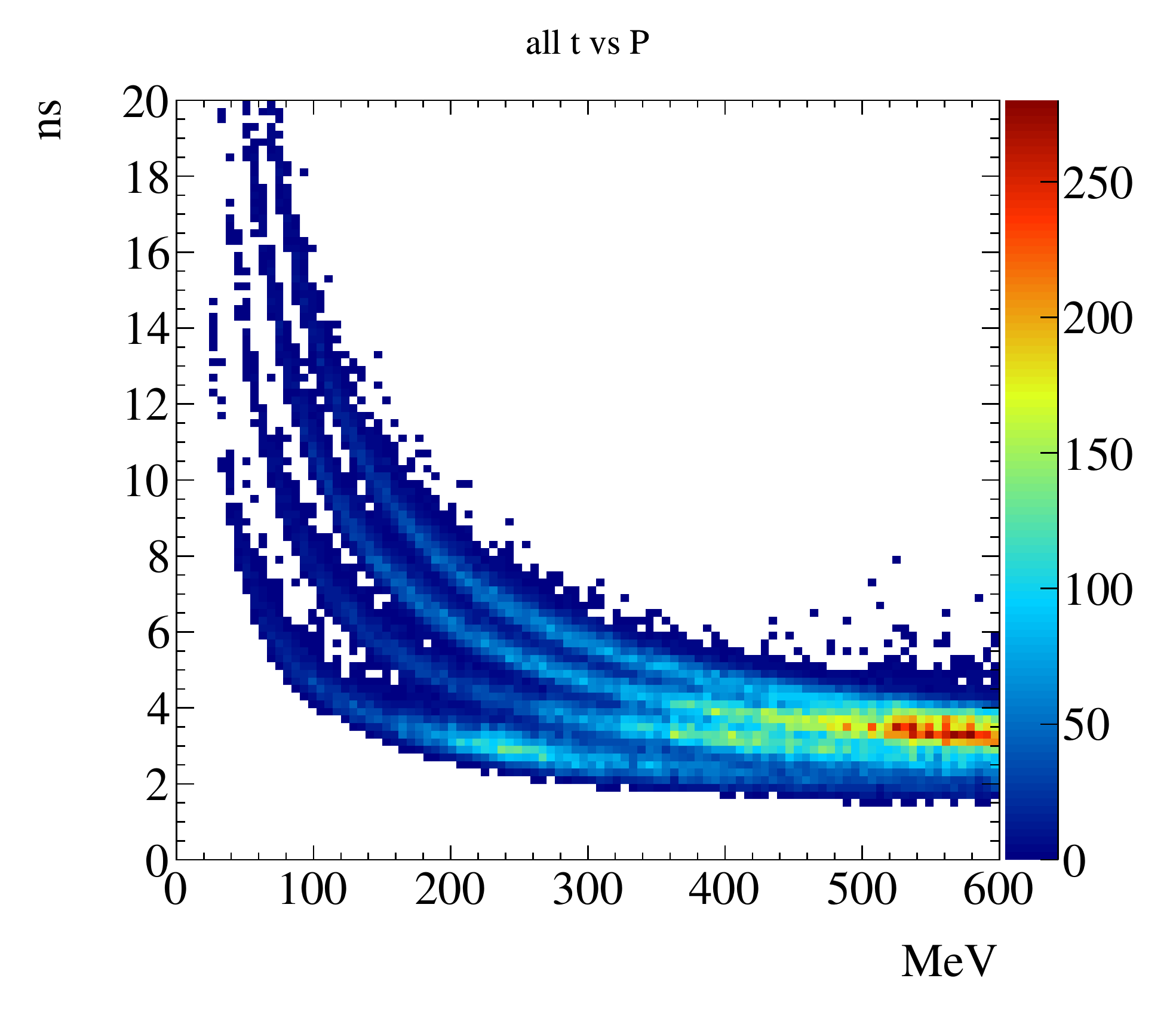}
    \includegraphics[angle=0, 
    width=0.48\textwidth]{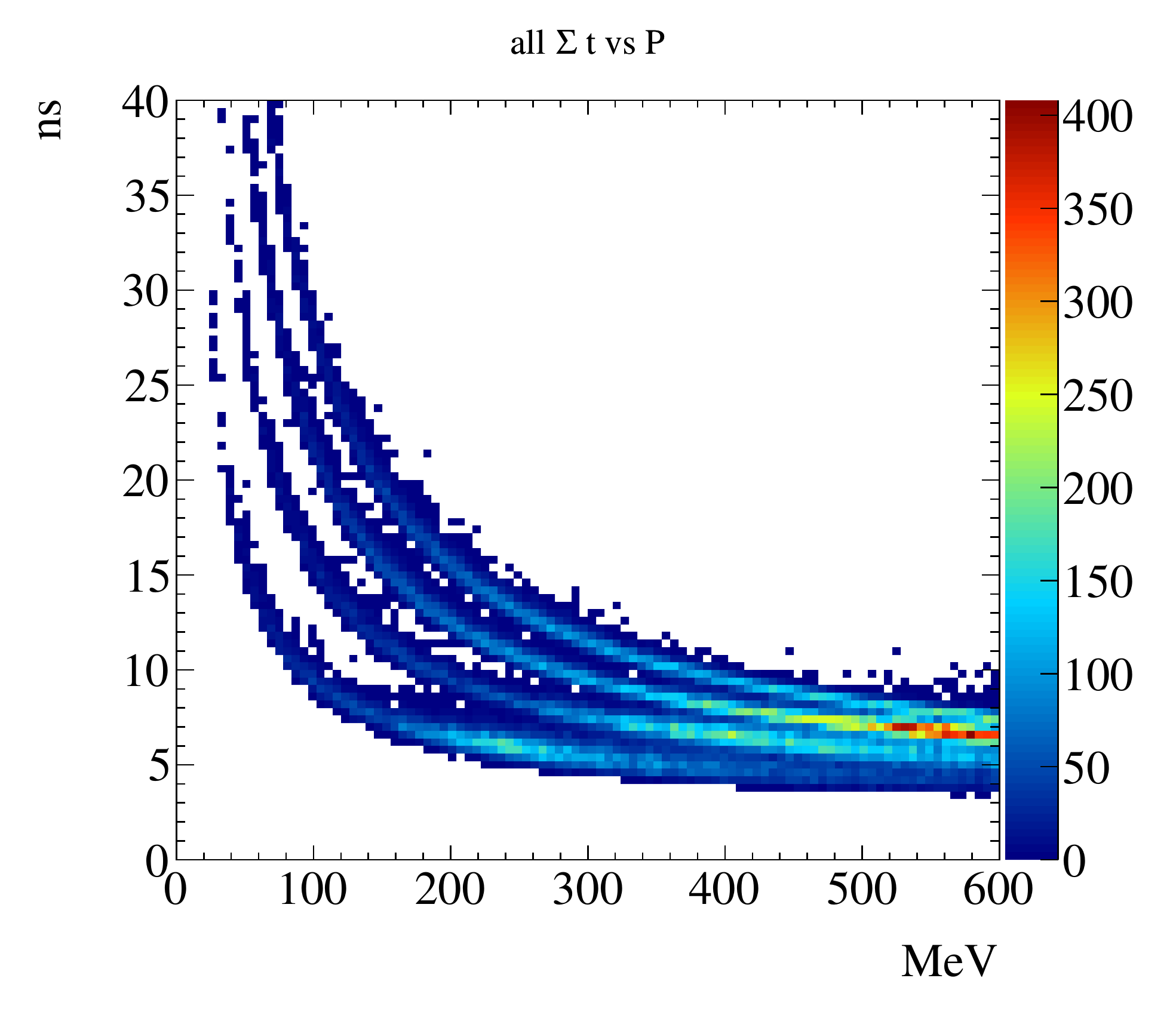}
    \caption{
       \label{fig:scintTimeVsP}Simulated TOF for the various recoil particles 
    vs Momentum. The TOF from just a single readout is shown on the left and 
 the sum of the dual ended readout is shown on the right.   }
  \end{center}
\end{figure}

The advantage of a dual ended readout is that the time sum is proportional to 
the TOF plus a constant. The improved separation of different particles can 
be seen in Figure~\ref{fig:scintTimeVsP}. Reconstructing the position of a hit 
along the length of a bar in the first layer is important for the doubly 
charged ions because they will not penetrate deep enough to reach the second 
layer of segmented scintillator.

The front-end electrons for the SiPMs will include preamplifiers and ASICs\footnote{ASIC: application-specific integrated circuit.}
which provide both TDC and ADC readouts. The PETIROC-2A\cite{PETIROC} ASIC 
provides excellent time resolution ($18$~ps on trigger output with 4 
photoelectrons detected) and a maximum readout rate at about 40k events/s.
Higher readout rates can be handled by using external digitizers by using the 
analog mode of operation and increase this rate by an order of magnitude. The 
ASIC also has the advantage of being able to tune the individual over-bias 
voltages with an 8-bit DAC.

The expected radiation damage to the SiPMs and scintillator material is found 
to be minimal over the length of the proposed experiment. We used the CLAS12 
forward tagger hodoscope technical design report~\cite{FThodo} as a very 
conservative baseline for this 
comparison. We arrived at an estimated dose of 1 krad after about 4.5 months of 
running. The damage to the scintillator at 100 times these radiation levels  
would not be problematic, even for the longest lengths of scintillator 
used~\cite{Zorn:1992ew}.
Accumulated dose on the SiPMs leads to an increased dark current. Similarly
than for scintillators, we do not expect it to be significant over the length of the 
experiment. The interested reader is referred to the work on
SiPMs for the Hall-D detectors~\cite{Qiang:2012zh,Qiang:2013uwa}. A front-end 
electronics prototype will be tested for radiation hardness but we expect  any 
damage to negligible~\cite{commPETIROC}.


\subsection{Target Cell}\label{sec:targetCell}

\begin{figure}
  \begin{center}
    \includegraphics[angle=0, trim={0 0 15cm 0}, clip,
    width=0.99\textwidth]{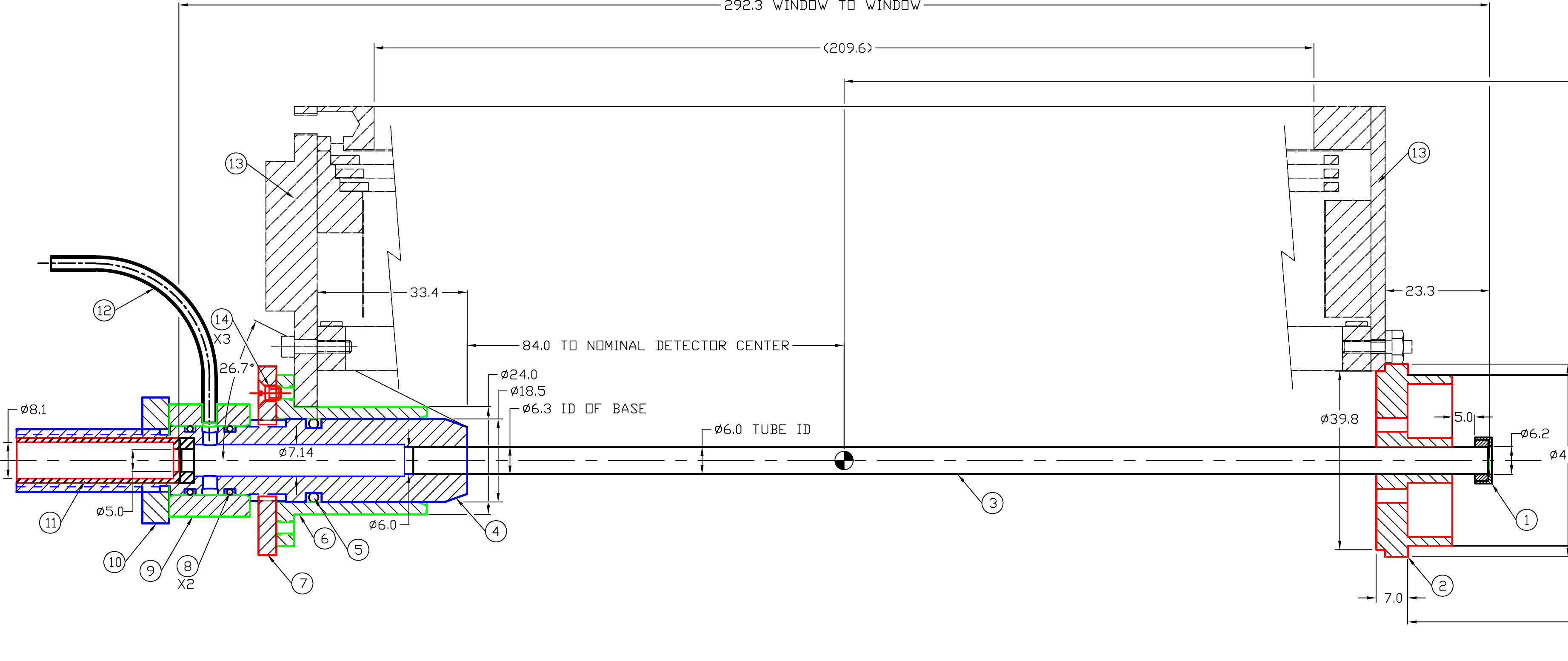}
    \caption{ \label{fig:eg6TargetDrawing}The eg6 target design drawing.}
  \end{center}
\end{figure}

The design of the proposed ALERT target will be very similar to the eg6 target shown 
in Figure~\ref{fig:eg6TargetDrawing}.
The target parameters are shown in Table~\ref{tab:target} 
with the parameters of other existing and PAC approved targets.
Note that, the proposed target has an increased radius of 6~mm compared to all the 
others which have 3~mm radius. This increase compared to the previous CLAS targets has been made
in order to compensate for the expected increase of beam size at 11~GeV. The BONuS12
target is still presently proposed to be 3~mm in radius, if such a target
is operated successfully in JLab, we will definitely consider using a
smaller radius as well, but we prefer to propose here a safer option that we know
will work fine.

\begin{table}
\centering
\caption{Comparison of various straw targets used at JLab.The 
"JLab test targets" correspond
to recent tests performed in JLab for the BONuS12 target, they have
been tested for pressure but have never been tested with beam.
}
\newcolumntype Y{S [ group-four-digits=true,
round-mode=places,
round-precision=1,
round-integer-to-decimal=true,
per-mode=symbol ,
detect-all]}
\tabucolumn Y
\label{tab:target}
\bgroup
\def\arraystretch{1.2}%
\tabulinesep=1mm
\begin{tabu}{l C{1.5cm}C{3.0cm}C{2cm}}
\tabucline[2pt]{-}
\textbf{Experiment} & \textbf{Length} & \centering \textbf{Kapton wall thickness} &
\textbf{Pressure} \\ \tabucline[1pt]{-}
CLAS target (eg6)           & 30~cm & \SI{27}{\um} & 6.0~atm   \\
BONuS12 (E12-06-113) target & 42~cm & \SI{30}{\um} & 7.5~atm \\
JLab test target 1          & 42~cm & \SI{30}{\um} & 3.0~atm   \\
JLab test target 2          & 42~cm & \SI{50}{\um} & 4.5~atm \\
JLab test target 3          & 42~cm & \SI{60}{\um} & 6.0~atm   \\
ALERT proposed target       & 35~cm & \SI{25}{\um} & 3.0~atm   \\
\tabucline[2pt]{-}
\end{tabu}
\egroup
\end{table}

%
\section{Simulation of ALERT and reconstruction} \label{sec:sim}
The general detection and reconstruction scheme for ALERT is as follows. We fit 
the track with the drift chamber and scintillator position information to
obtain the momentum over the charge. Next, using the 
scintillator time-of-flight, the particles are separated and identified by 
their mass-to-charge ratio, therefore leaving a degeneracy for the deuteron and 
$\alpha$ particles.
The degeneracy between deuteron and $\alpha$ particles can be resolved in a few 
ways.  The first and most simple way is to observe that an $\alpha$ will almost 
never make it to the second layer of scintillators and therefore the absence (presence) of a 
signal would indicate the particle is an $\alpha$~(deuteron). Furthermore, as 
will be discussed below, the measured dE/dx will differ for $^4$He and $^2$H, 
therefore, taking into account energy loss in track fitting alone can provide 
separation. Additionally taking further advantage of the measured total energy 
deposited in the scintillators can help separate the $\alpha$s and deuterons.

\subsection{Simulation of ALERT}
\begin{figure}[tbp]
    \begin{center}
        \includegraphics[width=0.45\textwidth]{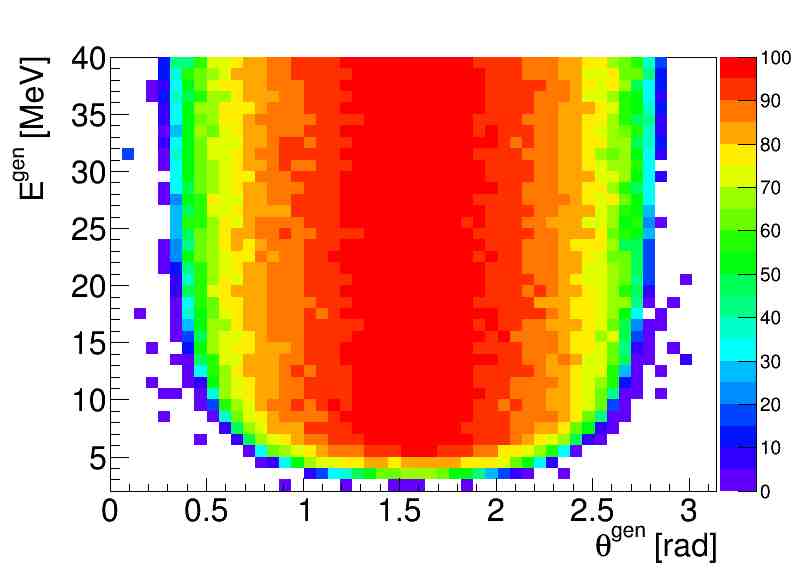}
        \includegraphics[width=0.45\textwidth]{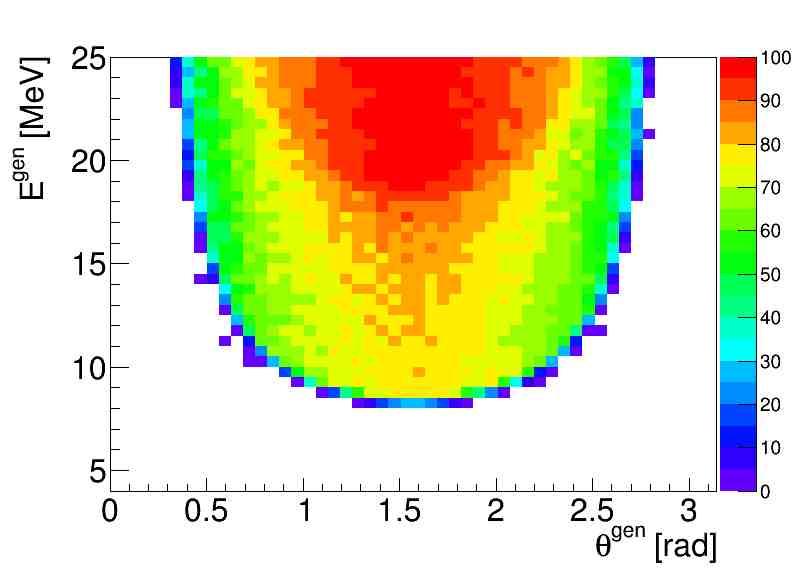}
        \caption{Simulated recoil detector acceptance percentage, for protons (left) and 
$^4$He (right), when requiring energy deposition in the scintillators arrays. 
\label{fig:acceptance}}
    \end{center}
\end{figure}

The simulation of the recoil detector has been implemented with the full 
geometry and material specifications in GEANT4. It includes a 5~Tesla homogeneous 
solenoid field and the entire detector filled with materials as described in the 
previous section. In this study all recoil species are generated with the same 
distributions: flat in momentum from threshold up to 40~MeV 
($\sim$~250~MeV/c) for protons and about 25~MeV for other particles; isotropic 
angular coverage; flat distribution in $z$-vertex; and a radial vertex 
coordinate smeared around the beam line center by a Gaussian distribution of 
sigma equal to the expected beam radius (0.2 mm).
For reconstruction, we require that the particle reaches the scintillator
and obtain the acceptance averaged over the $z$-vertex position shown in 
Figure~\ref{fig:acceptance}.

\subsection{Track Fitting}
\begin{figure}[tbp]
    \begin{center}
        \includegraphics[height=4.5cm, width=0.32\textwidth]{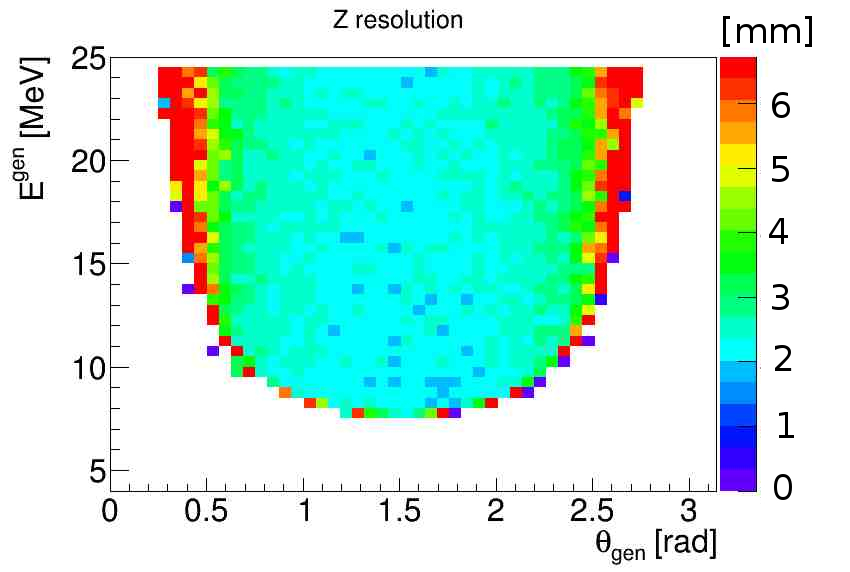}
        \includegraphics[height=4.5cm, width=0.32\textwidth]{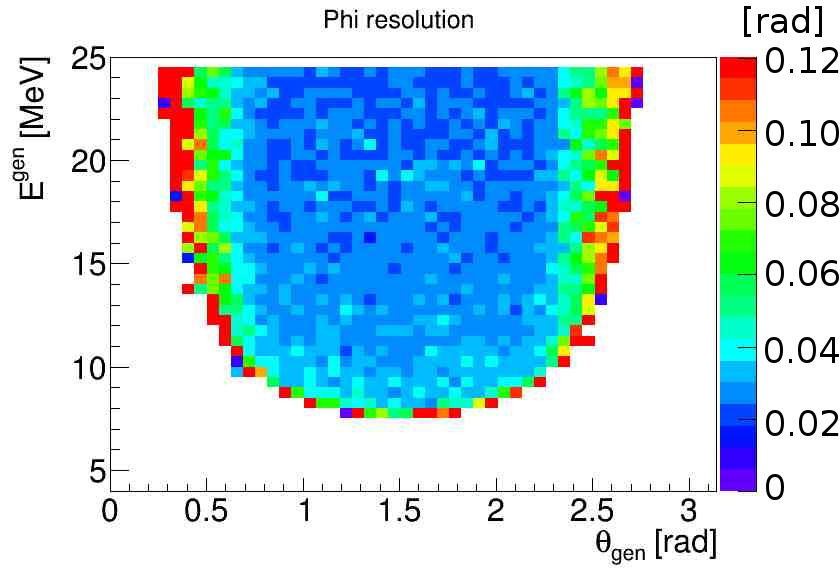}
        \includegraphics[height=4.5cm, width=0.32\textwidth]{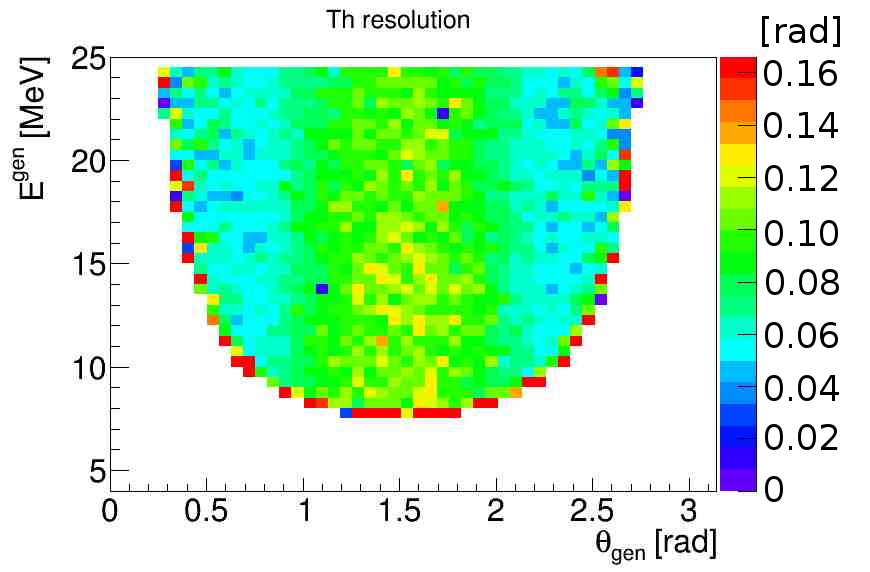}
        \caption{Resolutions for simulated $^4$He:  $z$-vertex resolution in mm (left), azimuthal (center) 
          and polar (right) angle resolutions in radians for the lowest energy
          regime when the recoil track reaches the scintillator.\label{fig:tracking}}
    \end{center}
\end{figure}

\begin{figure}[tbp]
    \begin{center}
        \includegraphics[width=0.45\textwidth]{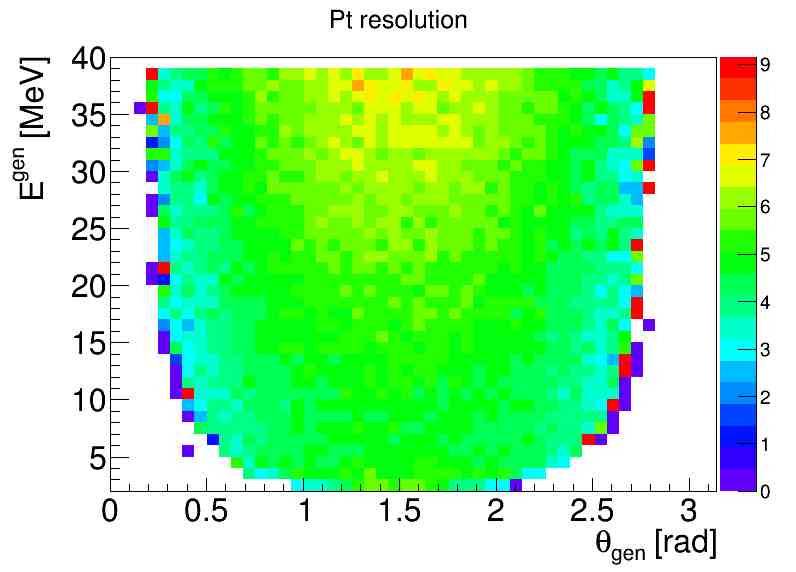}
        \includegraphics[width=0.45\textwidth]{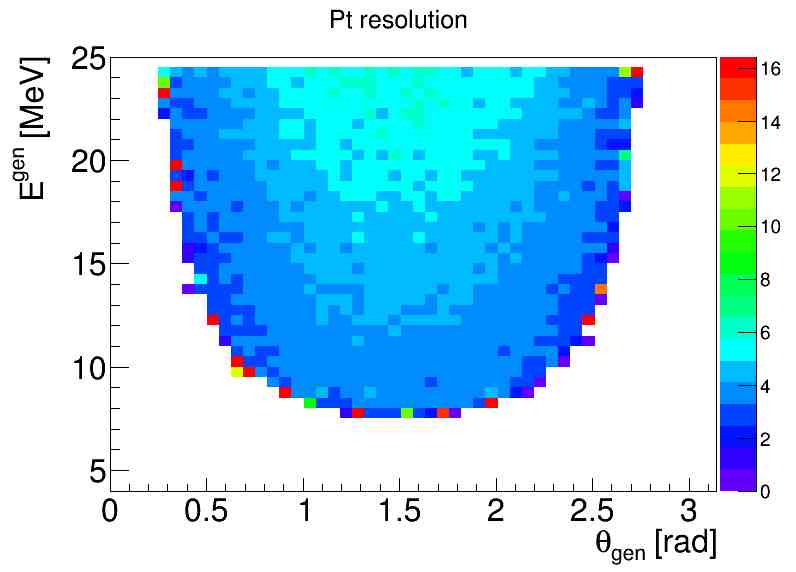}
        \caption{Simulated momentum resolutions (in \%) as a function of energy and 
                 polar angle for protons (left) and $^4$He 
                 (right) integrated over all $z$, when the recoil track reaches the scintillators 
                 array. \label{fig:presolution}}
    \end{center}
\end{figure}

The tracks are obtained using a helix fitter giving the coordinates of 
the vertex and the momentum of the particle. The energy deposited in 
the scintillators could also be used to help determine the kinetic energy of the 
nucleus, but is not implemented in the studies we performed here. 
The tracking capabilities of the recoil detector are investigated 
assuming a spatial resolutions of \SI{200}{\um} for the drift chamber. The wires 
are strung in the $z$-direction with a stereo angle of \ang{10}. The resulting difference between 
generated and reconstructed variables from simulation is shown in 
Figure~\ref{fig:tracking} for $^4$He particles. The momentum resolution for both protons and 
$^4$He is presented in Figure~\ref{fig:presolution}.


\subsection{Particle identification in ALERT}

\begin{figure}[tbp]
    \begin{center}
        \includegraphics[width=0.7\textwidth]{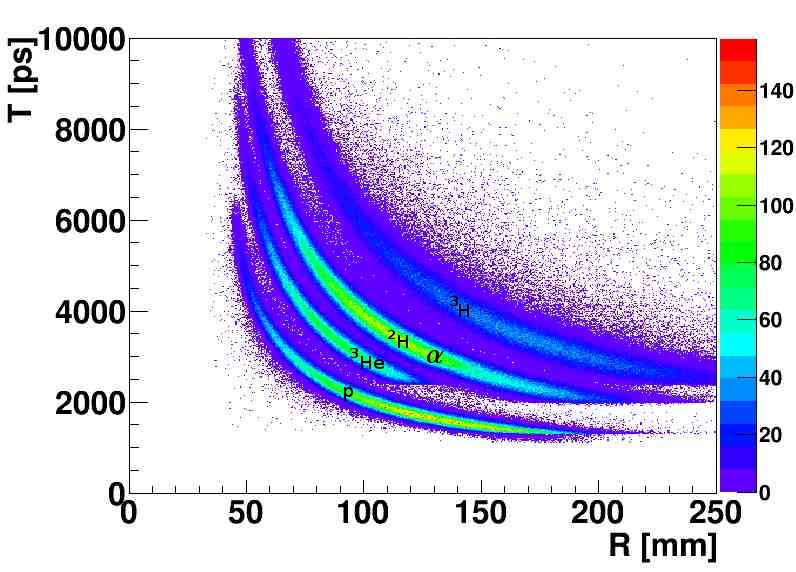}
        \caption{Simulated time of flight at the scintillator versus the 
reconstructed radius in the drift chamber. The bottom band corresponds to 
the proton, next band is the $^3$He nuclei, $^2$H and $\alpha$ are overlapping in 
the third band, the uppermost band is $^3$H\label{fig:SIMtof}. $^2$H and 
$\alpha$ are separated using dE/dx.}
    \end{center}
\end{figure}

The particle identification scheme is investigated using the GEANT4
simulation as well. The scintillators 
have been designed to ensure a 150~ps time resolution. To determine the dE/dx 
resolution, measurements will be necessary for the scintillators and for the 
drift chamber as this depends on the detector layout, gas mixture, 
electronics, voltages... Nevertheless, from \cite{Emi}, one can assume that 
with 8 hits in the drift chamber and the measurements in the 
scintillators, the energy resolution should be at least 10\%.
Under these conditions, a clean separation of three of the five nuclei is shown 
in Figure~\ref{fig:SIMtof} solely based on the time of flight measured by the 
scintillator compared to the reconstructed momentum from the drift chamber. 
We then separate $^2$H and $\alpha$ using dE/dx in the drift chamber and in the 
scintillators.

To quantify the separation power of our device, we simulated an equal quantity 
of each species. We obtained a particle identification efficiency of 99\% for 
protons, 95\% for $^3$He and 98\% for $^3$H and around 90\% for $^2$H and 
$\alpha$ with equally excellent rejection factors. It is important to note that 
for this analysis, only the energy deposited in the scintillators was used, not 
the energy deposited in the
drift chamber nor the path length in the scintillators, thus these numbers
are very likely to be improved when using the full information\footnote{The 
uncertainty remains important about the resolutions that will be achieved 
for these extra information. So we deemed more reasonable to ignore them
for now.}. This analysis indicates that the proposed reconstruction 
and particle identification schemes for this design are quite promising.  
Studies, using both simulation software and prototyping, are ongoing to 
determine the optimal detector parameters to minimize the detection threshold 
while maximizing particle identification efficiency. The resolutions presented 
above have been implemented in a fast Monte-Carlo used to evaluate their impact 
on our measurements.

\section{Drift chamber prototype}
\label{sec:proto}
Since the design of the drift chamber presents several challenges in term of
mechanical assembly, we decided to start prototyping early. The goal is to find a 
design that will be easy to install and to maintain if need be, while keeping the 
amount of material at a minimum. This section presents the work done in Orsay 
to address the main questions concerning the mechanics that needed to be answered:
\begin{itemize}
\item How to build a stereo drift chamber with a 2~mm gap between wires?
\item Can we have frames that can be quickly changed in case of a broken wire?
\item How to minimize the forward structure to reduce the multiple scattering,
while keeping it rigid enough to support the tension due to the wires?
\end{itemize}

\begin{figure}[tbp]
    \begin{center}
        \includegraphics[width=0.4\textwidth]{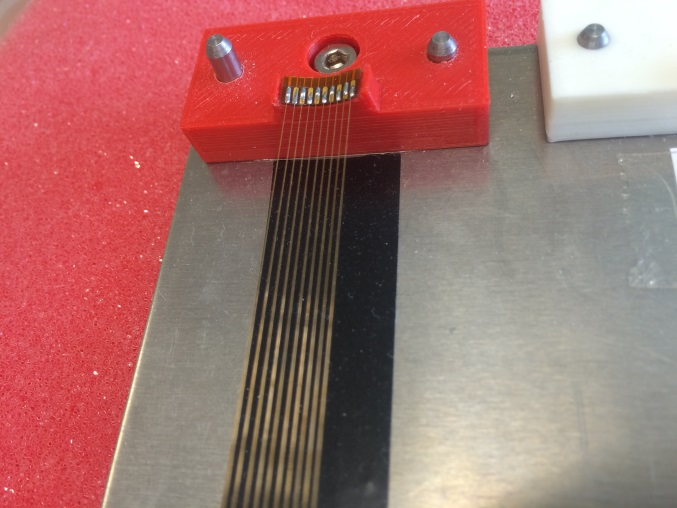}
        \caption{Welded wires on a curved structure with a 2~mm gap between each wire.}
        \label{soldOK}
    \end{center}
\end{figure}

For the first question, small plastic structures realized with a 3D printer 
were tested and wires welded on it, as shown in Figure \ref{soldOK}. This 
demonstrated our ability to weld wires with a 2~mm gap on a curved structure.
 
\begin{figure}[tbp]
    \begin{center}
        \includegraphics[width=0.40\textwidth]{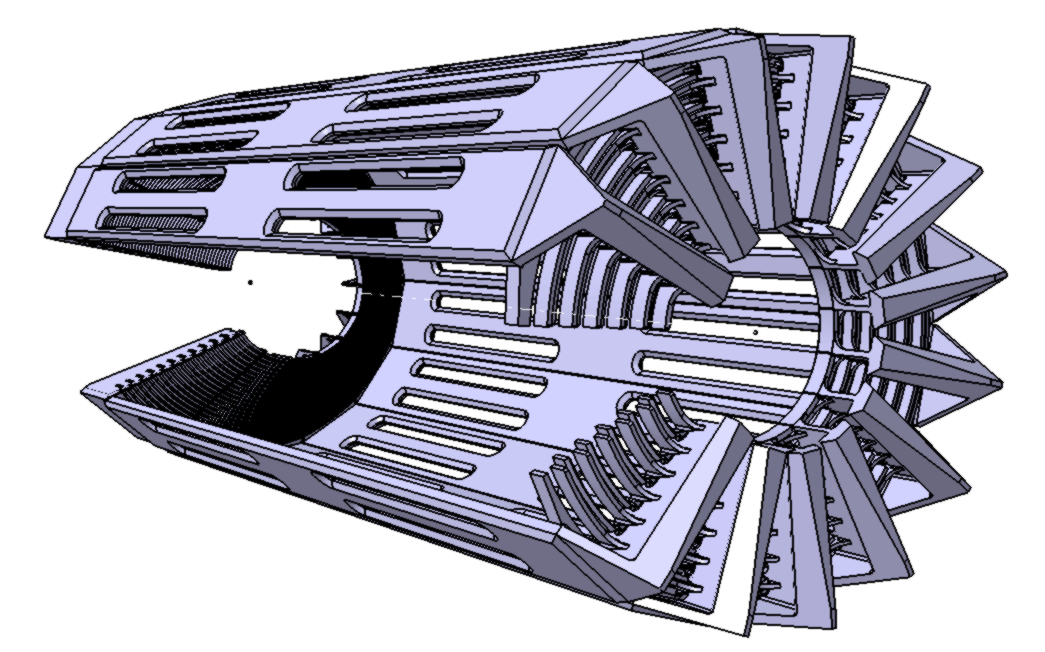}
        \includegraphics[width=0.40\textwidth]{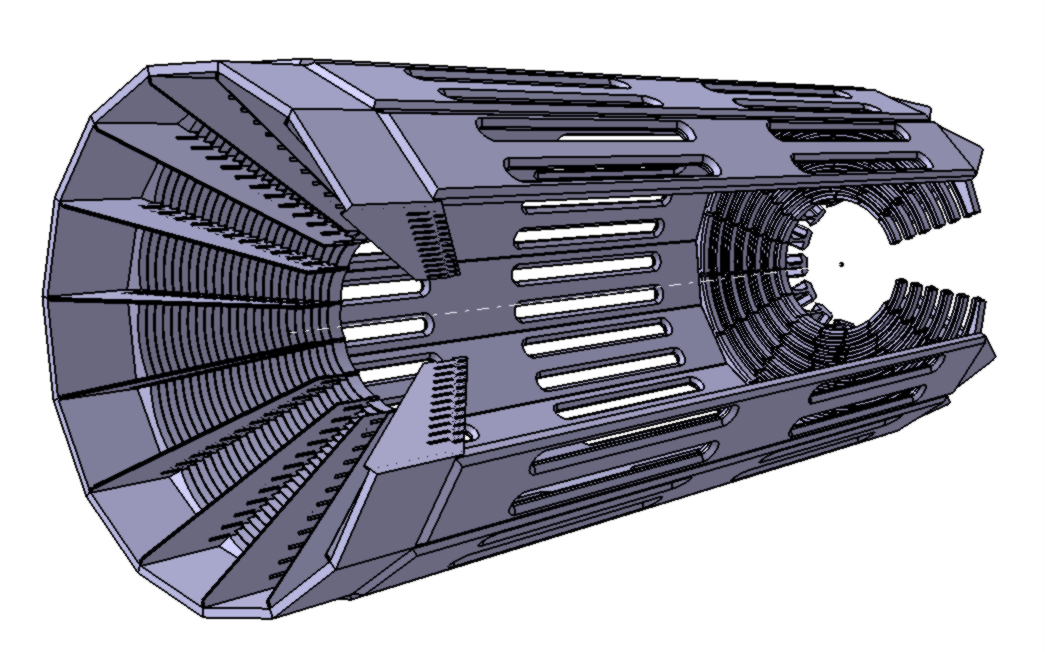}
        \caption{Upstream (left) and downstream (right) ends of the prototype 
        detector in computer assisted design (CAD) with all the sectors included.  \label{wholeView}}
    \end{center}
\end{figure}

\begin{figure}[tbp]
    \begin{center}
        \includegraphics[width=0.4\textwidth]{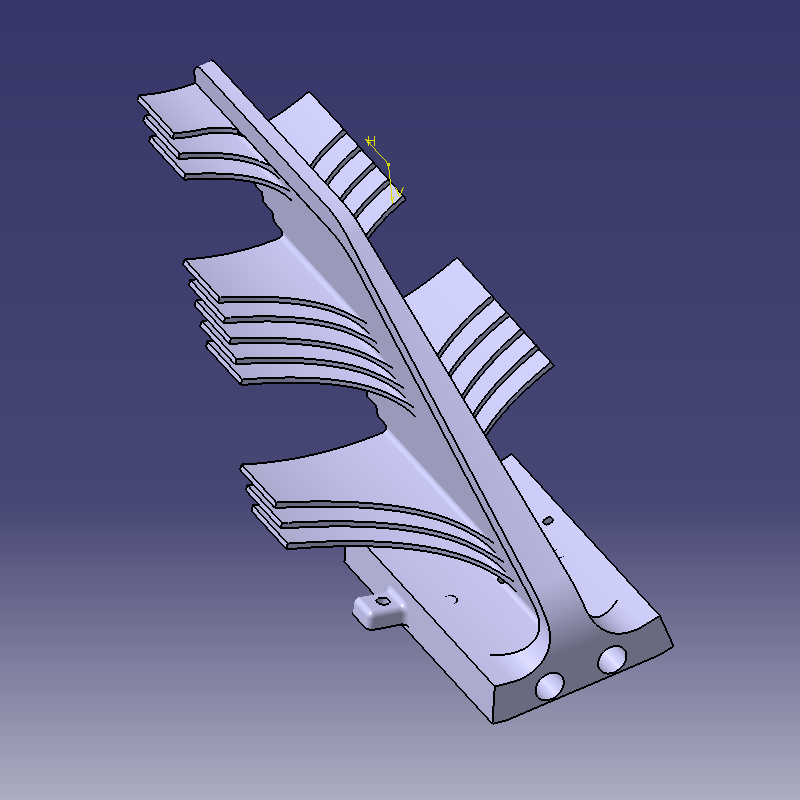}
        \includegraphics[width=0.4\textwidth]{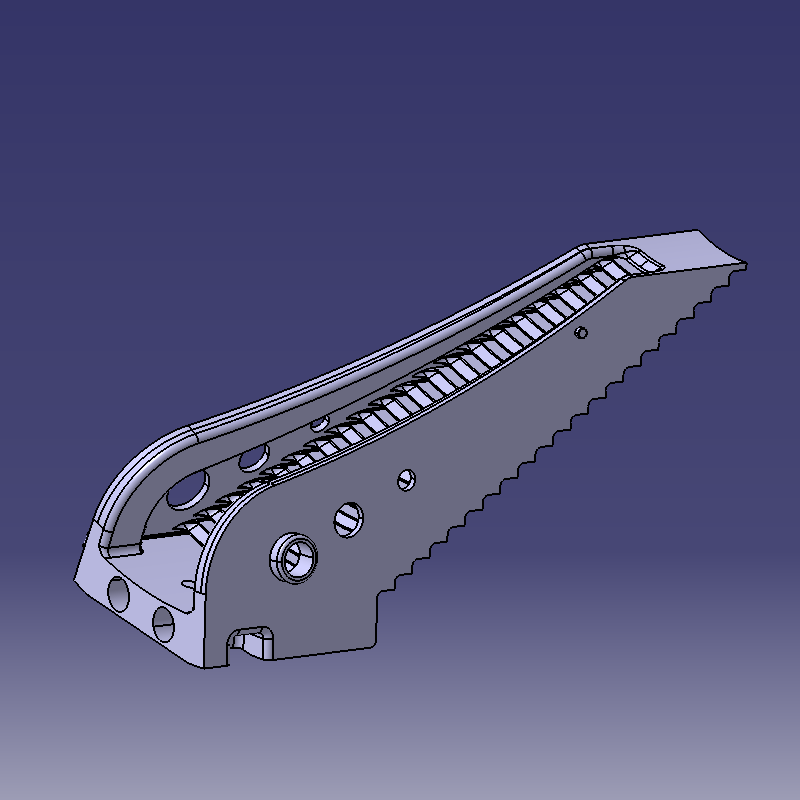}
        \caption{Close up on the CAD of the upstream piece (left) and downstream 
        piece(right) of the drift chamber. Note that the design of the pieces has been
        optimized in comparison of what is shown in Figure~\ref{wholeView}.}
        \label{fig:CAD}
    \end{center}
\end{figure}

To limit issues related to broken wires, we opted for a modular detector made of 
identical sectors. Each sector covers 20$^{\circ}$ of the azimuthal angle 
(Figure~\ref{wholeView}) and can be rotated around the beam axis to be separated 
from the other sectors. This rotation is possible due to the absence of one 
sector, leaving a 20$^{\circ}$ dead angle. Then, if a wire breaks, its sector 
can be removed independently and replaced by a spare. Plastic and metallic 
prototype sectors were made with 3D printers to test the assembling procedure and 
we have started the construction of a full size prototype of one sector.
The shape of each sector is constrained by the position of the wires. It has 
a triangular shape on one side and due to the stereo angle, the other side 
looks like a pine tree with branches alternatively going left and right from 
a central trunk (Figure~\ref{fig:CAD}).

\begin{figure}[tbp]
    \begin{center}
        \includegraphics[width=0.7\textwidth]{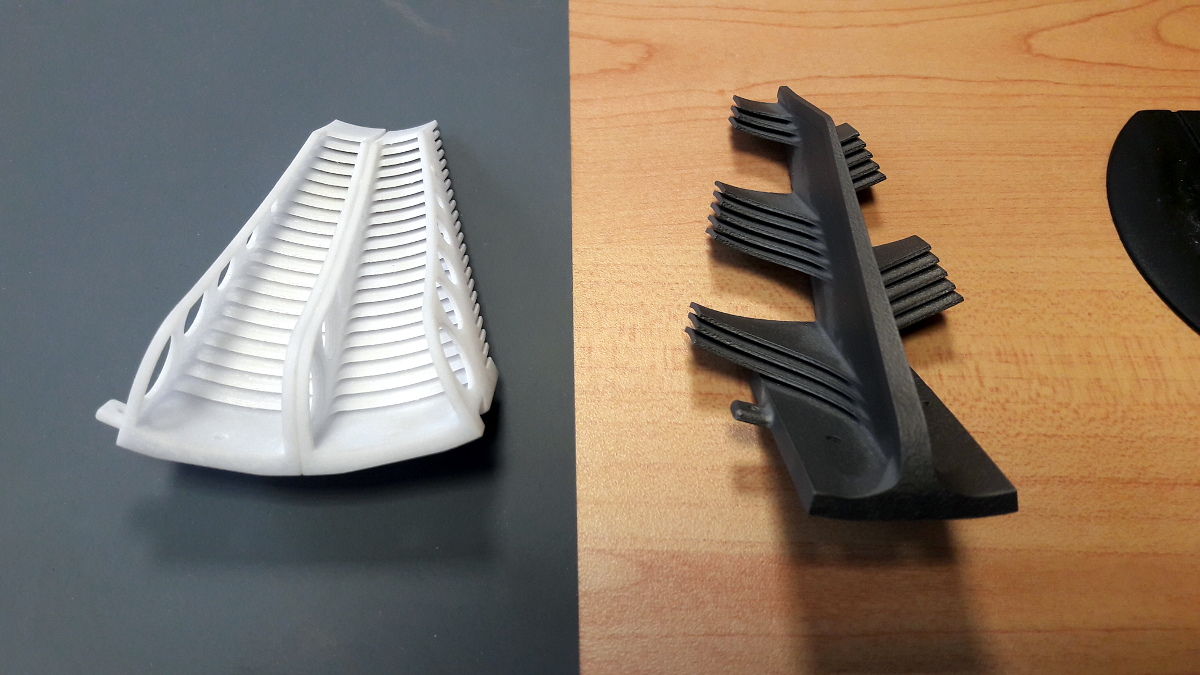}
        \caption{Prototypes for the mechanical parts of the drift chamber made out of plastic
        for the forward part and titanium for the backward.}
        \label{OneSector}
    \end{center}
\end{figure}

Finally, the material used to build the structure will be studied in details with 
future prototypes. Nevertheless, most recent plans are to use high rigidity plastic
in the forward region and metal for the backward structure (as in Figure~\ref{OneSector}). The 
prototypes are not only designed to check the mechanical requirements summarized above 
but also to verify the different cell configurations, and to test the DREAM 
electronics (time resolution, active range, noise). 

\section{Technical contributions from the research groups}
The effort to design, build and integrate the ALERT detector is led by four 
research groups, Argonne National Lab (ANL), 
Institut de Physique Nucl\'eaire d'Orsay (IPNO), Jefferson Lab and Temple 
University (TU). 

Jefferson Lab is the host institution. ANL, IPNO and TU have 
all contributed technically to CLAS12. ANL was involved in the construction of 
the high-threshold Cherenkov counters (HTCC) for CLAS12. ANL has a memorandum 
of understanding (MOU) with JLab on taking responsibility for the HTCC light 
collection system including testing the photomultipliers and the magnetic 
shielding. For the RICH detector for CLAS12, ANL developed full GEANT-4 
simulations in addition to the tracking software. ANL also developed the 
mechanical design of the detector support elements and entrance and exit 
windows in addition to the front-end electronics cooling system. IPNO took 
full responsibility for the design and construction of CLAS12 neutron detector 
(CND). The CND was successfully delivered to Jefferson Lab. TU played an 
important role in the refurbishment of the low threshold Cherenkov counters 
(LTCC), which was completed recently. All 216 photomultipliers have been coated 
with wavelength shifting material (p-Terphenyl) at Temple University, which 
resulted in a significant increase in the number of photoelectrons response.

The three institutions have already shown strong technical commitment to JLab 
12~GeV upgrade, with a focus on CLAS12 and this proposal is a continuation of 
this commitment.

\subsection{Argonne National Laboratory and Temple University}
The ANL medium energy group is responsible for the ALERT scintillator system, 
including scintillation material, light collection device and electronics. 
First results of simulations have led to the design proposed here. This work 
will continue to integrate the scintillator system with the wire chamber. ANL 
will collaborate closely with Temple University to test the light detection 
system. Both institutions will be responsible to assemble and test the detector.

Argonne will provide the electronics and technical support required to
integrate the scintillator detector system into the CLAS12 DAQ. The effort
will minimize the effort required on the part of the Hall B staff.

\subsection{Institut de Physique Nucl\'eaire d'Orsay}
The Institut de Physique Nucl\'eaire d'Orsay is responsible for the wire
chamber and the mechanical structure of the detector design and construction. 
As shown in the proposal, this work
has already started, a first prototype is being built 
to test different cell forms, wire material, wire thickness, 
pressure, etc. This experience will lead to a complete design of the ALERT detector 
integrating the scintillator built at ANL, the gas distribution system and the
electronic connections.

In partnership with {\it CEA Saclay}, IPN Orsay will also test the
use of the DREAM front-end chip for the wire chamber. Preliminary tests were
successful and will continue. The integration of the chip with CLAS12 is
expected to be done by the {\it CEA Saclay}, since they use the same chip to 
readout the CLAS12 MVT. Adaptations to the DAQ necessary when the MVT will be 
replaced by ALERT will be performed by the staff of IPN Orsay.

\subsection{Jefferson Laboratory}\label{sec:jlabContributions}
We expect Jefferson Lab to help with the configuration of the beam line.  
This will include the following items.

\paragraph{Beam Dump Upgrade}
The maximum beam current will be around 1000~nA for the production runs at 
$10^{35}$~cm$^{-2}$s$^{-1}$, which is not common for Hall-B.
To run above 500 nA the ``beam blocker'' will need to be upgraded to handle 
higher power. The beam blocker attenuates the beam seen by the Faraday cup.   
This blocker is constructed of copper and is water cooled. Hall B staff have 
indicated that this is a rather straightforward engineering task and has no 
significant associated costs~\cite{beamBlocker}. 

\paragraph{Straw Target}
We also expect JLab to design and build the target for the experiment as it 
will be a very similar target as the ones build for CLAS BONuS and eg6 runs.
See section \ref{sec:targetCell} for more details. 

\paragraph{Mechanical Integration}
We also expect Jefferson Laboratory to provide assistance in the detector 
installation in the Hall. This will include providing designers at ANL and IPNO 
with the technical drawings required to integrate ALERT with CLAS12. We will 
also need some coordination between designers to validate the mechanical 
integration. 

\paragraph{CLAS12 DAQ Integration}
We also will need assistance in connecting the electronics of ALERT to the 
CLAS12 data acquisition and trigger systems. This will also include help 
integrating the slow controls into the EPICs system.

\chapter{Proposed Measurements}
\label{chap:reach}
In light of the physics motivation presented and the capabilities 
of the new ALERT detector, we propose to measure the tagged 
deep-inelastic scattering off $^2$H and $^4$He and for a range of 
the recoiling spectator momenta 
$P_{A-1}$ from 70 to 400 MeV/$c$. We choose the helium target
for several reasons, first it is a light gas that can easily be used in
a very light gaseous target allowing to detect very low momentum spectators. 
Also, calculation for FSI are theoretically very challenging, keeping the
number of nucleon low is therefore of great help; moreover, as spectators
get heavier, their detection threshold increases, which explains why we want to use  
low $A$ target. The reactions we are going to study are:
\begin{itemize}
\item $^2$H$(e, e' p)X$ --- bound neutron;
\item $^4$He$(e, e'~^3\mathrm{H})X$ --- bound proton;
\item $^4$He$(e, e'~^3\mathrm{He})X$ --- bound neutron;
\end{itemize}
\subsection{Monte-Carlo Simulation}
To estimate the rates of our experiment and provide meaningful estimates of our
statistical error bars, we developed a Monte-Carlo simulation based on PYTHIA 
to which we added nuclear Fermi motion effects. The interaction on the nucleon is generated 
in a basic impulse approximation, neglecting the off-shellness of the target 
nucleon. We simulate the Fermi motion of the nucleons in the target nuclei 
according to the distribution provided by AV18+UIX potentials~%
\cite{Wiringa1995,Pudliner1997,Wiringa}. This leads to a target nucleon with 
momentum $\vec p_n$ and a nuclear spectator generated with a kinematic 
opposite to the interacting nucleon, -$\vec p_n$. The PYTHIA Monte-Carlo 
provides simulation for the DIS interaction and the fragmentation of the 
partons, we do not include nuclear effects such as FSI here. This should
not be an issue for our estimate as our key measurements are focused on
the parts of the phase space where the FSI are small.\\

In the simulation, we select DIS by requesting $Q^2 > 1.5$~GeV$^2$ and 
$W > 2$~GeV. These are the same for all figures, indication on the figures 
are for the theoretical predictions. In our experimental configuration and with 
the cut described above, we expect $\langle Q^2 \rangle \sim 3$~GeV$^2$.\\

The generated final-state particles undergo acceptance tests. Electrons, 
which will be detected by the forward detector, are treated by a GEANT4 
Monte-Carlo simulation of CLAS12. The recoiling nuclei (including protons) 
acceptance is based on the GEANT4 simulation described in section~\ref{sec:sim} 
and represented in Figure~\ref{fig:acceptance}. On top of these estimates, 
we apply an overall 75\% efficiency to this detection settings to account for 
the fiducial cuts and detector inefficiencies.
\subsection{Beam Time Request}
We estimate, based on past measurements with CLAS~\cite{Dupre:2011afa}, that 
the ratios we want to measure will be affected by systematic errors of $\sim3$ 
percents. The dominant factor being associated to acceptance corrections. Our beam time request, 
allows to have the statistical error bars of our key measurement 
(Figure~\ref{fig:ratio_a_proj} right) comparable to the systematic ones. Assuming the luminosity
of $3.10^{34}$~cm$^{-2}$s$^{-1}$, the beam time request for proposed measurements is of   
20 days for each target.
\section{Projections}
\begin{figure}[tbp]
  \begin{center}
    \includegraphics[angle=0, width=0.65\textwidth]{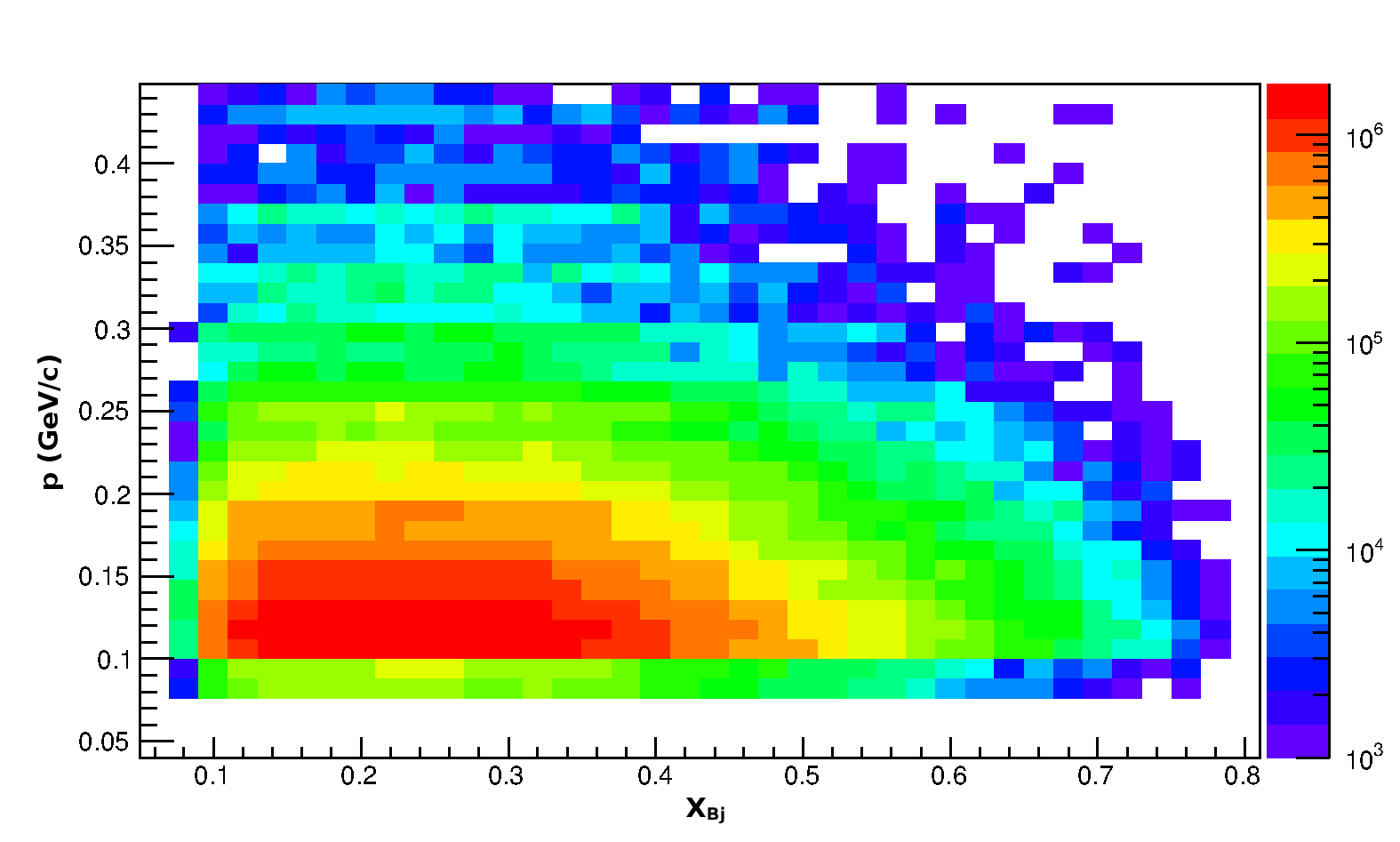}
    \caption{Expected event count as a function of $x_B$ and the recoil 
             momentum of the $^3$H from a $^4$He target.}
    \label{fig:xb_vs_p}
  \end{center}
\end{figure}
Based on our simulation, we determine the available kinematic range and 
production rates accessible for each channel. The $x_B$ and recoil momenta 
distributions are illustrated in Figure~\ref{fig:xb_vs_p} for tagged $^3H$ out 
of an $^4$He target. This figure shows the available phase space for a 
measurement of the bound proton structure function. 
\subsection{Testing the Spectator Model}
\begin{figure}[tbp]
  \begin{center}
    \includegraphics[angle=0, width=0.56\textwidth]{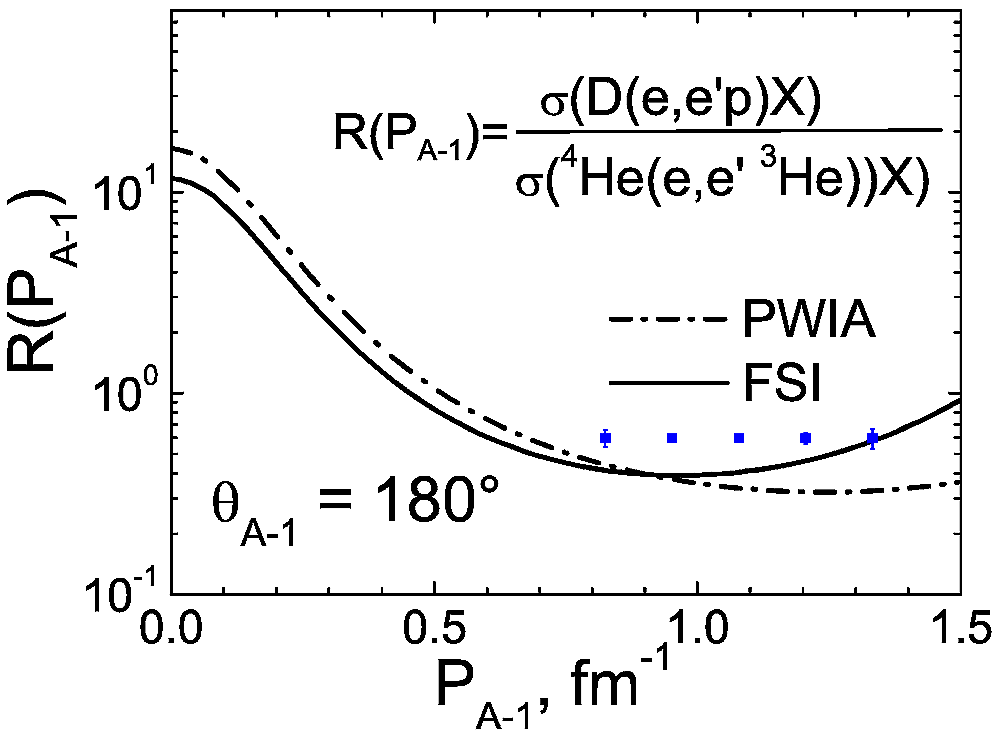}
    \caption{This figure is similar to Figure~\ref{fig:ratio_spec}, it shows 
the predictions for the ratio $R(A,A',|\vec P_{A-1}|)$ for CLAS12 
kinematic~\cite{CiofidegliAtti1999,CiofidegliAtti2012} 
compared to our projected statistical 
error bars (blue points).}
    \label{fig:ratio_spec_proj}
  \end{center}
\end{figure}

The projections presented in Figure~\ref{fig:ratio_spec_proj} show the capability 
of this experiment to measure cross section ratios of DIS on a bound nucleon in 
light nuclei and, therefore, our capability to check the validity of the 
spectator model used by the theoretical predictions. Statistical error bars in 
this figure are very modest and even smaller than the points on the logarithmic
scale. It is therefore clear that we have capability to test the spectator model 
in more details with multi-dimensional binning as was done in~\cite{Cosyn:2010ux} 
and for the first time perform similar studies on helium. However, we do not have 
specific model predictions for such measurement to compare with at the moment. \\ 
Other tests have been proposed to insure that outgoing pions are formed far
enough from the nuclei to limit FSI~\cite{Egiian:1994ey,Frankfurt:1994kt} and
understand the color transparency effect associated. This study is most 
sensitive to spectators emitted at 90$^\circ$, where the effect is larger,
and can then be extrapolated to lower angles. In Ref.~\cite{Cosyn:2016oiq}, testing 
that the $x_B$ scaling holds in tagged DIS is proposed as another way to confirm
the soundness of the method.\\

As shown above, the high statistics of the experiment will open the 
possibility to make several tests of the spectator mechanism and its limits.
In particular it will, for the first time, experimentally explore this
question for helium, opening the way for light nuclei tagged experiments.
\subsection{EMC effect in deuterium}
\begin{figure}[tbp]
  \begin{center}
    \includegraphics[angle=0, width=0.56\textwidth]{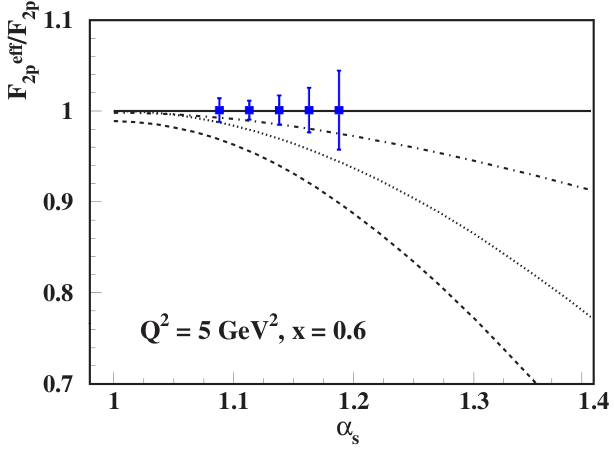}
    \caption{This figure is similar to Figure~\ref{fig:mel}, it shows 
the predictions from~\cite{Melnitchouk1997} 
of the ratio $F_{2p}^{bound}/F_{2p}$ compared to projected statistical 
error bars for the proposed experiment (blue points). Dashed line is a 
prediction for the PLC suppression model, dotted 
is for the $Q^2$-rescaling model, and dot-dashed for the binding/off-shell model.}
    \label{fig:resultmel}
  \end{center}
\end{figure}

As explained in the first chapter, it is possible to enhance the EMC effect in
the deuteron by selecting the highly off-shell nucleons. 
This prediction can be directly tested with our proposed experiment, as shown in
Figure~\ref{fig:resultmel} where we compare our measurement capabilities to Melnitchouk 
{\it et al.}~\cite{Melnitchouk1997} predictions for rescaling models as well as
the PLC suppression model. Note
than in our case, the proton being tagged, we are actually measuring 
$F_{2n}^{bound}/F_{2n}$, but this measurement can be interpreted similarly to
the proton case in regard to the EMC effect. The main limitation in this 
channel is the very fast decrease of the cross section for high $\alpha$ in 
deuterium.\\ 

We point out that such a measurement is also possible with the helium target, 
for either the proton or the neutron. This would allow to reach much higher
$alpha$ without running a prohibitively long experiment. We do not present 
these projections here because, at the moment, there is no theoretical 
predictions for these channels. The main reason 
is the difficulty to extend the calculations to the helium four-body system.
However, we have indications that these kind of studies are on-going in the 
theory community~\cite{Scopetta2016}.
\subsection{Testing the Rescaling Models}
\begin{figure}[tbp]
  \begin{center}
    \includegraphics[angle=0, width=0.46\textwidth]{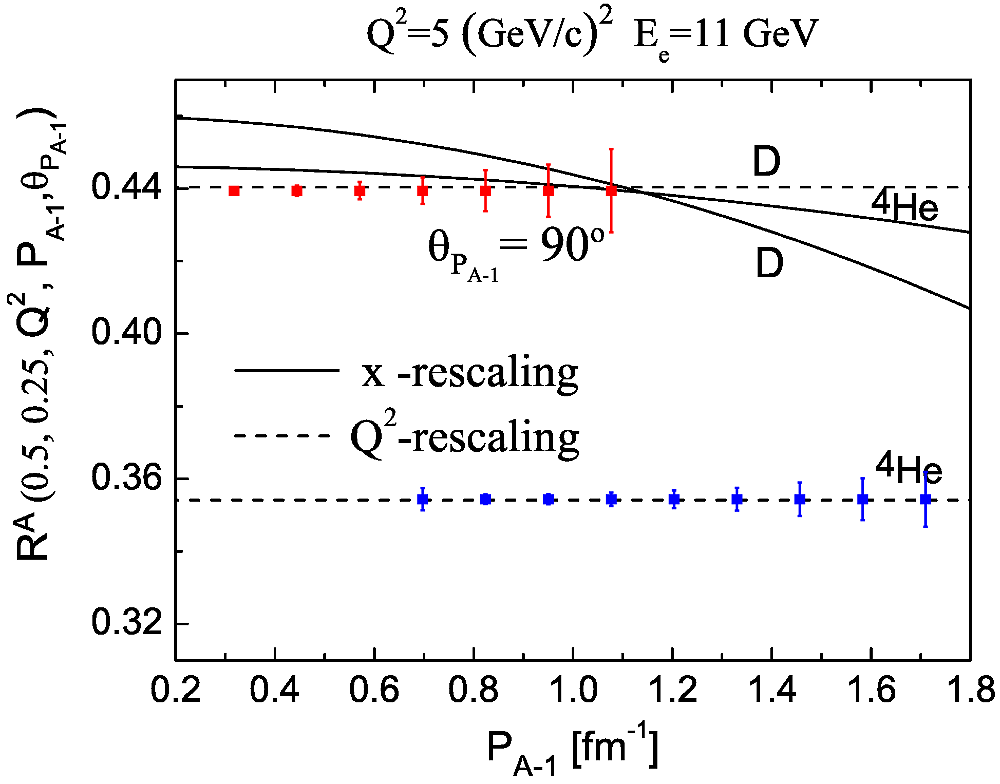}
    \includegraphics[angle=0, width=0.455\textwidth]{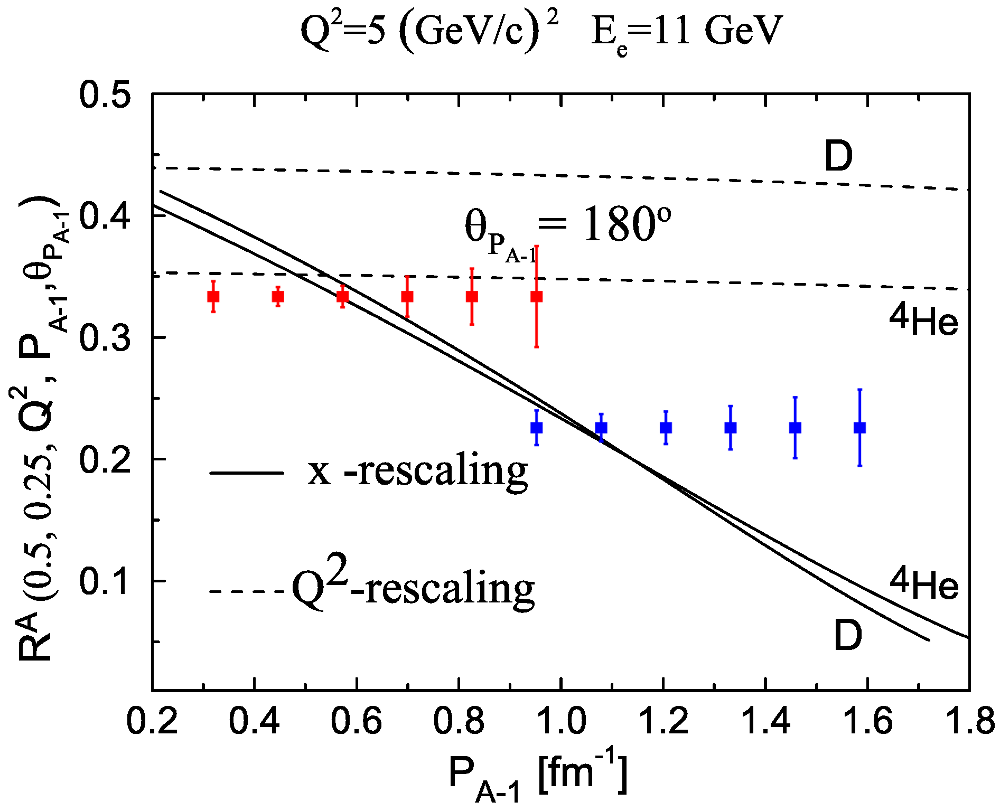}
    \caption{This figure is similar to figure \ref{fig:ratio_a}, it shows 
predictions of the ratio $R^A(x,x')$ for $A = 2$ and $A = 4$ as a function of 
the momentum of the recoil nucleus $A-1$ at perpendicular (left) and backward 
(right) angle. The full and dashed curves are predictions for CLAS12 kinematic~%
\cite{CiofidegliAtti1999,CiofidegliAtti2012} of the $x$-rescaling (binding) 
and $Q^2$-rescaling models, respectively, points are projections for $^2$H (red) 
and $^4$He (blue).}
    \label{fig:ratio_a_proj}
  \end{center}
\end{figure}

The main goal of our experiment is to discriminate decisively between models 
of EMC, Figure~\ref{fig:ratio_a_proj} illustrates this capability. We have here a 
high differentiation power between $x$-rescaling and Q$^2$-rescaling models. We 
note the good coverage and small error bars for $\theta_{P_{A-1}} = 90^\circ$ 
($75 < \theta_{P_{A-1}} <105^\circ$). This is due to the better acceptance for 
this angle. The measurement at backward angle ($\theta_{P_{A-1}} > 150^\circ$), 
however, is much more difficult and is the main constraint driving our beam 
time request. Still, in order to obtain our planned precision with a reasonable 
beam time request, the backward angles are selected from 150$^\circ$ and up instead of 
the 160$^\circ$ which is used for the theory predictions.\\

We notice the complementarity of our choice of targets in the phase space covered, this
is due to the fact that larger recoil nuclei are more absorbed by the target
material and have higher detection threshold. At the same time, the Fermi
momentum is larger in helium allowing better statistics at high $p_{A-1}$. Using
helium is then also an opportunity to explore higher spectator momentum with a 
reasonable beam time request.
\subsection{Tagged EMC Ratio}
\begin{figure}
  \begin{center}
    \includegraphics[angle=0, width=0.7\textwidth]{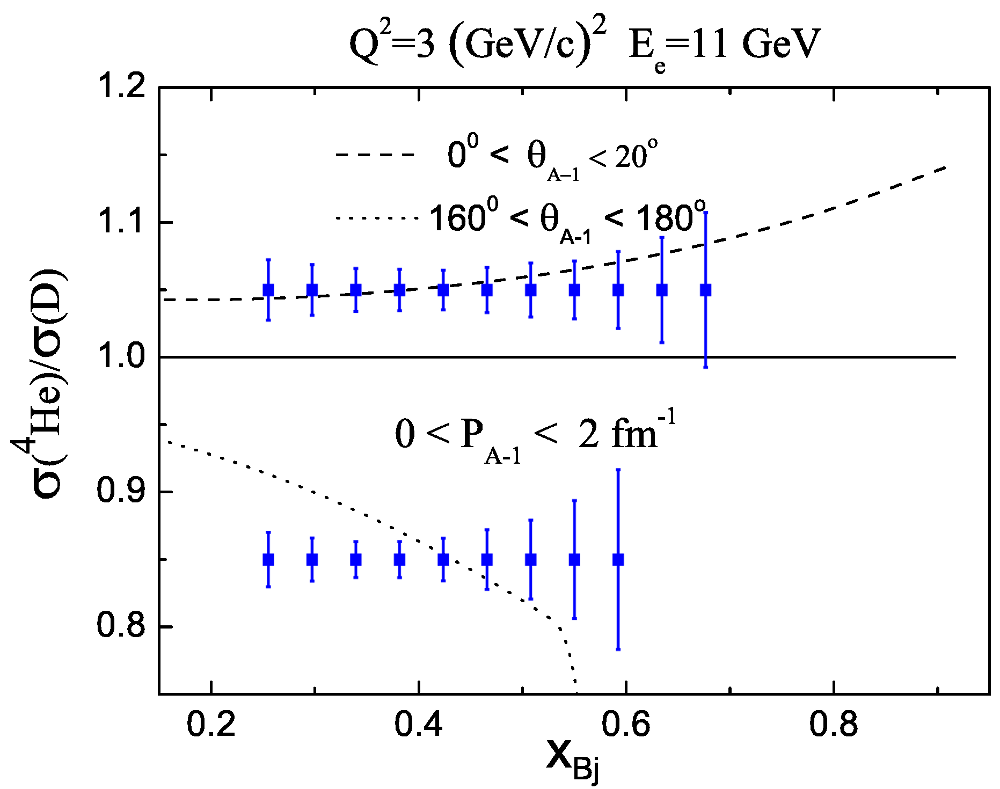}
    \caption{This figure is similar to figure \ref{fig:r0}, it shows the 
semi-inclusive EMC ratio $R_0(x,Q^2)$ as a function of $x_B$ with recoils 
emitted forward and backward, the dashed and dotted curves are predictions for 
the local EMC effect for CLAS12 kinematic~\cite{CiofidegliAtti1999,CiofidegliAtti2012} 
the blue points are statistical error bar projections for our measurement.}
    \label{fig:fig:r0_proj}
  \end{center}
\end{figure}

The experiment can also confront the striking predictions for backward versus 
forward tagged EMC in binding models, as illustrated in the 
Figure~\ref{fig:fig:r0_proj}. We see that the model prediction will be clearly 
tested, however the reach in $x_B$ for the backward recoils is also
strongly constrain by the beam time available for the experiment. Indeed, the 
strongest effect is expected at $x_B \sim 0.5$ for which we need high statistics.\\

The measurement of the tagged EMC ratio is a very good observable even for other
kinds of model, in the low momentum regime one should be able to reproduce very 
nicely the classic EMC effect and then be able to study its dependence to the 
spectator angle and momentum. In general, models based only on off-shellness 
predict no differences between nuclei at a given spectator kinematic. This prediction
can be tested nicely with the measurement presented here.
\subsection{The Flavor Dependent Nuclear Effects}
\begin{figure}
  \begin{center}
    \includegraphics[angle=0, width=0.7\textwidth]{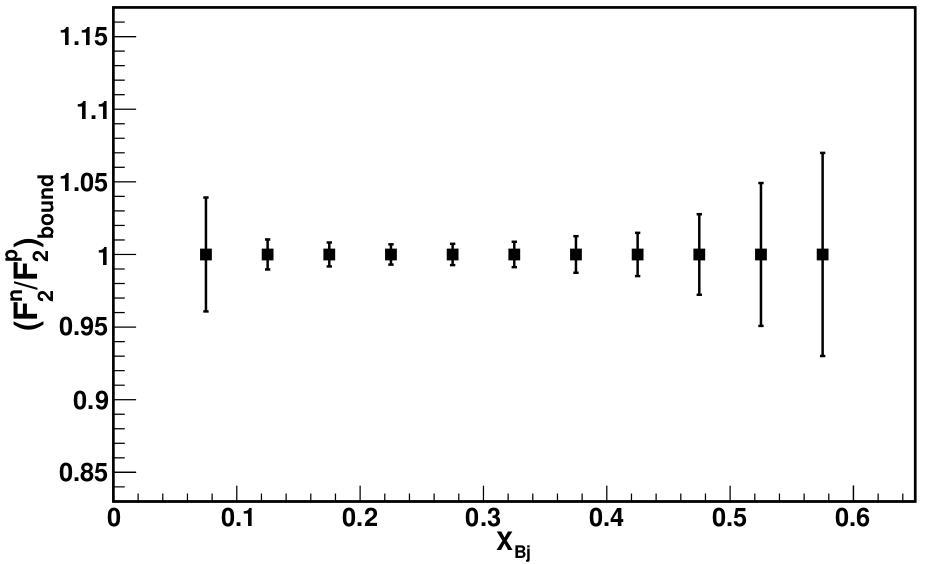}
    \caption{Statistical error bar projections for the ratio $F_2^n/F_2^p$ for 
bound nucleons as a function of $x_{Bj}$ using an $^4$He target.}
    \label{fig:flavor_proj}
  \end{center}
\end{figure}

Finally with our experimental setup, we will explore the flavor dependence of nuclear PDFs. 
Figure~\ref{fig:flavor_proj} illustrates our capabilities, for $^4$He the isovector 
model predicts a ratio of 1~\cite{Cloet2009}, but others 
predict that nuclear effects change the d/u ratio and should therefore be 
observed here~\cite{Brodsky:2004qa}. We will be able to explore any variation 
in bound nucleons with 1 to 2\% statistical error bars -- {\it i.e.} 4\% when including
expected systematic error bars -- from $x_B$ of $0.1$ to $0.5$.

\chapter*{Summary and Answers to PAC44\markboth{\bf Summary and Answers to PAC44}{}}
\label{chap:conclusion}
\addcontentsline{toc}{chapter}{Summary and Answers to PAC44}

\setlength\parskip{\baselineskip}%
\section*{Answers to PAC44 issues\markboth{\bf Answer to PAC44 issues}{}}
\addcontentsline{toc}{section}{Answer to PAC44 issues}

{ \it \textbf{Issues:}}

{ \it 
The Drift Chamber/scintillator technology needs to be demonstrated. We observe 
that a strong program of prototype studies is already underway. }

{\bf Answer:} We feel the technology has no major unknowns, wire chambers and
scintillators have been used for decades as detectors of low energy nuclei
and their properties have been well established. We present in the
proposal a conceptual design demonstrating the feasibility of the detector,
it is common practice to work on the optimization of a certain number of 
parameters after the proposal is approved. In particular, because it is easier to
fund and man a project that has an approved status than a future proposal.
Nevertheless, we remain open to discuss 
the topic in more depth if the committee has any concerns.

{ \it 
The TAC report voiced concerns about the length of the straw cell target and 
the substantial effort needed to integrate the DAQ for this detector into the 
CLAS12 DAQ. }

{\bf Answer:} The TAC and PAC44 raised concerns about the target cell. We have 
added extra discussion  in section \ref{sec:targetCell}, which includes a table 
of existing or planned targets that are similar to the one we proposed. In summary,
our proposed target is twice as wide as the ones used in the 6 
GeV era for the BONuS and eg6 run and should therefore cause no issues. Note 
that the experiment 12-06-113 (BONuS12) is approved with a longer and thinner 
target. Their design will be reviewed by JLab for their experiment readiness 
review (ERR) before the PAC45 meeting. The result of this review should settle 
the question, but in any case, we propose a safer solution based on the 
successful experiments of the 6 GeV era. 

The TAC and PAC44 raised issues regarding integration of ALERT into the CLAS12 
DAQ. First, they raised a concern that the resources necessary for this integration 
are not clearly identified. We have added text in section 
\ref{sec:jlabContributions} outlining the resources provided by each group 
and the technical support they are expected to provide. Secondly, they mentioned 
a concern about the ``substantial effort needed to integrate the DAQ for this 
detector into the CLAS12 DAQ''. We want to emphasize that the read-out systems for 
ALERT are already being used in the CLAS12 DAQ to readout Micromegas detectors. 
Therefore, we will use and build on the experience gained from these systems.

{ \it 
The proposal does not clearly identify the resources (beyond generic 
JLAB/CLAS12 effort) necessary for DAQ integration which may be a substantial 
project. }

{\bf Answer:} As mentioned above, we do not feel this contribution is major,
nevertheless we made this part clearer in the proposal.

{ \it 
During review the collaboration discovered an error in converting the 
luminosity to beam current. This resulted  in a revision that will either 
require doubling the current or the target density. The beam current change 
would require changes to the Hall B beam dump, while raising the target density 
could impact the physics reach of the experiment by raising the minimum 
momentum threshold. }

{\bf Answer:} During the PAC44 proposal submission process the wrong beam 
current was requested. It was a factor of 2 too low. This increased beam 
current brought into contention the issue of possible Hall B beam current 
limits. We chose to use the higher beam current in this new version. 
Based on discussions with the Hall-B and accelerator staff, the only
necessary upgrade necessary to run at \SI{1}{\uA} is with the Hall-B beam blocker.

{ \it 
The precise interplay between final state interactions (FSI) and the tails of 
the initial state momentum distribution in DVCS on 4He was a topic of some 
debate. The collaboration makes an argument that the excellent acceptance of 
the apparatus allows novel constraints that allow selection of kinematic ranges 
where FSI is suppressed. While the originally suggested method to unambiguously 
identify areas of FSI was revised during the review, the committee remains 
unconvinced that the new kinematic selections suggested do not also cut into 
interesting regimes for the initial state kinematics. The committee believes 
that this is model dependent and would like to see more quantitative arguments 
than were provided in this version of the proposal. }

{\bf Answer:} We acknowledge there was an overstatement of the possibilities 
of the Tagged-DVCS proposal on this topic, this has been corrected. We now show 
a reduction, in opposition to the complete suppression previously claimed, in 
events that differ from the PWIA result. This finding is based on a simulation 
using a simple model of FSIs together with a Monte-Carlo event generator.

{ \it 
\textbf{Summary:}

The committee was generally enthusiastic about the diverse science program 
presented in this proposal; in particular the tagged EMC studies and the unique 
study of coherent GPD's on the 4He nucleus. However, the substantial 
modifications made in the proposal during review indicate that it could be 
substantially improved on a reasonably short time scale. We would welcome a new 
proposal that addresses the issues identified by the committee and by the 
collaboration. }

{\bf Answer:} We hope that the new proposals will answer all the questions 
raised by the PAC44 and will make the physics case even more compelling.

{ \it 
We also note that there are multiple experiments, proposed and 
approved, to study the EMC effect, including several with novel methods of 
studying the recoil system. We appreciate the comparisons of recoil 
technologies in this proposal and would welcome a broader physics discussion of 
how the proposed measurements contribute to a lab-wide strategy for exploring 
the EMC effect. }

{\bf Answer:} While no strategy document has been drafted after them, we want to point out
to the PAC that the community of physicist interested by the partonic 
structure of nuclei meets regularly, with often a large focus on what can be done at JLab
(see workshops at Trento\footnote{New Directions in Nuclear Deep Inelastic Scattering \url{http://www.ectstar.eu/node/1221}}, 
Miami\footnote{Next generation nuclear physics with JLab12 and EIC \url{https://www.jlab.org/indico/event/121/}}, 
MIT\footnote{Quantitative challenges in EMC and SRC Research and Data-Mining \url{http://web.mit.edu/schmidta/www/src_workshop/}}, 
and Orsay\footnote{Partons and Nuclei \url{https://indico.in2p3.fr/event/14438/}} for example). 
Nonetheless, we added
in the tagged EMC proposal summary an extension about the 12~GeV approved experiments  
related to the EMC effect. This short annex will hopefully clarify the context 
and the uniqueness of the present experiments.

\section*{Relation to other EMC related proposals\markboth{\bf Relation to other EMC related proposals}{}}
\addcontentsline{toc}{section}{Relation to other EMC related proposals}

As noted by the PAC44, there is a wide program of experiments focused on the 
EMC effect that are already planned
for the 12~GeV era of Jefferson Lab. We will try here to summarize this existing
effort and explain why we think our approach is unique and essencial to a full
understanding of the EMC effect. 

The first, most obvious, class of experiments approved to study the EMC effect
(experiments 12-10-008 and 12-10-103) are dedicated to its direct measurement 
in its traditional form, by measuring the inclusive ratio $\sigma_A/\sigma_D$. 
We note that these proposals concentrate on light nuclei, which is common with 
our proposal and many others. The motivation for this choice is to look at 
nuclei which nuclear structure in term of nucleons is well understood to 
isolate partonic effects. These mesurements should provide high precision data
allowing to understand the dependence of the EMC effect on local density and
isospin symmetry.

Another approche is focused on deepening our understanding of the EMC effect
(experiments 12-14-001 and 12-14-002) by measuring other inclusive observables 
like the spin structure functions
or extracting the ratio $R = \sigma_L/\sigma_T$ in nuclei. These measurement 
will tell if the EMC effect is a general effect affecting similarly all structure
functions or a more complex effect with different impact on these different 
observables. These are focused on very specific questions that once answered can
have a major impact on our understanding of the EMC effect.

The short range correlations (SRC) approach is the focus of two
EMC experiments (12-11-107 and 12-11-003A). These are exploring the 
link between SRC and the EMC effect in deuterium. This
is motivated by the recent finding that the strength of the EMC effect seems
correlated to the number of SRC pairs in a nucleus. A series of other 
experiments (12-06-105, 12-11-112 and 12-14-011) are focused specificly on SRC, while
these can be partially motivated by the possible links to EMC, they do not
address it directly. They probably should not be considered as part of the 
EMC program pursued in JLab.

Our proposal has a different approach to the EMC effect than all previously
approved experiments. Our goal is to understand, in general, the link between the EMC 
effect and the internal nuclear dynamic using tagging. To achieve this
goal, we decided to use two nuclear targets to compare the effect observed 
in different nuclei with similar spectator kinematics and in the same nuclei
with different spectator kinematics. In our view this is necessary to ensure that we
are really understanding the measured effect and we feel that a measurement on
one of these targets only will always be subject to diverging interpretations. 
Also, we decided to focus on the mean field energies rather than high energies
linked to SRC, but in a heavier helium nuclei. In this way we observe spectators 
of similar momentum classified SRC in deuterium but mean field in helium and can 
also assess if the effect depends purely on the momentum or other factors as well.

\section*{Summary and Beam Time Request\markboth{\bf Summary and Beam Time Request}{}}
\addcontentsline{toc}{section}{Summary and Beam Time Request}

In summary, we proposed a tagged DIS measurement on light nuclear targets 
($^2$H and $^4$He) by detecting the backward recoiling spectators. By taking the 
advantage of the high luminosity and large kinematic coverage of CLAS12, we 
will be able to cover a wide range in spectator kinematic insuring a good
control over FSI effects.\\ 

In order to make this measurement, we propose to use a new  
recoil detector to fit our experimental needs in term of low energy nuclei 
detection. The detector is designed such that it will provide good timing 
resolution and particle identification. Prototyping of this detector is
currently underway in Orsay as part of a larger R\&D program on drift chambers.\\

We propose to measure various tagged ratios and double ratios with their 
dependencies on the recoil kinematics. These measurements will provide very
stringent tests of numerous models for the EMC and anti-shadowing effects
and, more importantly, a model independent insight into the origin of the EMC 
effect in term of $x_B$ or $Q^2$-rescaling.\\

In order to achieve all the goals presented in this proposal, we need
45 days of running. With 11~GeV electron beam at 
$3.10^{34}$~cm$^{-2}$s$^{-1}$ ($= 500$~nA) with helium and deuterium targets 
(20 days each) and 5 days of commissioning of the ALERT detector at
2.2~GeV at various luminosities with helium and hydrogen 
targets.

\bibliographystyle{ieeetr}
\bibliography{biblio}

\begin{thebibliography}{10}

\bibitem{Aubert:1983xm}
J.~J. Aubert {\em et~al.}, ``{The ratio of the nucleon structure functions
  $F2_n$ for iron and deuterium},'' {\em Phys. Lett.}, vol.~B123, pp.~275--278,
  1983.

\bibitem{Ashman:1988bf}
J.~Ashman {\em et~al.}, ``{Measurement of the Ratios of Deep Inelastic Muon -
  Nucleus Cross-Sections on Various Nuclei Compared to Deuterium},'' {\em
  Phys.Lett.}, vol.~B202, p.~603, 1988.

\bibitem{Arneodo:1988aa}
M.~Arneodo {\em et~al.}, ``{Shadowing in Deep Inelastic Muon Scattering from
  Nuclear Targets},'' {\em Phys.Lett.}, vol.~B211, p.~493, 1988.

\bibitem{Arneodo:1989sy}
M.~Arneodo {\em et~al.}, ``{Measurements of the nucleon structure function in
  the range $0.002-{\rm GeV}^2 < x < 0.17-{\rm GeV}^2$ and $0.2-GeV^2 < q^2 <
  8-GeV^2$ in deuterium, carbon and calcium},'' {\em Nucl.Phys.}, vol.~B333,
  p.~1, 1990.

\bibitem{Gomez:1993ri}
J.~Gomez {\em et~al.}, ``{Measurement of the A-dependence of deep inelastic
  electron scattering},'' {\em Phys.Rev.}, vol.~D49, pp.~4348--4372, 1994.

\bibitem{Allasia:1990nt}
D.~Allasia {\em et~al.}, ``{Measurement of the neutron and the proton F2
  structure function ratio},'' {\em Phys.Lett.}, vol.~B249, pp.~366--372, 1990.

\bibitem{Seely2009}
J.~Seely {\em et~al.}, ``{New measurements of the EMC effect in very light
  nuclei},'' {\em Phys. Rev. Lett.}, vol.~103, p.~202301, 2009.

\bibitem{Geesaman1995}
D.~F. Geesaman, K.~Saito, and A.~W. Thomas, ``{The nuclear EMC effect},'' {\em
  Ann. Rev. Nucl. Part. Sci.}, vol.~45, pp.~337--390, 1995.

\bibitem{Norton2003}
P.~R. Norton, ``{The EMC effect},'' {\em Rept. Prog. Phys.}, vol.~66,
  pp.~1253--1297, 2003.

\bibitem{Malace:2014uea}
S.~Malace, D.~Gaskell, D.~W. Higinbotham, and I.~Cloet, ``{The Challenge of the
  EMC Effect: existing data and future directions},'' {\em Int.J.Mod.Phys.},
  vol.~E23, p.~1430013, 2014.

\bibitem{Ericson1983}
M.~Ericson and A.~W. Thomas, ``{Pionic Corrections and the EMC Enhancement of
  the Sea in Iron},'' {\em Phys. Lett.}, vol.~B128, p.~112, 1983.

\bibitem{Dunne1985}
G.~V. Dunne and A.~W. Thomas, ``{DEEP INELASTIC SCATTERING AS A PROBE OF
  NUCLEON AND NUCLEAR STRUCTURE},'' {\em Nucl. Phys.}, vol.~A446,
  pp.~437c--443c, 1985.

\bibitem{Akulinichev1985}
S.~V. Akulinichev, S.~A. Kulagin, and G.~M. Vagradov, ``{The Role of Nuclear
  Binding in Deep Inelastic Lepton Nucleon Scattering},'' {\em Phys. Lett.},
  vol.~B158, pp.~485--488, 1985.

\bibitem{Jung1988}
H.~Jung and G.~A. Miller, ``{NUCLEONIC CONTRIBUTION TO LEPTON NUCLEUS DEEP
  INELASTIC SCATTERING},'' {\em Phys. Lett.}, vol.~B200, pp.~351--356, 1988.

\bibitem{Close1983}
F.~E. Close, R.~G. Roberts, and G.~G. Ross, ``{The Effect of Confinement Size
  on Nuclear Structure Functions},'' {\em Phys. Lett.}, vol.~B129, p.~346,
  1983.

\bibitem{Nachtmann1984}
O.~Nachtmann and H.~J. Pirner, ``{Color Conductivity in Nuclei and the EMC
  Effect},'' {\em Z. Phys.}, vol.~C21, p.~277, 1984.

\bibitem{Jaffe1984}
R.~L. Jaffe, F.~E. Close, R.~G. Roberts, and G.~G. Ross, ``{On the Nuclear
  Dependence of Electroproduction},'' {\em Phys. Lett.}, vol.~B134, p.~449,
  1984.

\bibitem{Close1988}
F.~E. Close, R.~G. Roberts, and G.~G. Ross, ``{Factorization Scale
  Independence, the Connection between Alternative Explanations of the EMC
  Effect and QCD Predictions for Nuclear Properties},'' {\em Nucl. Phys.},
  vol.~B296, p.~582, 1988.

\bibitem{Frankfurt:1985cv}
L.~Frankfurt and M.~Strikman, ``{POINT - LIKE CONFIGURATIONS IN HADRONS AND
  NUCLEI AND DEEP INELASTIC REACTIONS WITH LEPTONS: EMC AND EMC LIKE
  EFFECTS},'' {\em Nucl.Phys.}, vol.~B250, pp.~143--176, 1985.

\bibitem{Frankfurt:1981mk}
L.~Frankfurt and M.~Strikman, ``{High-Energy Phenomena, Short Range Nuclear
  Structure and QCD},'' {\em Phys.Rept.}, vol.~76, pp.~215--347, 1981.

\bibitem{Kumano1990}
S.~Kumano and F.~E. Close, ``{DEPENDENCE OF THE EMC EFFECT ON NUCLEAR
  STRUCTURE},'' {\em Phys. Rev.}, vol.~C41, pp.~1855--1858, 1990.

\bibitem{ciofiliuti1991}
C.~Ciofi~degli Atti and S.~Liuti, ``{Can nuclear binding explain the classical
  EMC effect?},'' {\em Nucl.Phys.}, vol.~A532, pp.~241--248, 1991.

\bibitem{CiofidegliAtti1999}
C.~Ciofi~degli Atti, L.~P. Kaptari, and S.~Scopetta, ``{Semi-inclusive deep
  inelastic lepton scattering off complex nuclei},'' {\em Eur. Phys. J.},
  vol.~A5, pp.~191--207, 1999.

\bibitem{Higinbotham:2010ye}
D.~Higinbotham, J.~Gomez, and E.~Piasetzky, ``{Nuclear Scaling and the EMC
  Effect},'' 2010.

\bibitem{Weinstein:2010rt}
L.~Weinstein {\em et~al.}, ``{Short Range Correlations and the EMC Effect},''
  {\em Phys.Rev.Lett.}, vol.~106, p.~052301, 2011.

\bibitem{Geesaman:2015fha}
A.~Aprahamian {\em et~al.}, ``{Reaching for the horizon: The 2015 long range
  plan for nuclear science},'' 2015.

\bibitem{Frankfurt:1993sp}
L.~Frankfurt, M.~Strikman, D.~Day, and M.~Sargsian, ``{Evidence for short range
  correlations from high Q**2 (e, e-prime) reactions},'' {\em Phys.Rev.},
  vol.~C48, pp.~2451--2461, 1993.

\bibitem{Egiyan:2003vg}
K.~Egiyan {\em et~al.}, ``{Observation of nuclear scaling in the A(e, e-prime)
  reaction at x(B) greater than 1},'' {\em Phys.Rev.}, vol.~C68, p.~014313,
  2003.

\bibitem{Egiyan:2005hs}
K.~Egiyan {\em et~al.}, ``{Measurement of 2- and 3-nucleon short range
  correlation probabilities in nuclei},'' {\em Phys.Rev.Lett.}, vol.~96,
  p.~082501, 2006.

\bibitem{Hen:2014nza}
O.~Hen {\em et~al.}, ``{Momentum sharing in imbalanced Fermi systems},'' {\em
  Science}, vol.~346, pp.~614--617, 2014.

\bibitem{CiofidegliAtti:2007vx}
C.~Ciofi~degli Atti {\em et~al.}, ``{On the dependence of the wave function of
  a bound nucleon on its momentum and the EMC effect},'' {\em Phys. Rev.},
  vol.~C76, p.~055206, 2007.

\bibitem{Baillie:2011za}
N.~Baillie {\em et~al.}, ``{Measurement of the neutron F2 structure function
  via spectator tagging with CLAS},'' {\em Phys.Rev.Lett.}, vol.~108,
  p.~142001, 2012.

\bibitem{eg6_note}
{M. Hattawy {\it et al.} (EG6 Working Group)}, ``{Deeply Virtual Compton
  Scattering off $^4$He},'' {\em CLAS internal analysis note}, 2016.

\bibitem{CiofidegliAtti2003}
C.~Ciofi~degli Atti and B.~Z. Kopeliovich, ``{Final state interaction in
  semi-inclusive DIS off nuclei},'' {\em Eur. Phys. J.}, vol.~A17,
  pp.~133--144, 2003.

\bibitem{ciofi2004}
C.~Ciofi~degli Atti, L.~Kaptari, and B.~Kopeliovich, ``{Final state interaction
  effects in semiinclusive DIS off the deuteron},'' {\em Eur.Phys.J.},
  vol.~A19, pp.~145--151, 2004.

\bibitem{Alvioli:2006jd}
M.~Alvioli, C.~Ciofi~degli Atti, and V.~Palli, ``{Slow proton production in
  semi-inclusive DIS off nuclei: The Role of final state interaction},'' {\em
  Nucl.Phys.}, vol.~A782, pp.~175--178, 2007.

\bibitem{Palli2009}
V.~Palli {\em et~al.}, ``{Slow Proton Production in Semi-Inclusive Deep
  Inelastic Scattering off Deuteron and Complex Nuclei: Hadronization and Final
  State Interaction Effects},'' {\em Phys.Rev.}, vol.~C80, p.~054610, 2009.

\bibitem{frankfurt1988}
L.~Frankfurt and M.~Strikman, ``{Hard Nuclear Processes and Microscopic Nuclear
  Structure},'' {\em Phys.Rept.}, vol.~160, pp.~235--427, 1988.

\bibitem{CiofidegliAtti:1993ep}
C.~Ciofi~degli Atti and S.~Simula, ``{Slow proton production in semiinclusive
  deep inelastic lepton scattering off nuclei},'' {\em Phys.Lett.}, vol.~B319,
  pp.~23--28, 1993.

\bibitem{Melnitchouk1997}
W.~Melnitchouk, M.~Sargsian, and M.~I. Strikman, ``{Probing the origin of the
  EMC effect via tagged structure functions of the deuteron},'' {\em Z. Phys.},
  vol.~A359, pp.~99--109, 1997.

\bibitem{CiofidelgiAtti:2007qu}
C.~Ciofi~delgi Atti and L.~Kaptari, ``{A Non factorized calculation of the
  process He-3(e,e-prime p) H-2 at medium energies},'' {\em Phys.Rev.Lett.},
  vol.~100, p.~122301, 2008.

\bibitem{Atti:2010yf}
C.~Ciofi~degli Atti and L.~Kaptari, ``{Semi-inclusive Deep Inelastic Scattering
  off Few-Nucleon Systems: Tagging the EMC Effect and Hadronization Mechanisms
  with Detection of Slow Recoiling Nuclei},'' {\em Phys.Rev.}, vol.~C83,
  p.~044602, 2011.

\bibitem{Melnitchouk:1993nk}
W.~Melnitchouk, A.~W. Schreiber, and A.~W. Thomas, ``{Deep inelastic scattering
  from off-shell nucleons},'' {\em Phys. Rev.}, vol.~D49, pp.~1183--1198, 1994.

\bibitem{klimenko2006}
A.~Klimenko {\em et~al.}, ``{Electron scattering from high-momentum neutrons in
  deuterium},'' {\em Phys.Rev.}, vol.~C73, p.~035212, 2006.

\bibitem{Kaptari:2013dma}
L.~P. Kaptari, A.~Del~Dotto, E.~Pace, G.~Salme', and S.~Scopetta, ``{Distorted
  spin-dependent spectral function of an A=3 nucleus and semi-inclusive deep
  inelastic scattering processes},'' {\em Phys. Rev.}, vol.~C89, no.~3,
  p.~035206, 2014.

\bibitem{Griffioen:2015hxa}
K.~A. Griffioen {\em et~al.}, ``{Measurement of the EMC Effect in the
  Deuteron},'' {\em Phys. Rev.}, vol.~C92, no.~1, p.~015211, 2015.

\bibitem{bonus6}
H.~Fenker {\em et~al.}, ``{BoNus: Development and use of a radial TPC using
  cylindrical GEMs},'' {\em Nucl. Instrum. Meth.}, vol.~A592, pp.~273--286,
  2008.

\bibitem{Sargsian:2005rm}
M.~Sargsian and M.~Strikman, ``{Model independent method for determination of
  the DIS structure of free neutron},'' {\em Phys. Lett.}, vol.~B639,
  pp.~223--231, 2006.

\bibitem{Cosyn:2016oiq}
W.~Cosyn, V.~Guzey, M.~Sargsian, M.~Strikman, and C.~Weiss,
  ``{Electron-deuteron DIS with spectator tagging at EIC: Development of
  theoretical framework},'' {\em EPJ Web Conf.}, vol.~112, p.~01022, 2016.

\bibitem{Brodsky:2004qa}
S.~J. Brodsky, I.~Schmidt, and J.-J. Yang, ``{Nuclear antishadowing in neutrino
  deep inelastic scattering},'' {\em Phys. Rev.}, vol.~D70, p.~116003, 2004.

\bibitem{Cloet2009}
I.~C. Cloet, W.~Bentz, and A.~W. Thomas, ``{Isovector EMC effect explains the
  NuTeV anomaly},'' {\em Phys. Rev. Lett.}, vol.~102, p.~252301, 2009.

\bibitem{CD}
``{CLAS12 Technical Design Report},'' 2008.

\bibitem{bonus12}
M.~Amaryan {\em et~al.}, ``{The Structure of the Free Neutron at Large
  x-Bjorken (PR12-06-113)},'' {\em A proposal to PAC 30}, 2006.

\bibitem{AliceMuonArmChamber}
J.~Peyr\'e, B.~Genolini, and J.~Poutas, ``{A Full-Scale Prototype for the
  Tracking Chambers of the ALICE Muon Spectrometer},'' 1998.

\bibitem{BelleIItdr}
T.~Abe {\em et~al.}, ``{Belle II Technical Design Report},'' 2010.

\bibitem{ATLASChamber}
E.~Etzion {\em et~al.}, ``{The Certification of ATLAS Thin Gap Chambers
  Produced in Israel and China},'' 2004.

\bibitem{Magboltz}
S.~Biagi, ``{Monte Carlo simulation of electron drift and diffusion in counting
  gases under the influence of electric and magnetic fields},'' {\em
  Nucl.Instrum.Meth.}, vol.~A421, pp.~234--240, 1999.

\bibitem{7097517}
C.~Ciofi~degli Atti {\em et~al.}, ``The readout system for the clas12
  micromegas vertex tracker,'' in {\em 2014 19th IEEE-NPSS Real Time
  Conference}, pp.~1--11, May 2014.

\bibitem{FThodo}
T.~C. Collaboration, ``Clas12 forward tagger (ft) technical design report.''
  \url{https://www.jlab.org/Hall-B/clas12-web/docs/ft-tdr.2.0.pdf}, 2012.
\newblock Online; accessed 29 January 2016.

\bibitem{PETIROC}
``Petiroc-2a.'' \url{http://www.weeroc.com/en/products/petiroc-2}.
\newblock Accessed: 2017-05-15.

\bibitem{Zorn:1992ew}
C.~Zorn, ``{A pedestrian's guide to radiation damage in plastic
  scintillators},'' {\em Nucl. Phys. Proc. Suppl.}, vol.~32, pp.~377--383,
  1993.

\bibitem{Qiang:2012zh}
Y.~Qiang, C.~Zorn, F.~Barbosa, and E.~Smith, ``{Radiation Hardness Tests of
  SiPMs for the JLab Hall D Barrel Calorimeter},'' {\em Nucl. Instrum. Meth.},
  vol.~A698, pp.~234--241, 2013.

\bibitem{Qiang:2013uwa}
Y.~Qiang, C.~Zorn, F.~Barbosa, and E.~Smith, ``{Neutron radiation hardness
  tests of SiPMs},'' {\em AIP Conf. Proc.}, vol.~1560, pp.~703--705, 2013.

\bibitem{commPETIROC}
S.~Ahmad. Private communication, May 2017.
\newblock Weeroc SAS.

\bibitem{Emi}
K.~Emi {\em et~al.}, ``{Study of a dE/ dx measurement and the gas-gain
  saturation by a prototype drift chamber for the BELLE-CDC},'' {\em Nucl.
  Instrum. Meth.}, vol.~A379, pp.~225--231, 1996.

\bibitem{beamBlocker}
S.~Stepanyan and A.~P. Freyberger. Private communication, May 2017.

\bibitem{Wiringa1995}
R.~B. Wiringa, V.~Stoks, and R.~Schiavilla, ``{An Accurate nucleon-nucleon
  potential with charge independence breaking},'' {\em Phys.Rev.}, vol.~C51,
  pp.~38--51, 1995.

\bibitem{Pudliner1997}
B.~S. Pudliner {\em et~al.}, ``{Quantum Monte Carlo calculations of nuclei with
  A <= 7},'' {\em Phys.Rev.}, vol.~C56, pp.~1720--1750, 1997.

\bibitem{Wiringa}
R.~B. Wiringa {\em {, Private communication}}.

\bibitem{Dupre:2011afa}
R.~Dupré, {\em {Quark Fragmentation and Hadron Formation in Nuclear Matter}}.
\newblock PhD thesis, Lyon, IPN, 2011.

\bibitem{CiofidegliAtti2012}
C.~Ciofi~degli Atti {\em , Private communication}.

\bibitem{Cosyn:2010ux}
W.~Cosyn and M.~Sargsian, ``{Final-state interactions in semi-inclusive deep
  inelastic scattering off the Deuteron},'' {\em Phys. Rev.}, vol.~C84,
  p.~014601, 2011.

\bibitem{Egiian:1994ey}
K.~Egiian, L.~Frankfurt, W.~R. Greenberg, G.~A. Miller, M.~Sargsian, and
  M.~Strikman, ``{Searching for color coherent effects at intermediate Q**2 via
  double scattering processes},'' {\em Nucl. Phys.}, vol.~A580, pp.~365--382,
  1994.

\bibitem{Frankfurt:1994kt}
L.~L. Frankfurt, W.~R. Greenberg, G.~A. Miller, M.~M. Sargsian, and M.~I.
  Strikman, ``{Color transparency effects in electron deuteron interactions at
  intermediate Q**2},'' {\em Z. Phys.}, vol.~A352, pp.~97--113, 1995.

\bibitem{Scopetta2016}
S.~Scopetta {\em , Private communication}.

\end{thebibliography}

\end{document}